\documentclass{aa}
\usepackage[varg]{txfonts}

\usepackage{graphicx}   
\usepackage{hyperref}
\usepackage{natbib} 
\bibpunct{(}{)}{;}{a}{}{,} 
\hypersetup{colorlinks=true,linkcolor=blue,citecolor=blue,filecolor=blue,urlcolor=blue}
\usepackage{orcidlink}

\newcommand{\orcid}[1]{\orcidlink{#1}}
\newcommand{\msun}{\mbox{$M_{\odot}$}}
\newcommand{\msunyr}{\mbox{$M_{\odot}\,\mathrm{yr^{-1}}$}}
\newcommand{\mstar}{\mbox{$M_{\star}$}}
\newcommand{\DeltaMfit}{\mbox{$\Delta \log M_{\star}$}}

\def\sphinxs{{\sc Sphinx}}
\def\sphinx{{\sc Sphinx$^{20}$}}
\def\bagpipes{{\sc Bagpipes}}
\def\prospector{{\sc Prospector}}
\def\cloudy{{\sc Cloudy}}
\def\HII{{\sc H ii}}

\definecolor{pink}{RGB}{255,51,153}

\definecolor{green}{RGB}{34,139,34}

\begin{document}

\title{Uncertainties in high-$z$ galaxy properties inferred from spectral energy distribution fittings using JWST NIRCam photometry}

\author{Jiyoung Choe\inst{1}
          \and
        Taysun Kimm\inst{1}\thanks{E-mail:tkimm@yonsei.ac.kr}\orcid{0000-0002-3950-3997}
          \and
        Harley Katz\inst{2,3}\orcid{0000-0003-1561-3814}
          \and
        Maxime Rey\inst{1}\orcid{0009-0007-7943-0378}
          \and
        Daniel Han\inst{1}\orcid{0000-0002-2624-3129}
          \and
        J. K. Jang\inst{1}
           \and
        Joki Rosdahl\inst{4}\orcid{0000-0002-7534-8314}
          }

\institute{Department of Astronomy, Yonsei University, 50 Yonsei-ro, Seodaemun-gu, Seoul 03722, Republic of Korea
\and
Department of Astronomy \& Astrophysics, University of Chicago, 5640 S Ellis Avenue, Chicago, IL 60637, USA
\and
Kavli Institute for Cosmological Physics, University of Chicago, Chicago IL 60637, USA
\and
Centre de Recherche Astrophysique de Lyon UMR5574, Univ Lyon, Univ Lyon1, Ens de Lyon, F-69230 Saint-Genis-Laval, France
}

\date{Received XXX; accepted YYY}

\abstract{ 
Numerous high-$z$ galaxies have recently been observed with the James Webb Space Telescope (JWST), providing new insights into early galaxy evolution. Their physical properties are typically derived through spectral energy distribution (SED) fitting, but the reliability of this approach remains uncertain owing to limited constraints on star formation histories (SFHs) and on the  contribution from emission for such early systems. Applying \bagpipes\ on simulated SEDs with $\mathrm{SFR}_{10}>0.3\,\msunyr$ at $z=6$ from the \sphinx\ cosmological simulation, we examine the uncertainties related to the recovery of stellar masses, star formation rates ($\mathrm{SFR}_{10}$), and stellar metallicities from mock JWST/Near-Infrared Camera photometry, spanning F115W--F444W. Even without dust or emission lines, fitting the intrinsic stellar continuum overestimates the stellar mass by about 60\%, on average (and by up to a factor of five for low-mass galaxies with recent starbursts). It also underestimates the $\mathrm{SFR}_{10}$ by a factor of 2, due to inaccurate SFHs and age–metallicity degeneracies. In full SED-fitting models that include dust attenuation and nebular emission, stellar mass estimates are primarily affected by age–metallicity degeneracy and emission lines. Short-term SFRs are most sensitive to dust attenuation and nebular emission, while long-term SFRs additionally depend on the assumed SFHs. Incorporating bands that are free of strong emission lines, such as F410M, helps mitigate stellar mass overestimation by disentangling line emission from older stellar populations. We also find that best fit or likelihood-weighted estimates are generally more accurate than median posterior values. Although stellar mass functions are reproduced reasonably well (particularly when the minimum-$\chi^2$ estimates are used), the slope of the main sequence of  star formation acutely depends  on the adopted fitting model. Overall, these results underscore the importance of careful modelling when interpreting high-$z$ photometry, particularly for galaxies with recent star formation burst and/or strong emission lines, to minimise systematic biases in derived physical properties.
}


\titlerunning{Uncertainties in SED fitting}

\maketitle

\section{Introduction}

Understanding how galaxies form and evolve at high redshift is vital for elucidating the earliest phases of cosmic history. After recombination, gas accumulates along the cosmic web of dark matter filaments, leading to the formation of the first stars and galaxies \citep[e.g.][]{Abel2002,Bromm2002}. These primordial systems mark the onset of galaxy formation and they drive the reionisation of the intergalactic medium by emitting
ionising photons \citep[e.g.][]{Dayal2018,Robertson2022}. As they evolve hierarchically through mergers and baryonic processes, these high-redshift galaxies preserve crucial information regarding the interplay among structure formation, feedback, and the thermal and ionisation states of the Universe \citep[e.g.][]{Somerville2015}. Consequently, examining their properties is fundamental to building a coherent picture of galaxy evolution across cosmic time.

Several approaches are typically employed to investigate the evolution of galaxies in these early stages, including semi-analytic models of galaxy formation \citep[e.g.][]{Yung2019,Mauerhofer2025} and analytic methods \citep[e.g.][]{Dekel2023}. In particular, recent advances in computational power and numerical methods have improved the fidelity of simulations modelling the interstellar medium (ISM) in greater detail \citep{Naab2017}. By incorporating strong stellar feedback, numerical simulations are able to successfully reproduce observed ultraviolet (UV) luminosity functions, stellar mass functions, and mass-metallicity relations at high redshift  \citep[e.g.][]{Ma2016,Rosdahl2018,Finlator2018,Kannan2020}. Such simulations provide theoretical frameworks for generating mock observations of high-$z$ galaxies, enabling more direct comparisons with existing data and enhancing our understanding of early galaxy formation \citep{Marshall2022,Katz2023}.

In parallel, photometric and spectroscopic data continue to provide essential observational constraints on the nature of high redshift galaxies. For instance, oxygen abundances can be derived from strong-line diagnostics and the direct $T_\mathrm{e}$ (electron temperature) method \citep{Curti2023,Nakajima2023,Chemerynska2024}, while star formation rates (SFRs) are inferred from H$\alpha$ or [CII] line luminosities \citep{delooze2014,Smit2018,Bethermin2020,Rinaldi2023Ha}. While such spectroscopic observations provide direct and reliable insights into galaxy properties, they require substantial observation time and resources. Therefore, as a more feasible alternative for large samples, spectral energy distribution (SED) fittings based on photometric data are widely employed. For instance, \citet{Song2016} performed SED fitting of ultraviolet (UV)-selected samples at $4<z<8$ and obtained UV luminosity functions and their stellar mass functions (SMFs). Building on this and using photometry spanning the rest-frame UV to near-infrared (NIR), \citet{Leung2025} investigated the observed 'little red dots' and demonstrated that including James Webb Space Telescope/Mid-Infrared Instrument (JWST/MIRI) bands and an active galactic nucleus (AGN) emission component substantially reduced stellar mass estimates, alleviating the apparent tension with respect to current cosmological models. \citet{Harvey2025} measured stellar masses for galaxies at $6.5<z<13.5$ using various  stellar population synthesis (SPS) models and initial mass functions (IMFs); they proposed that the low-mass slope of the SMFs steepens toward higher redshift. Collectively, these empirical constraints are central to understanding galaxy assembly and serve as critical benchmarks for validating theoretical models \citep[e.g.][]{Mauerhofer2025}.

Numerous publicly available SED fitting codes support this approach \citep{Carnall2018, Chevallard2016, daCunha2015, Leja2017}, enabling the estimation of posterior distributions for physical parameters such as stellar mass by fitting observed fluxes across multiple bandpasses. In a recent comparative analysis, \citet{Pacifici2023} observed that while stellar mass estimates are generally robust, the SFR and dust attenuation ($A_\mathrm{V}$) vary markedly with modelling assumptions, including the choice of star formation history (SFH), dust law, and nebular emission treatment. Among these factors, the assumed parametric SFH model proves particularly important for inferring the properties of quenched galaxies, as demonstrated using both mock and observed photometric datasets. In line with this approach, \citet{Carnall2018} reported that a double power-law SFH model yields an unbiased quenching timescale, whereas an exponentially declining SFH model underestimates both the stellar age and the quenching timescale. The reliability of the parametric SFH model was further tested through SED fitting of mock photometry derived from simulated galaxies \citep{Haskell2024}, revealing that such models typically underestimate the SFR of recently quenched systems by $\sim0.4$ dex, but they do tend to overestimate the SFR of starburst galaxies. The choice of stellar population synthesis template likewise influences stellar mass estimates. For instance, using different combinations of stellar population and dust emission models for galaxies from the Cosmic Evolution Survey \citep{Scoville2007}, \citet{Jones2022} reported that stellar masses are, on average, $0.14$ dex lower when the Binary Population and Spectral Synthesis (BPASS) model was employed, instead of the model introduced by \citet[][2016 version\footnote{This updated version is available at: \url{https://www.bruzual.org/bc03}.}]{Bruzual2003}. In a related study, \citet{Meldorf2024} investigated whether a flexible dust attenuation law could recover the input dust parameters. They found that the correlation between the true slope of the dust attenuation curve and the residual is difficult to eliminate, even with the inclusion of IR bands. The importance of adopting non-parametric or flexible SFH models has also been increasingly emphasised over the past few years. Although they are computationally intensive, such models achieve greater accuracy in recovering the true SFH and associated physical parameters, especially for galaxies with stochastic SFHs or bursty young stellar populations \citep{Lower2020, Jain2024}.

In previous studies, SED fittings conducted on high-$z$ galaxies were primarily confined to either highly luminous systems or those in lensed fields \citep{McLure2011,Laporte2017,Jolly2021}. However, with the successful launch of the James Webb Space Telescope (JWST), unprecedented volumes of data are now available at much fainter magnitudes. Nevertheless, as JWST observations accumulate, new challenges appear, with the most notable being the apparent excess of luminous galaxies \citep{Ferrara2023,Finkelstein2023,Carniani2024,Harikane2024,Sabti2024}. Current theoretical frameworks suggest that the observed number of UV-bright galaxies is too high to align with the classical picture of galaxy formation models \citep{Lovell2022}. A plausible explanation is that star formation in these galaxies proceeds in a highly bursty manner \citep[e.g.][]{Sun2023,Shen2023}. As spectroscopic confirmations of such galaxies increase \citep{Carniani2024}, the aforementioned tension with the canonical framework becomes more pronounced. Due to the high optical fluxes observed in these systems, several studies have either sought to attribute some of the emission to AGNs or refined the modelling of young stellar populations \citep{Leung2025, Harvey2025}. Nevertheless, quantifying the influence of AGNs and bursty star formation remains difficult, underscoring the importance of carefully assessing uncertainties in SED fitting. Recently, \citet{Cochrane2025} examined the impact of SED fitting on SMFs using mock JWST photometry of \sphinx\ galaxies \citep{Rosdahl2022,Katz2023}. They concluded that stellar masses of low-mass galaxies are consistently overestimated because of inadequate modelling of strong emission lines.

While accurately inferring galaxy properties is essential, the accuracy of SED fitting and its associated uncertainties in high-$z$ galaxies have received relatively little attention \citep[c.f.][]{Narayanan2024,Cochrane2025}. Galaxies formed during the epoch of reionisation are believed to differ notably from those in the local Universe, frequently displaying active and bursty star formation \citep{Caputi2017, Rinaldi2023, Dressler2023, Dressler2024, Hu2023}. Accordingly, the nebular emission is expected to contribute substantially to their total luminosity \citep{Wilkins2020, Esther2015}. Because these systems are at an early evolutionary stage, metallicity and dust extinction are generally low \citep{Traina2024, Heintz2023}, whereas the escape fraction of LyC radiation can be high \citep{Hayes2011,Kimm2017}. Determining whether SED fitting, which is validated primarily with local galaxies, can reliably recover the properties of high-$z$ systems is therefore crucial. To this end, we use mock observations from the SEDs of simulated galaxies at $z=6$ in the \sphinx\ cosmological simulation, using JWST/NIRCam photometry. Furthermore, we applied SED fitting with \bagpipes\ to evaluate the reliability of the derived physical quantities, such as stellar mass and SFR, and to assess how these affect the determination of the SMFs and the star-forming main sequence.

This paper is structured as follows. Section~\ref{sec:methods} briefly introduces the \sphinx\ simulation, outlines the construction of mock SEDs for simulated galaxies, and describes the fitting procedure with \bagpipes. Section~\ref{sec:results} presents an evaluation of the accuracy of galaxy property recovery under different fitting configurations and an investigation of the sources of associated uncertainties and biases. Section~\ref{sec:discussion} presents strategies for improving fitting accuracy and an assessment of discrepancies between the true and fitted statistical observables. Finally, Sect.~\ref{sec:summary} summarises our results and conclusions, while the appendix presents detailed fitting results.

\section{Methods}
\label{sec:methods}

To assess how emission lines influence observed photometry and introduce uncertainties in stellar mass and SFH estimations, we analysed mock spectra from the \sphinx\ cosmological simulation. In this section, we outline the construction of mock SEDs for each galaxy and the generation of photometric data for the JWST Near-Infrared Camera (NIRCam) bands. 

\subsection{Simulation}
\label{sec:simulation} 

The \sphinxs\ suite comprises cosmological radiation-hydrodynamic simulations designed to investigate galaxy properties during the epoch of reionisation \citep{Rosdahl2018,Rosdahl2022}. With a spatial resolution of 10.9 pc (physical) at $z=6$ and a dark matter resolution of $2.5\times10^5\,\msun$, the \sphinx simulation captures the evolution of roughly $32,000$ star-forming halos at $z=6$\footnote{The \sphinx\ simulation is run down to $z=4.64$, where it contains approximately $32,500$ star-forming halos.} and resolves galaxies hosted by dark matter halos with masses down to the atomic cooling limit ($M_\mathrm{vir}\sim 10^8\,\msun$). Dark matter halos are identified using the {\sc AdaptaHOP} halo finder \citep{Tweed2009} and star particles are assigned to the nearest (sub)halos. Galaxies are defined as the associations of gas cells and star particles within each (sub)halo. The large volume of the \sphinx\ $(20\,\mathrm{cMpc})^3$  allows an analysis of the photometric and spectroscopic properties of galaxies spanning stellar masses up to $\mstar\approx 10^{10}\,\msun$ \citep{Katz2023}. 

In \sphinx, gas cooling from hydrogen and helium is computed by solving non-equilibrium chemistry coupled with ionising radiation \citep{Rosdahl2013}, while metal and molecular cooling are incorporated, following the approach of \citet{Rosen1995}. As described by \citet{Kimm2017,Rosdahl2018}, stellar particles form when the gas collapses and becomes gravitationally unstable, with the star formation efficiency regulated by local turbulence. When a stellar particle with a mass as low as $m_\star=400\,\msun$ forms, ionising radiation is injected into its host cell by interpolating SEDs of different ages and metallicities from the BPASS model \citep[][version 2.2.1]{Stanway2018}. This radiation then propagates through the interstellar medium (ISM), heating nearby gas and transferring momentum to the absorbing cells \citep{Rosdahl2013}. After 4--50 Myr of stellar evolution, supernovae occur and enrich both the ISM and the circumgalactic medium \citep{Kimm2014,Kimm2015}. Additional details regarding the physical modelling in \sphinx\ are provided by \citet{Rosdahl2018,Rosdahl2022}.

\begin{figure}
\includegraphics[width=0.47\textwidth]{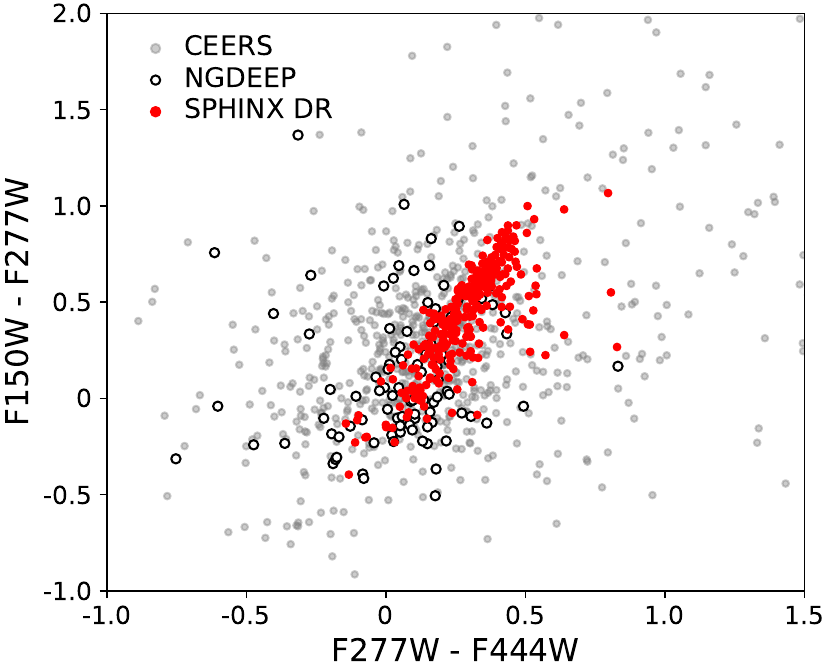}
    \caption{Observed NIRCam colour–colour distribution of galaxies at $5.8<z_\mathrm{phot}<6.2$ in the CEERS and NGDEEP fields (grey points and open circles), compared with the \sphinx\ DR $z=6$ galaxies used in our analysis (red points). The \sphinx\ galaxies trace the same region in the F150W-F277W versus F277W-F444W plane as the observed population, indicating that their broad-band colours are consistent with those of typical star-forming galaxies at $z\approx6$.}
    \label{fig:color-color}
\end{figure}

The \sphinxs\ simulations reproduce the UV and Ly$\alpha$ luminosity functions at $z\ge 6$ \citep{Garel2021}. To further demonstrate that the simulations provide a realistic sample whose photometric properties resemble those of galaxies at $z\approx6$, we compared the \sphinx\ galaxies with NIRCam-selected sources in the CEERS and NGDEEP fields. As shown in Fig.~\ref{fig:color-color}, the \sphinx\ DR galaxies occupy nearly the same region as the observed population in the F150W–F277W versus F277W–F444W colour–colour plane, indicating that their broad-band colours are representative of typical galaxies at this epoch.

\subsection{Mock SEDs}

\citet{Katz2023} post-processed 1,380 star-forming galaxies from \sphinx\ with $\mathrm{SFR_{10}}>0.3\,\msunyr$ at $4.64<z<10$, where $\mathrm{SFR_{10}}$ is the star formation rate (SFR) averaged over the last 10 Myr, deriving dust-attenuated SEDs for ten viewing angles per galaxy. The sample was restricted to post-processed galaxies that are likely to be observable with  JWST, adopting a limiting UV magnitude of $M_\mathrm{UV}\approx-17$ \citep{Katz2023,Rieke2023}. In particular, using the same IMF parameters as in the simulation --- a mass cutoff of 0.1-100\,\msun\ with slopes of -1.3 at low mass and -2.35 at high mass --- the authors first computed the intrinsic stellar continua with BPASS v2.2.1. They subsequently incorporated the contribution of 52 emission lines and the nebular continuum (free-free, free-bound, and two-photon) using \cloudy\ \citep[][v17.03]{Ferland2017}. These emission lines include strong lines such as Ly$\alpha$ 1216 \text{\AA}, [OIII] 4959/5007 \text{\AA}, and H$\alpha$ 6563 \text{\AA}, which strongly influence broad-band photometry \citep{Zackrisson2008,Schaerer2009}. For the H and He emission lines, recombinative and collisional rates were derived from non-equilibrium ionisation fractions directly extracted from the simulations when the local Stromgren sphere was resolved. Conversely, in under-resolved cells, the line luminosities were estimated from pre-computed \cloudy\ model grids under the assumption of spherical geometry. The resulting intrinsic SEDs were then propagated with the Monte Carlo radiative transfer code {\sc Rascas} \citep{Michel-Dansac2020}, which self-consistently models resonant and non-resonant line transfer together with absorption and scattering by Small Magellanic Cloud (SMC) type dust. The dust content was assumed to scale with the neutral hydrogen density, following the model proposed by \citet{Laursen_2009}. Importantly, the attenuation curve of a galaxy does not necessarily resemble the SMC curve, as it is also determined by the spatial distribution of dust relative to the stars and by the total dust content \citep{Narayanan2018, Trayford2019}.

\begin{figure*}
\sidecaption
  \includegraphics[width=12cm]{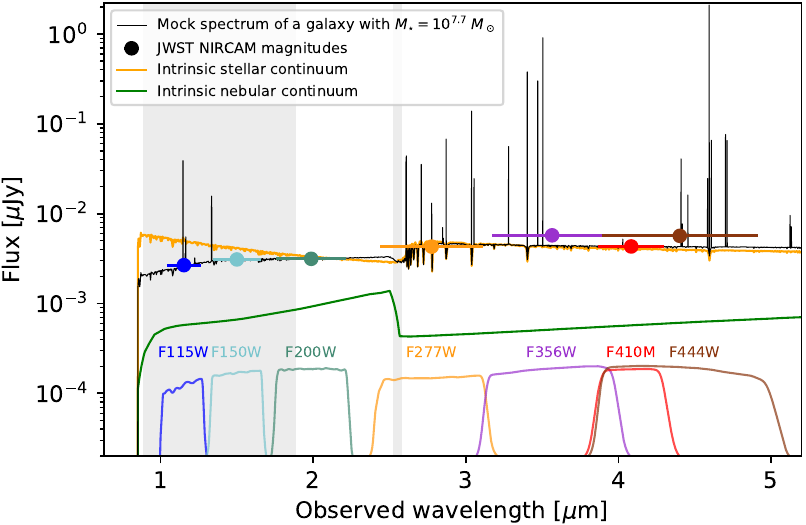}
    \caption{Example SED of a \sphinx\ galaxy at $z=6$. The black line represents the attenuated spectrum along a random line of sight. Coloured points mark the pivot wavelengths and bandwidths of the seven JWST NIRCam filters, with corresponding throughput curves shown below. The orange and green lines indicate the intrinsic stellar and nebular continua, respectively. The shaded regions indicate the wavelength range used to measure the UV slope ($\beta$) and the position of the Balmer break, respectively.}
    \label{fig:sed_ex}
\end{figure*}

Figure~\ref{fig:sed_ex} presents an SED of a \sphinx\ dwarf galaxy with $\mstar=10^{7.7}\,\msun$ (black line). The solid orange and yellow lines denote contributions from the intrinsic stellar and nebular continua, respectively, while the corresponding magnitudes in the JWST NIRCam filter system are plotted as filled circles with error bars. We find that dust attenuation notably reddens the UV slope relative to the intrinsic stellar continuum \citep[e.g.][]{Katz2023}. The nebular continuum further modifies the SED \citep[e.g.][]{Katz2024,Narayanan2025}, especially near the Balmer break, and can become bright enough to surpass the stellar continuum depending on the galaxy's star formation history. Moreover, certain band magnitudes (e.g. F356W) are occasionally brighter than the continuum, suggesting that emission lines can significantly influence SED fitting.

To ensure both a sufficiently high redshift and an adequate sample size, we based our analysis on all 276 galaxies in the \sphinx\ data release at $z=6$\footnote{We also examined SED-fitting uncertainties for 66 galaxies at $z\approx 9$ and confirmed that our conclusions remain unchanged (Sect.~\ref{Sec:discussion_medium_band}).}, each represented by ten spectra derived from ten distinct viewing angles. The stellar masses and mass-weighted stellar metallicities of these galaxies span $10^{6.6}$--$10^{9.8}\msun$ and $0.002 - 0.309\, Z_\odot$, respectively, where the solar metallicity is $Z_\odot=0.02$. For the photometric coverage, we employed six JWST NIRCam filters (F115W, F150W, F200W, F277W, F356W, and F444W), which have also been used in the Next Generation Deep Extragalactic Exploratory Public (NGDEEP) survey \citep{Bagley2024}. At $z=6$, the filters cover the rest-frame far-UV to optical wavelength range of galaxies. The role of the medium-band filter is addressed in a later section.

\subsection{SED fitting} 

We employed the widely used Bayesian inference code \bagpipes\ \citep{Carnall2018} to infer galaxy properties from SED fitting. \bagpipes\ utilises the MultiNest sampling algorithm \citep{Feroz2008,Feroz2009}, which efficiently explores a wide range of parameter spaces. In the following, we outline the input models for SFH, nebular emission, and dust adopted in the SED modelling.

We adopted the BPASS v2.2.1 stellar population synthesis model coupled with the \citet{Kroupa2002} IMF. This template is the same as that employed to generate the mock simulated SEDs in \citet{Katz2023}\footnote{The IMF may deviate from the canonical Kroupa form in high-redshift, metal-poor galaxies \citep[e.g.][]{Cameron2024,Kroupa2026}. For example, adopting a top-heavy IMF would lower the inferred stellar masses by approximately 0.5 dex \citep{Harvey2025}. While a qualitative exploration of IMF variations is beyond the scope of this study, a systematic assessment of their impact is warranted and is left to future work.}. For the SFHs, we explored two common parametric models (constant and double-power law), a flexible non-parametric model, and the normalised true SFHs. The constant SFH model prescribes a fixed SFR over a defined interval, with the SED shape governed by the initial and final stellar ages. Although straightforward and intuitive, the constant SFH cannot capture rising or declining trends, which are more effectively described by the double-power law,  
\begin{equation}
    \mathrm{SFR}(t) \propto \left[\left(\frac{t}{\tau}\right)^\alpha+\left(\frac{t}{\tau}\right)^{-\beta}\right]^{-1},
        \label{eq:dbplw}
\end{equation}
where $\tau$ denotes the time of peak star formation and $\alpha>0$ and $\beta>0$ represent free parameters defining the rising and declining phases, respectively. For the flexible model, we adopted the method of \citet{Leja2019}, which prescribes a multi-component SFH with a constant SFR in each time bin. Rather than permitting fully unconstrained SFRs, we imposed a continuity prior defined as $x=\mathrm{log(SFR_n/SFR_{n+1})}$, which follows a Student's-t distribution, expressed as
\begin{equation}
    \mathrm{PDF}(x,\nu)=\frac{\Gamma(\frac{\nu+1}{2})}{\sqrt{\nu\pi}\space\Gamma(\frac{\nu}{2})}\left(1+\frac{(x/\sigma)^2}{\nu}\right)^{-\frac{\nu+1}{2}}.
        \label{eq:flexible}
\end{equation}
Here, $\Gamma$ denotes the gamma function, $\sigma$ represents the scale parameter, and $\nu$ is a parameter controlling the tail width of the distribution. We adopted $\sigma=0.3$ and $\nu=2$, following \citet{Leja2019}. The SFH was discretised into seven age bins, with the first bin covering 0--10 Myr. The remaining bins are evenly spaced in log time from 10 Myr to $t_\mathrm{max}-100$ Myr, where $t_\mathrm{max}$ denotes the age of the universe. The normalised true SFH model employs the intrinsic SFH obtained directly from the simulated galaxy and normalised by the total stellar mass. For all SFH models, we adopted uniform priors for the stellar mass formed ($M_{\star}$) and stellar metallicity ($Z_\star$), spanning $[10,\space10^{13}]\space\msun$ and $[0.001,\space1]\space Z_\odot$, respectively. We can use
\bagpipes\ to model the nebular continuum and line emission with the precomputed grids provided by \citet{Byler2017}. These authors employed the photoionisation code \cloudy\ to compute the nebular emission from \HII\ regions using input spectra with varying simple stellar population ages and ionisation parameter, $U$, under the assumption that all hydrogen-ionising photons from young stars end up absorbed. We assumed that stars younger than 10 Myr contribute to nebular emission. We confirm that treating this age threshold as a free parameter has a negligible impact on the fitting results. The prior range for $\log U$ was selected as $[-4,-1]$. 

For the dust model, we adopted three different attenuation laws: SMC \citep{Gordon2003}, Calzetti \citep{Calzetti2000}, and Salim \citep{Salim2018}. Each law is characterised by $A_{\lambda}/A_{V}$, with $A_{\lambda}$ denoting the attenuation at wavelength $\lambda$. The SMC and Calzetti laws can be widely applied to high-$z$ galaxies, which are generally metal-poor or actively star-forming. In contrast to the SMC and Calzetti laws, which fix the slope ($A_{\lambda}/A_{V}$) for each $\lambda$, the Salim dust law introduces a tunable slope through the parameter $\delta$, as
\begin{equation}
(A_\lambda/A_V)_\mathrm{Sal}=\left(\frac{\lambda}{5500\,\text{\AA}}\right)^\delta(A_\lambda/A_V)_\mathrm{Cal}+D_\lambda,
        \label{eq:Eq2}
\end{equation}
where $(A_\lambda/A_V)_\mathrm{Sal}$ and $(A_\lambda/A_V)_\mathrm{Cal}$ represent the slopes of the Salim and Calzetti laws, respectively. Furthermore, $D_\lambda$ regulates the strength of the UV bump at $2175\,\text{\AA}$, which is set to zero in this study. The Salim law simplifies to the Calzetti law when $\delta=0$, while positive (negative) $\delta$ values yield a flatter (steeper) slope. For all dust models, we used a uniform prior of $[0,2]$ for $A_V$. 
For the Salim law, we impose a Gaussian prior on $\delta$ with mean of $\mu=0.35$ and standard deviation of $\sigma=0.446$, motivated by the actual $\delta$ distribution of the simulated \sphinx\ galaxies.

During the SED fitting, we fixed the redshift to the true values of the sample galaxies to minimise any uncertainties arising from the redshift estimation. We assumed a signal-to-noise ratio of S/N=10 for all filter magnitudes, consistent with the approach of \citet{Narayanan2024}\footnote{We  verified that our inferred physical quantities are largely insensitive to this assumption: adopting the NGDEEP limiting magnitudes, as well as the NGDEEP limiting magnitudes combined with CEERS Poisson noise, yields consistent results.}. Throughout this study, the fitted value of each parameter is considered as the median of its posterior distribution. Additional uncertainties arising from redshift determination and from the effects of the summary statistic are discussed separately in Appendix~\ref{sec:appendix_C} and Sect.~\ref{sec:discussion}.

\begin{table*}
\caption{Description of the fitting models.}
\label{tab:fitting_sets}
\centering
\begin{tabular}{lcccc}
\hline \hline
Fitting label & $Z_\mathrm{\star,fit}$ & Input SED & Model SED & Remark\\      
        \hline  
        \texttt{intS\_Z$_\mathrm{true}$} & $Z_\mathrm{\star,true}$ & intrinsic S & intrinsic S \\
        \texttt{intS\_Z} & free param & intrinsic S & intrinsic S \\
        \texttt{attS\_Z$_\mathrm{true}$} & $Z_\mathrm{\star,true}$ & attenuated S & attenuated S \\
        \texttt{attS\_Z} & free param & attenuated S & attenuated S \\
        \texttt{attSNE\_Z$_\mathrm{true}$} & $Z_\mathrm{\star,true}$ & attenuated S+N+E & attenuated S+N+E \\
        \texttt{attSNE\_Z} & free param & attenuated S+N+E & attenuated S+N+E \\ 
        \hline
        \texttt{attSNE\_Z\_F410M} & free param & attenuated S+N+E & attenuated S+N+E & F410M added\\
        \texttt{attSN\_Z\_SNE} & free param & attenuated S+N & attenuated S+N+E & no E in input SED \\
        \texttt{attSNE\_Z\_SN} & free param & attenuated S+N+E & attenuated S+N & no E in model SED \\  
        \hline
 \end{tabular}
 \tablefoot{Summary of fitting models. From left to right, the columns specify the model name, stellar metallicity adopted in the fit, input SED obtained from \sphinx, and the model SED used in \bagpipes. $Z_{\star,\mathrm{fit}}$ is set to the mass-weighted true stellar metallicity ($Z_{\star,\mathrm{true}}$) unless treated as a free parameter. S, N, and E indicate the stellar continuum, nebular continuum, and nebular emission lines, respectively. The \texttt{attSNE\_Z\_F410M} model additionally incorporates medium-band photometry from the F410M filter. The SMC dust law is adopted as the default, with the Salim and Calzetti laws considered as alternatives.}
\end{table*}

\begin{figure*}
    \centering
\includegraphics[width=\textwidth]{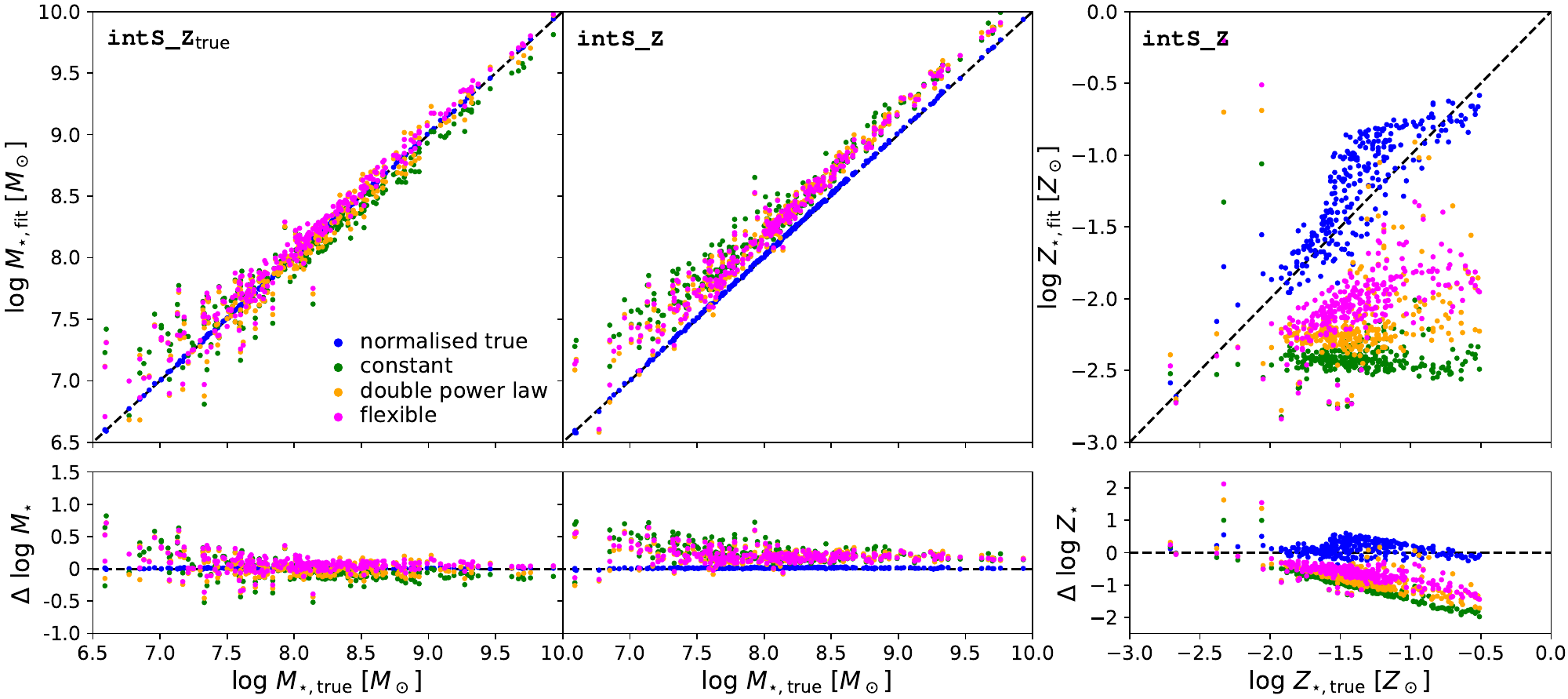}
    \caption{Comparison of the true and derived stellar masses with (left; \texttt{intS\_Z$_\mathrm{true}$}) and without (right; \texttt{intS\_Z}) fixing the stellar metallicity. Different SFH models are shown in different colours, as indicated in the legend. The right panel also displays the fitted $Z_\star$ values for each SFH model, while the lower panels show the differences relative to the true quantities. When metallicity is treated as a free parameter, $M_\star$ is overestimated by 0.2 dex for both parametric and non-parametric SFH models. This bias reflects the age--metallicity degeneracy, which tends to yield posteriors with underestimated metallicity and overly high fractions of old stars relative to the true values.}
    \label{fig:intr_stellar}
\end{figure*}

\section{Results}
\label{sec:results}

This study quantifies how different physical assumptions in SED fitting-such as SFH, dust attenuation, and nebular emission-affect uncertainties in the recovered properties of high-$z$ galaxies. We systematically evaluated the influence of each component on the accuracy and bias of SED-derived quantities, with a particular focus on the sparse photometric coverage typical of broad-band surveys. We first examined how model SFHs and stellar metallicity affect the posterior distributions of stellar mass and SFR, disregarding nebular emission and dust attenuation. We then gradually added further model components to assess their impact on the fitting results. Finally, we incorporated the nebular emission and quantitatively compared the inferred values with the simulated ones. The fitting sets and their configurations are summarised in Table~\ref{tab:fitting_sets}.

\subsection{Impact of the model SFH and metallicity}
\label{sec:3.1}

A long-standing challenge in stellar population studies is the age-metallicity degeneracy \citep[e.g.][]{Worthey1994}, wherein older stellar ages and higher metallicities yield similar broad-band colours, making them difficult to disentangle using photometry alone. Previous studies have shown that this degeneracy can be partly alleviated by incorporating spectroscopic features such as absorption lines \citep[e.g.][]{Gallazzi2005, Conroy2013}. However, in the absence of such spectral information, quantifying how this degeneracy influences stellar properties inferred from SED fitting remains essential.

\begin{figure*}
\includegraphics[width=0.476\linewidth]{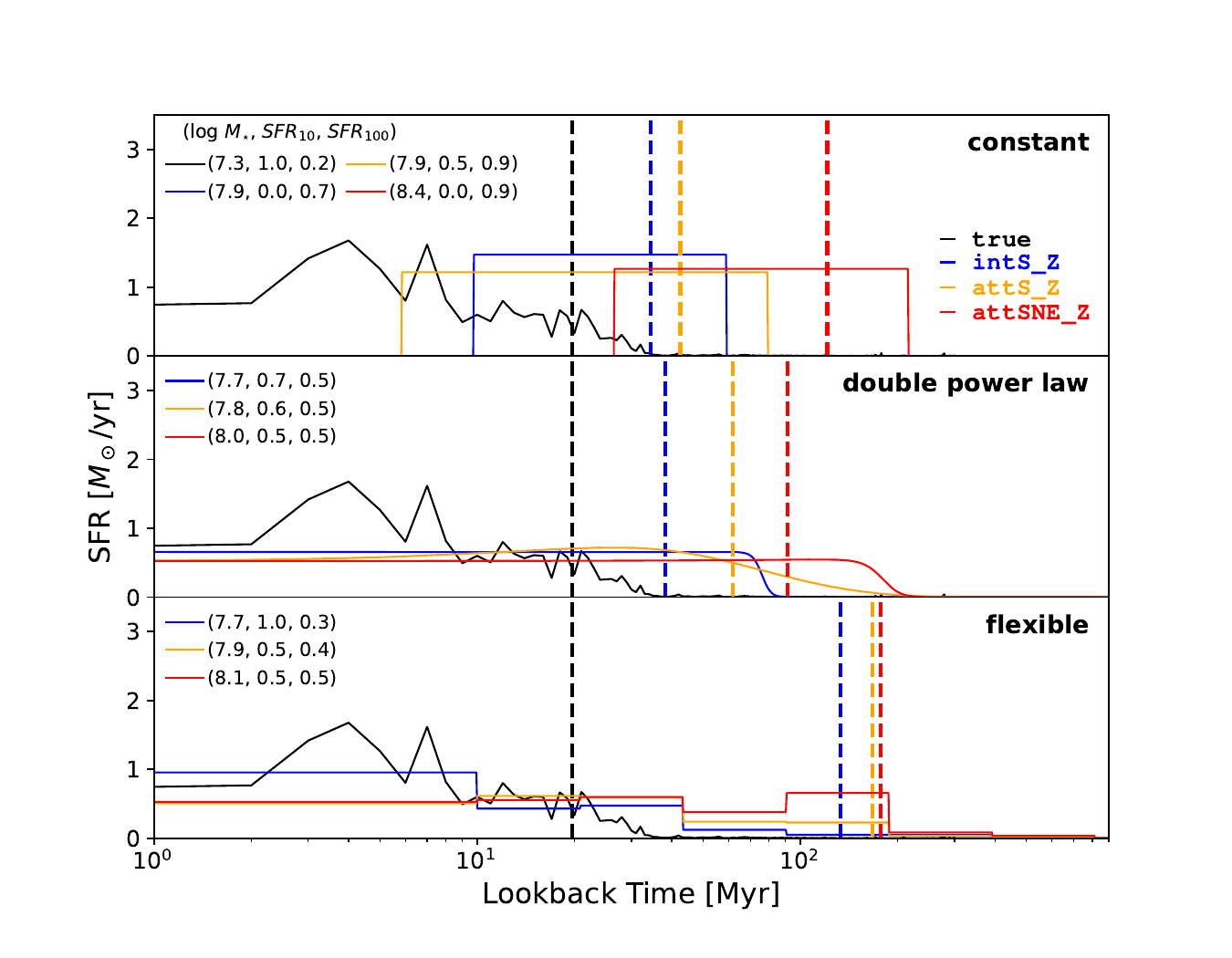}
\includegraphics[width=0.49\linewidth]{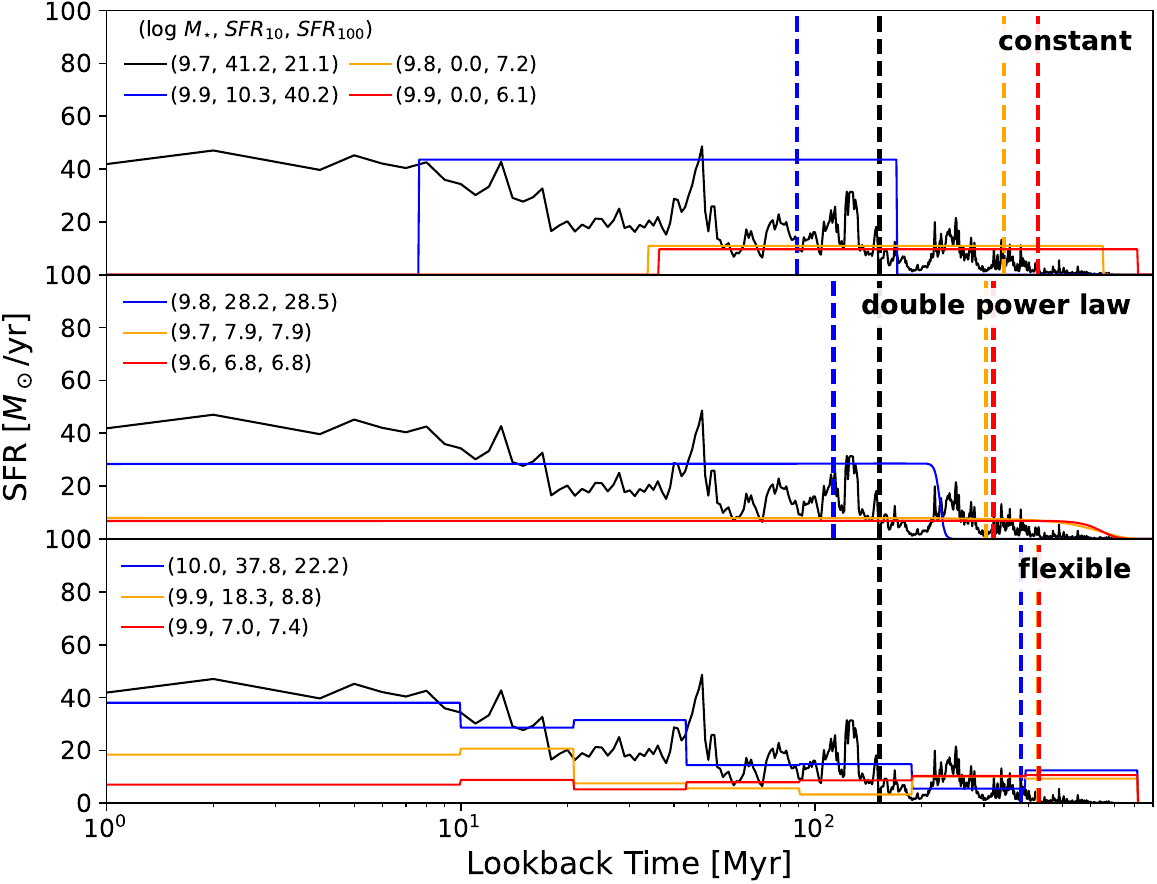}
    \caption{Posterior SFHs corresponding to the median SFR$_\mathrm{100}$ values are displayed together with the true SFH (solid black line) for two representative galaxies: one low-mass ($\log\,M_\star/M_\odot=7.32$, left panels) and one high-mass ($\log\,M_\star/M_\odot=9.67$, right panels). Each row corresponds to a different SFH model used in the fitting, and each coloured line denotes a separate fitting configuration (blue: \texttt{intS\_Z}, orange: \texttt{attS\_Z}, red: \texttt{attSNE\_Z}), with the SMC attenuation law applied when a dust model is included. Dashed vertical lines indicate the mass-weighted age. Under \texttt{intS\_Z}, the constant SFH model generates the fewest young stars (< 10 Myr) among all models and yields the lowest mass-weighted age. Consequently, its strongest underestimation is SFR$_\mathrm{10}$ and it also overestimates SFR$_\mathrm{100}$, as most of the stellar mass is concentrated in intermediate-age populations. When dust and nebular components are added, all three SFH models predict larger old stellar populations, increasing the mass-weighted age and leading to an underestimation of SFR$_\mathrm{100}$, particularly in high-mass galaxies.}
    \label{fig:sfh_median}
\end{figure*}

\subsubsection{Overestimation of stellar masses in the presence of recent starbursts}

We began by fitting photometric data derived from the intrinsic stellar continua of simulated galaxies at $z = 6$ to evaluate how SFHs and metallicities affect inferred galaxy properties. For each SFH model, the SED fitting was performed twice: once with $Z_\mathrm{\star,true}$, the true mass-weighted stellar metallicity (\texttt{intS\_Z$_\mathrm{true}$}) and once with metallicity treated as a free parameter (\texttt{intS\_Z}). Both fits assumed a single metallicity for all stellar populations within each galaxy. Figure~\ref{fig:intr_stellar} illustrates the offsets between the fitted and true $M_\star$ and $Z_\star$. When stellar metallicity is fixed (\texttt{intS\_Z$_\mathrm{true}$}), the model with the normalised true SFH recovers $M_\star$ almost perfectly\footnote{If an alternative input SED is employed in the fitting, such as \citet{Bruzual2003}, the stellar mass is overestimated by 0.04 dex because the BPASS SED generates greater optical flux (see Appendix~\ref{sec:appendix_A}).}, whereas the constant, double power law, and flexible SFH models show deviations of 0.09, 0.07, and 0.08 dex, respectively. The magnitude of overestimation is closely tied to the contribution from young stellar populations. For instance, in SEDs with recent star formation bursts, the blue spectrum can be reproduced either by lowering the contribution of old stars or by boosting the numbers of both old stars and young stars together. Because the posterior distribution favours older ages owing to the larger number of possible combinations, fits to SEDs with recent SF bursts frequently overestimate the stellar mass. Under the adopted selection criterion in the \sphinx\ data release (i.e. SFR$_\mathrm{10}$ $\geq$ 0.3 $M_\odot$/yr), our low-mass galaxies are typically dominated by young populations, resulting in intrinsically blue SEDs with UV slopes as steep as $-2.77$ (without nebular emission). This can drive stellar mass overestimates of up to an order of magnitude. Conversely, in galaxies with substantial old stellar populations, the constant SFH model tends to underestimate \mstar\ because the fit is biased toward the luminosity-weighted age, which is lower than the mass-weighted age. These biases do not occur when using the normalised true SFH is applied, as the ratios of young to old stars are fixed across all posteriors. Accurately capturing recent star formation bursts is therefore crucial for recovering stellar masses in dwarf galaxies with diverse SFHs.

\subsubsection{Underestimation of stellar metallicities in parametric fitting models}

When stellar metallicity is instead treated as a free parameter (\texttt{intS\_Z}), the model with the normalised true SFH still recovers $M_\star$ almost perfectly and $Z_{\star}$ with reasonable accuracy. The fitted $Z_{\star}$ tends to exceed the true value at intermediate metallicities ($\log\, Z_\star/Z_\odot\sim-1.3$), although the offset is not severe. This upward bias reflects the posterior's preference for young, metal-enriched populations that dominate the UV fluxes. Because not all stars in the simulation share the same $Z_{\star}$, the fitting outcome depends jointly on the SFH and the stellar metallicity distribution.

For the parametric SFH models, the age--metallicity degeneracy significantly impacts the posterior distribution, since  lowering $Z_{\star}$ or increasing the mass of young stars does end up yielding similarly blue SEDs. Our findings indicate that the photometric properties of the simulated galaxies are frequently reproduced by reducing $Z_\star$, while simultaneously enhancing the contribution from old populations, which, in turn, elevates $M_\star$. This outcome reflects the stronger degeneracy in SED modelling at low metallicities. Specifically, low-$Z_{\star}$ SEDs permit a wider range of mass ratios between young and old stars, whereas high-$Z_{\star}$ models necessitate the inclusion of young populations to offset redder colours. Consequently, the posterior distributions display a systematic bias towards low metallicities, with the constant SFH model often predicting values near the lower limit of the prior range (upper right panel of Fig.~\ref{fig:intr_stellar}). Although this effect is weaker in the double-power law and non-parametric SFH models, high median offsets of 0.78 and 0.6 dex, respectively, are observed, indicating that the fitted metallicities are unreliable.

\begin{figure}
\includegraphics[width=\linewidth]{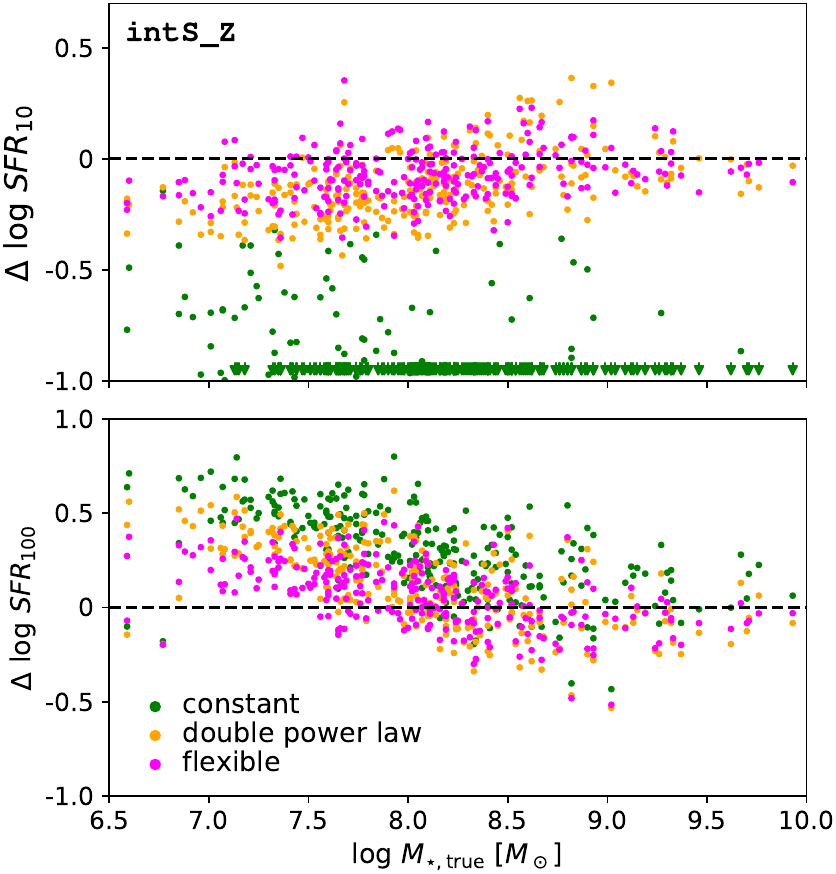}
    \caption{Difference between the true and recovered SFRs from the intrinsic SEDs as a function of true $M_\star$ (\texttt{intS\_Z}). The upper and lower panels present SFRs averaged over 10 Myr (SFR$_\mathrm{10}$) and 100 Myr (SFR$_\mathrm{100}$), respectively. In the fitting, stellar metallicity is treated as a free parameter. The constant SFH model substantially underestimates SFR$_\mathrm{10}$ but overestimates SFR$_\mathrm{100}$, whereas the double power-law and flexible models recover SFRs to within $\sim$0.5 dex.}
    \label{fig:sfr_intr}
\end{figure}

\begin{figure*}
    \centering
\includegraphics[width=\textwidth]{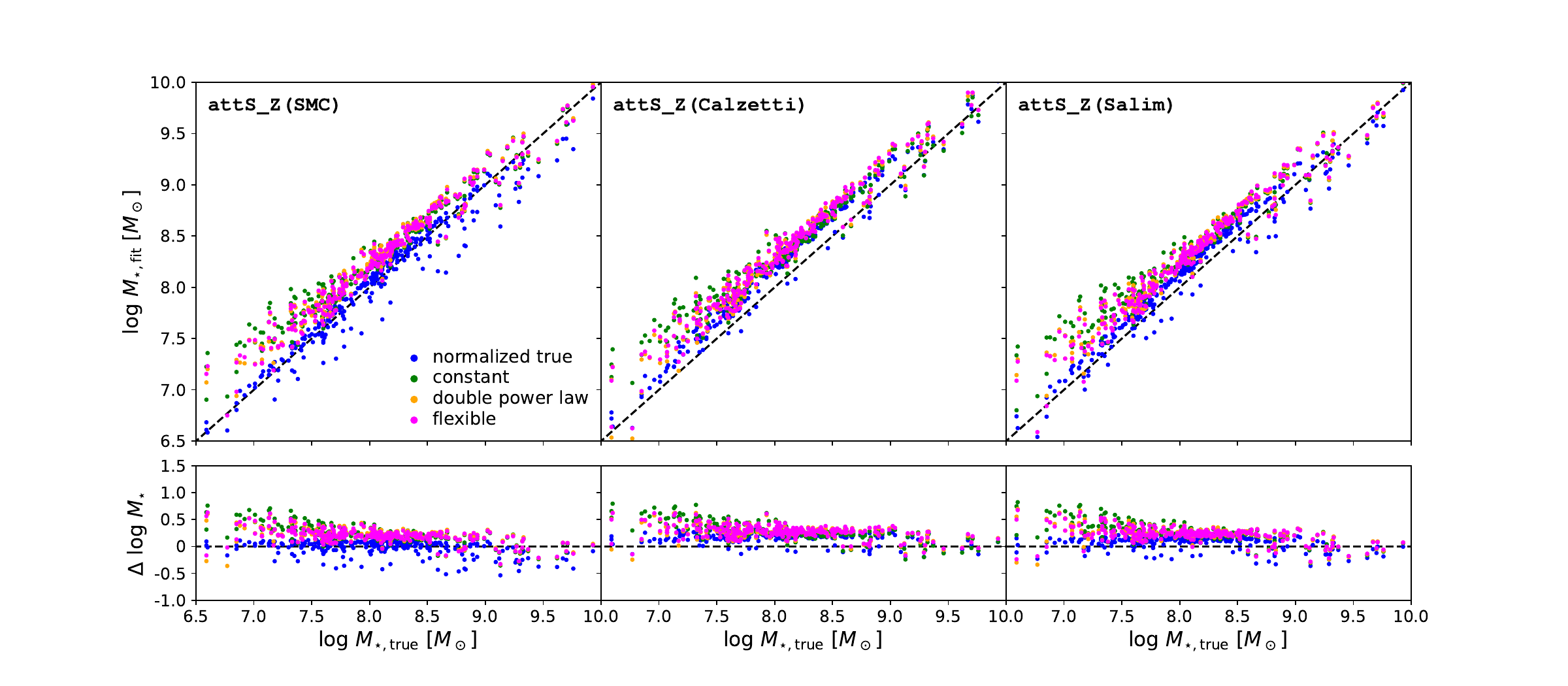}
    \caption{Fitted $M_\star$ and its offset from the true value for fits to the attenuated stellar continuum, with metallicity treated as a free parameter and three dust models applied: SMC (left), Calzetti (middle), and Salim (right). Considerable offsets in $M_\star$ persist even when the normalised true SFH is used, driven mainly by slope mismatches between the true normalised attenuation curve and the assumed model curve. Parametric and non-parametric SFH models yield comparable results. Across all four SFH models, the SMC law yields the closest agreement in $M_\star$, whereas the Calzetti law produces the largest offset.}
    \label{fig:att_stellar}
\end{figure*}

\subsubsection{Underestimation of short-term SFRs and overestimation in long-term SFRs by recent starbursts}

Our analysis also reveals notable biases in SFR$_\mathrm{100}$; for instance, the star formation rate averaged over the past 100 Myr, with its values overestimated when recent starbursts are present and underestimated when they are absent. Figure~\ref{fig:sfh_median} presents the posterior SFHs for two representative galaxies, with stellar masses of $\log M_{\star}/M_\odot=7.32$ (left panels) and $9.67$ (right panels). In each panel, the blue, orange and red curves correspond to posterior samples from the \texttt{intS\_Z}, \texttt{attS\_Z} and \texttt{attSNE\_Z} fits, respectively, selected to have SFR$_\mathrm{100}$ values near the median of their distributions. The constant SFH model (top panels) with \texttt{intS\_Z} (blue lines) predicts the smallest stellar mass formed in the last 10 Myr and the lowest contribution from stars older than 100 Myr, concentrating most of the stellar mass in intermediate-age populations. Consequently, $\mathrm{SFR}_\mathrm{100}$ is the highest among all SFH models. The double power-law and flexible SFH models similarly underestimate $\mathrm{SFR}_\mathrm{10}$, while overestimating $\mathrm{SFR}_\mathrm{100}$ in this low-mass galaxy (and generally in the low-mass galaxies in our sample; left panels). For the more massive galaxy (right panels), the posteriors contain larger fractions of old stellar populations, occasionally leading to an underestimation of $\mathrm{SFR}_\mathrm{100}$ relative to the true value. This effect becomes more pronounced when dust and nebular components are included.

Figure~\ref{fig:sfr_intr} illustrates the differences between the true and recovered $\mathrm{SFR}_\mathrm{10}$ and $\mathrm{SFR}_\mathrm{100}$ across the full sample. $\mathrm{SFR}_\mathrm{10}$ is consistently underestimated in all models, with the largest bias observed in the constant SFH model, where many galaxies are fitted with values of zero. This bias arises because the posteriors fail to reproduce the contribution from young stellar populations, while overproducing intermediate-to-old populations. In contrast, $\mathrm{SFR}_\mathrm{100}$ exhibits a clear mass-dependent pattern: it is overestimated in low-mass galaxies, but this bias weakens with increasing mass and can even turn into underestimation in some high-mass systems. This outcome reflects the shift in the dominant stellar population age with increasing mass. In low-mass galaxies, where most stars form within the last 100 Myr, any overestimation of stellar mass directly translates into an overestimation of $\mathrm{SFR}_\mathrm{100}$. Conversely, in massive galaxies, the low fraction of young stars leads SED fitting to underpredict recent star formation, resulting in lower inferred values of $\mathrm{SFR}_\mathrm{100}$.

\subsection{Impact of dust attenuation}
\label{sec:dust_impact}

The age-metallicity-dust degeneracy is well known to complicate reliable inference of galaxy properties from SED fitting \citep[e.g.][]{Papovich2001,Csizi2024}. Although far-IR photometric data can help mitigate this effect by constraining dust content through the IR excess, considerable degeneracy in dust parameters may remain \citep[e.g.][]{Qin2022}. To isolate the effect of dust attenuation on stellar property derivations, especially in the absence of IR data, we used the attenuated stellar continuum from \sphinx\ and performed an SED fitting with three widely used dust laws: SMC \citep{Gordon2003}, Calzetti \citep{Calzetti2000}, and Salim \citep{Salim2018}. The stellar metallicity is again treated as a free parameter during the fitting.

Figure~\ref{fig:att_stellar} displays the fitted versus true values of $M_\star$. Offsets remain evident even when the normalised true SFH model is applied. The smallest median offset is obtained with the SMC law ($\Delta \log M_\star \equiv \log M_{\star,\mathrm{fit}}-\log M_{\star,\mathrm{true}} \approx 0.04$ dex), whereas larger offsets arise with the Salim ($0.12$ dex) and Calzetti laws ($0.19$ dex). A similar trend appears across the other three SFH models, yielding combined offsets of 0.20, 0.24, and 0.27 dex for the SMC, Salim, and Calzetti laws, respectively. In what follows, we first examine how the dust attenuation curve affects the SED fitting results and its influence in conjunction with different SFH models.

\begin{figure}
\includegraphics[width=0.47\textwidth]{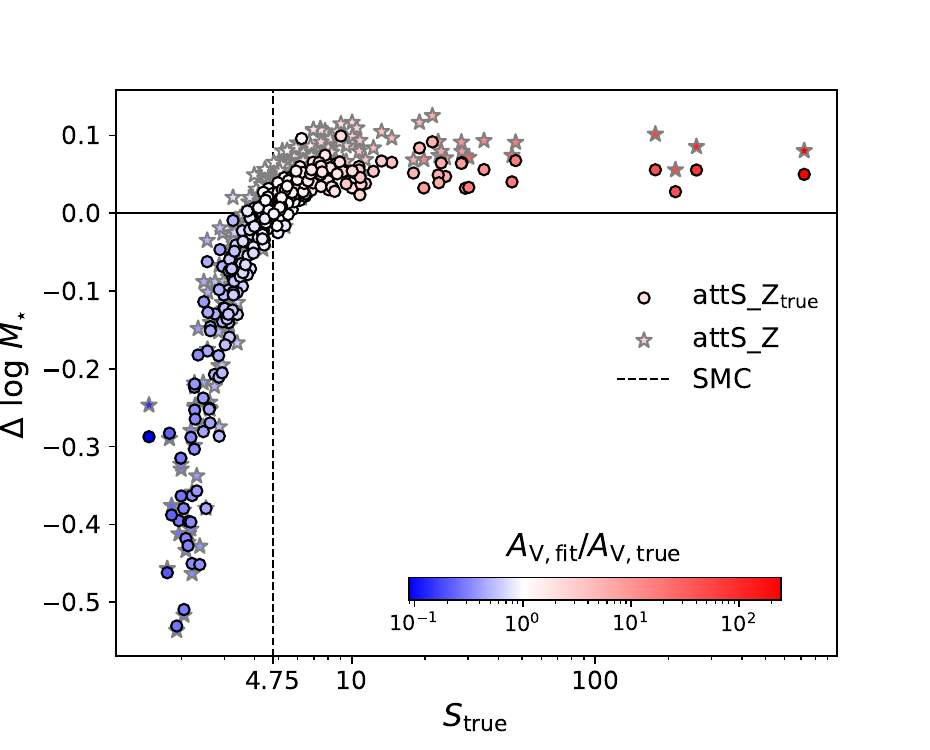}
    \caption{Relation between the true UV--optical slope $S$ ($\equiv A_\mathrm{1500}/A_\mathrm{V}$) and the stellar mass offset from the true value, derived by fitting the normalised true SFH with the SMC-type dust law under \texttt{attS\_Z$_\mathrm{true}$} and \texttt{attS\_Z}. The degree of over- or underestimation in $M_\star$ and $A_\mathrm{V}$ depends on the steepness of the true normalised attenuation curve relative to the model curve. This behaviour persists even when metallicity is treated as a free parameter, although the fitted values of $A_\mathrm{V}$ and $M_\star$ increase slightly.}
    \label{fig:S-Av}
\end{figure}

\subsubsection{Biases in stellar mass and dust attenuation driven by attenuation-curve slope mismatches}

To isolate the effect of the dust attenuation curve, we first examine the case with the 'normalised true SFH'. The influence of alternative SFH models under different dust laws is addressed in the next sub-section.

Figure~\ref{fig:att_stellar} (left-most panel, blue circles) illustrates that while most galaxies fitted with the SMC-type dust law exhibit a slight overestimation of $M_\star$, a sub-set of galaxies (particularly the more massive ones) display significant underestimations, with $\Delta \log M_{\star} \approx -0.6$ dex. This discrepancy mainly arises from differences in the attenuation curve slope between the simulated galaxies and the assumed SMC-type law during SED fitting, reflecting variations in dust geometry. This effect is further depicted in Fig.~\ref{fig:S-Av}, which presents the ratio of fitted to true $A_V$ and the stellar mass offset as functions of the true UV--optical slope, defined as $S\equiv A_\mathrm{1500}/A_V$, where $A_\mathrm{1500}$ and $A_V$ are the attenuations at $1500\,\text{\AA}$ and $5500\,\text{\AA}$, respectively. Notably, $S$ values greater than 4.75 indicate stronger UV attenuation than the SMC law, corresponding to a steeper attenuation curve. Our findings indicate that when the attenuation curve is steeper than the SMC, both $A_V$ and $M_\star$ are overestimated, whereas a shallower curve leads to underestimation.

\begin{figure*}
    \centering
\includegraphics[width=\textwidth]{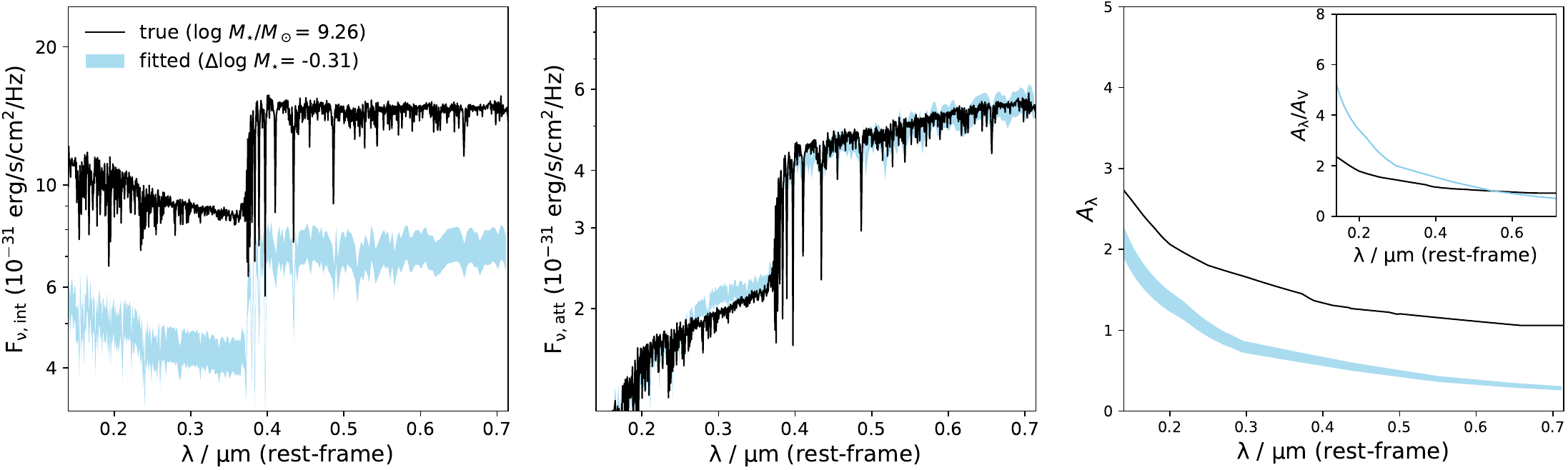}
\includegraphics[width=\textwidth]{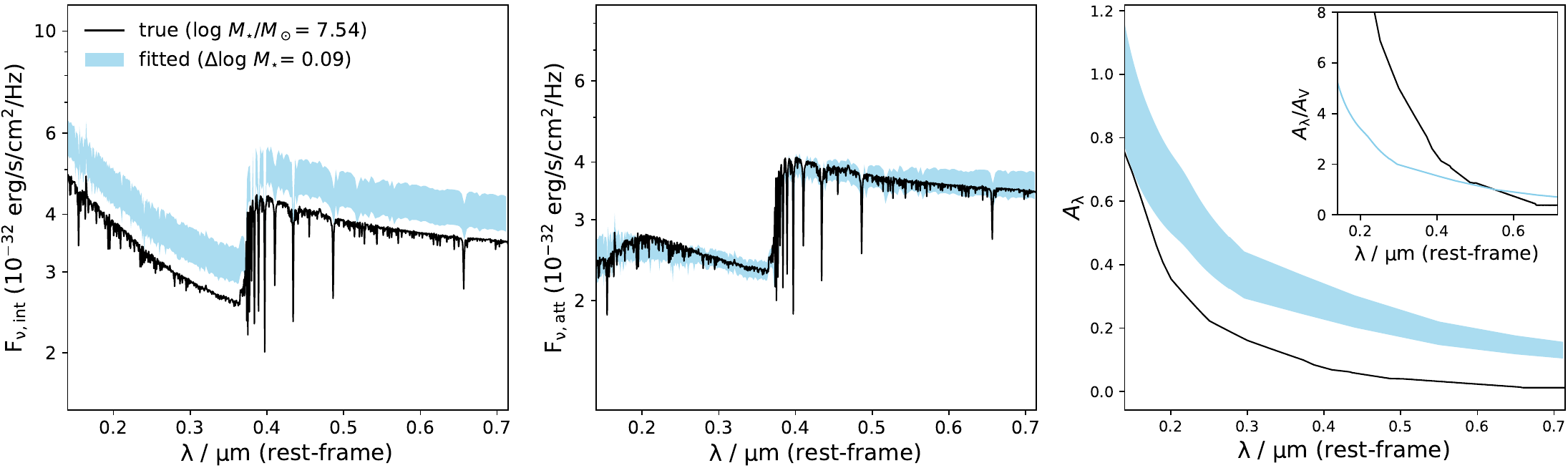}
    \caption{16th--84th percentile range of the posterior SEDs (blue shaded regions) is shown for two example galaxies, together with their true attenuated SEDs (black lines). Stellar metallicities are fixed to the true mass-weighted values. The left and middle panels display the intrinsic and attenuated stellar continua, respectively, and the right panels show the attenuation curves, with insets presenting their normalised forms (scaled by $A_V$). Here, $A_\lambda$ is scaled to reproduce the attenuated SEDs while preserving the intrinsic SED shape ($\Upsilon(\lambda)$). The galaxy in the upper panel has a flatter dust slope than the model, yielding a fitted $A_V$ lower than the true value, whereas the galaxy in the lower panel has a steeper dust slope, resulting in a fitted $A_V$ higher than the true value.}
    \label{fig:att_example}
\end{figure*}

The physical origin of these correlations can be understood as follows. For the normalised SFH model fitted with the SMC-type dust law, the stellar mass offset  (see Appendix~\ref{sec:appendix_B}) can be expressed as
\begin{equation}
    \Delta\log M_{\star}=
    \frac{A_V}{2.5}\left[\frac{A_{V,\mathrm{fit}}}{A_V}\, s_\mathrm{SMC}(\lambda)-s(\lambda)\right],
        \label{eq:mass_offset_1}
\end{equation}
where $s$ denotes the normalised dust attenuation curve (or slope), defined as 
$s(\lambda) \equiv A(\lambda)/A_V$.
Because the stellar mass-to-light ratio $\Upsilon(\lambda)$ is known, the right-hand side becomes fixed for any $\lambda$ once $\Delta\log M_\star$ is determined for a given posterior. 
At $\lambda=1500\,\text{\AA}$, $s(\lambda=1500\,\text{\AA})\equiv S$ and Eq.~\ref{eq:mass_offset_1} is simplified to
\begin{equation}
    \Delta\log M_{\star}=\frac{A_V}{2.5}(\frac{A_{V, \mathrm{fit}}}{A_V}S_\mathrm{SMC}-S),
        \label{eq:mass_offset_3}
\end{equation}
where $S_\mathrm{SMC}$ denotes the UV--optical slope derived from the SED attenuated by the SMC-type dust law. Because $\DeltaMfit$ is positively correlated with $A_\mathrm{V,fit}/A_\mathrm{V}$, Eq.~\ref{eq:mass_offset_3} implies that if the normalised dust attenuation curve is steeper (flatter) than the model curve used in the SED fitting, the fitted $A_V$ and $M_\star$ will be overestimated (or underestimated, respectively). Representative examples of both cases are presented in Fig.~\ref{fig:att_example}, which compares the simulated stellar continua with the fitted results for two galaxies displaying different dust slopes. In the upper (lower) panel, the assumed dust slope (i.e. SMC, blue lines) is steeper (or shallower) than the actual extinction curve of the simulated galaxy (black lines), resulting in an underestimation (or overestimation) of the intrinsic fluxes and, consequently, the stellar mass.

\begin{figure*}
    \centering
\includegraphics[width=\textwidth]{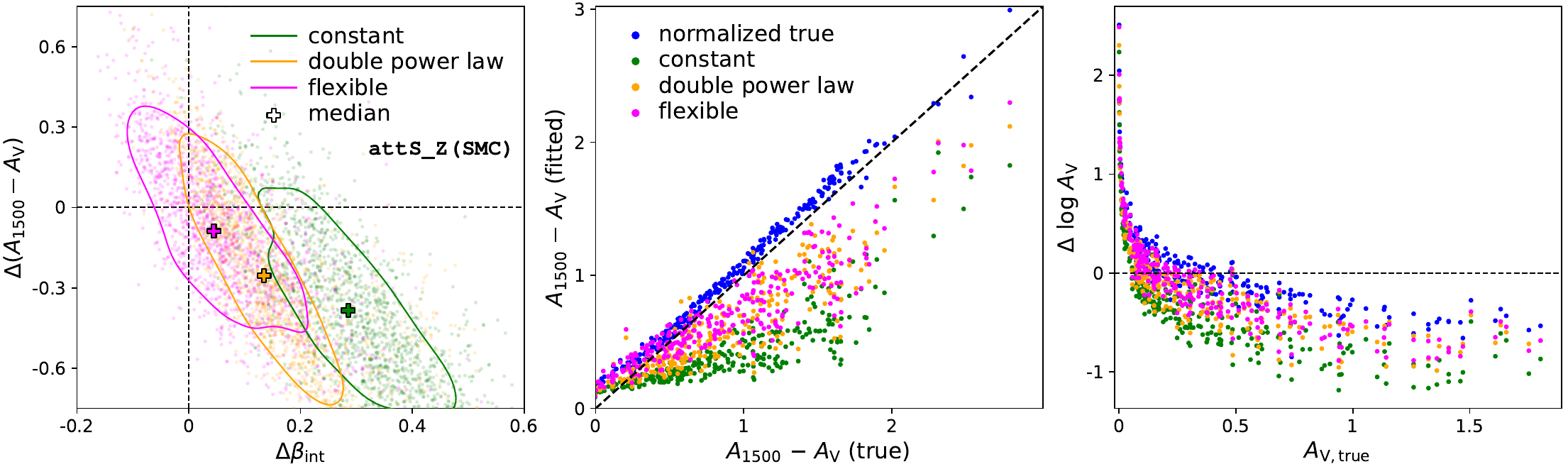}
    \caption{Distribution of the difference between the fitted and true values of $\beta_\mathrm{int}$ and the colour excess for a randomly selected galaxy, shown in the left panel. Crosses mark the medians, and contours enclose the 1 $\sigma$ confidence regions of the posterior distributions for each SFH model. A negative correlation is evident between the two quantities. Among the models, the constant SFH model yields the largest overestimation of $\beta_\mathrm{int}$, followed by the double power-law and flexible models. This trend appears in most galaxies in our sample, independent of the dust model adopted. The middle panel presents the resulting distribution of the recovered colour excess, while the right panel displays the distribution of the $A_V$ offset, which is directly related to the colour excess attributed to their positive correlation.}
    \label{fig:beta_and_color}
\end{figure*}

Another important feature of Eq.~\ref{eq:mass_offset_1} concerns the dust properties of a galaxy. For any pair of wavelengths $\lambda_\mathrm{1}$ and $\lambda_\mathrm{2}$, we have
\begin{align}
    A_V[s(\lambda_\mathrm{2})-s(\lambda_\mathrm{1})] 
    &= A_{V, \mathrm{fit}}[s_\mathrm{SMC}(\lambda_\mathrm{2})-s_\mathrm{SMC}(\lambda_\mathrm{1})], \notag \\
    E(\lambda_\mathrm{2}-\lambda_\mathrm{1}) 
    &= E_\mathrm{fit}(\lambda_\mathrm{2}-\lambda_\mathrm{1}),
        \label{eq:Efit}
\end{align}
where $E(\lambda_\mathrm{2}-\lambda_\mathrm{1})$ and $E_\mathrm{fit}(\lambda_\mathrm{2}-\lambda_\mathrm{1})$ denote the colour excesses between $\lambda_\mathrm{1}$ and $\lambda_\mathrm{2}$ for the input attenuated SED of the simulated galaxy and the corresponding model fit, respectively. This relation indicates that SED fitting primarily recovers the colour excess rather than the absolute dust content ($A_V$) or the slope of the attenuation curve ($S$). The data in Fig.~\ref{fig:att_example} (right panels) support this interpretation: the fitted $A_\lambda$ curves closely follow those of the simulated galaxy, even when the corresponding stellar masses may diverge from the true values.

The trends obtained with \texttt{attS\_Z$_\mathrm{true}$}, depicted in Fig.~\ref{fig:S-Av}, persist even when metallicity is allowed to vary (\texttt{attS\_Z}). In this case, both the fitted $A_V$ and $M_\star$ increase slightly relative to those using \texttt{attS\_Z$_\mathrm{true}$}, by $+0.04$ dex and $+0.03$ dex, respectively. These increases result from the underestimation of $Z_\star$ ($-0.11$ dex), which renders the intrinsic model stellar continuum bluer. To match the observed colour of the input attenuated stellar continuum, the fit  requires a steeper attenuation curve or equivalently a larger colour excess. Because the dust law in the SED fitting is fixed to the SMC curve, this steepening can only be achieved by raising $A_V$. This, in turn, enhances the attenuation across all wavelengths, resulting in an increase in intrinsic model fluxes and, thus, in the fitted $M_\star$.

The correlation between $S_\mathrm{true}$ and the stellar mass offset (Fig.~\ref{fig:S-Av}) remains when alternative dust attenuation laws are used in the SED fitting, although the absolute values differ slightly. For instance, applying the Calzetti law yields larger deviations in $M_\star$ ($\Delta\log M_\star \approx 0.19$) and $A_V$ ($\Delta A_V \approx 0.42$ dex) than applying SMC law ($\Delta\log M_\star \approx 0.07$ dex and $\Delta A_V \approx 0.20$ dex). This occurs because the normalised attenuation curves of simulated galaxies in \sphinx\ are generally steeper than the Calzetti law, but more closely resemble the SMC law. Consequently, reproducing the true colour excess requires a higher $A_V$ when the Calzetti law is adopted. The stellar mass, $M_\star$, is then overestimated to compensate for the stronger attenuation at a fixed observed flux. By contrast, the Salim law model allows the slope to vary through an additional parameter, $\delta$. Through simultaneous adjustment of $\delta$ and $A_V$, the Salim law can more flexibly recover the colour excess, reducing the systematic biases in $A_V$ ($\Delta A_V \approx 0.26$ dex) and in $M_\star$ ($\Delta\log M_\star \approx 0.12$ dex) relative to the Calzetti law; however, it still does not surpass the SMC law despite its greater flexibility.

In summary, the slope of the dust attenuation curve has a significant impact on the inferred stellar mass and $A_V$ in the SED fitting, with shallower (steeper) true attenuation curves than the model leading to underestimation (overestimation). This bias arises because SED fitting primarily retrieves the colour excess rather than the absolute dust content or the attenuation slope. Although flexible models such as the Salim law alleviate this bias more effectively than the Calzetti law, the SMC law continues to provide the most accurate fits for galaxies with steep attenuation curves, such as those in the \sphinx\ simulation.

\begin{figure}
    \centering
\includegraphics[width=0.47\textwidth]{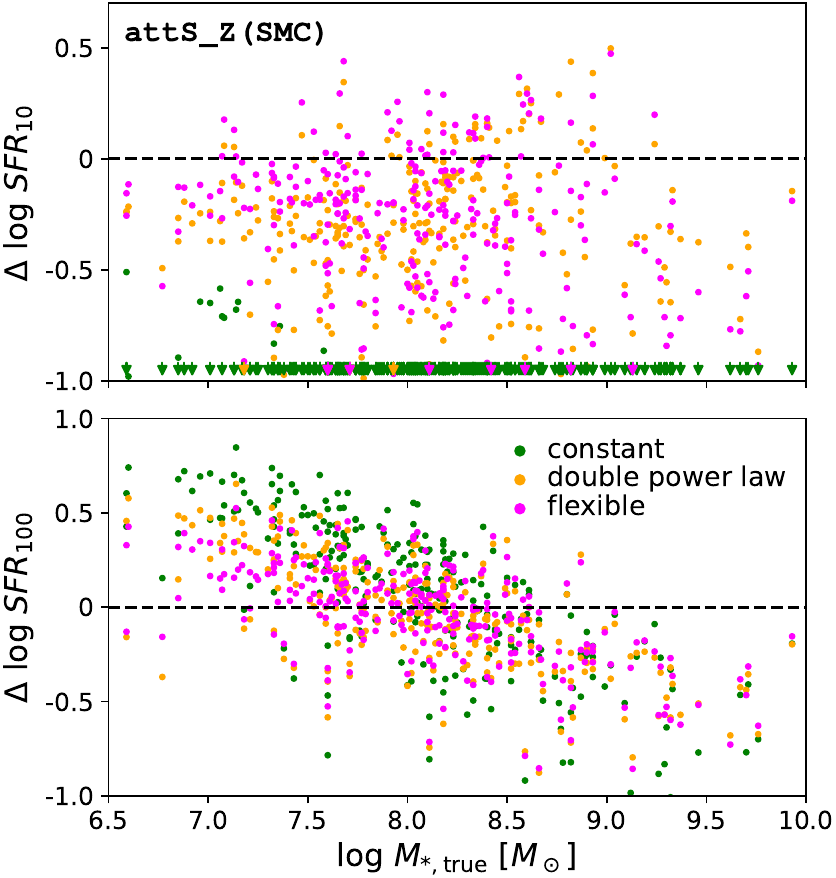}
    \caption{Same as Fig.~\ref{fig:sfr_intr}, but for the \texttt{attS\_Z} model adopting an SMC-type dust attenuation curve. Compared to Fig.~\ref{fig:sfr_intr}, the underestimation of SFR$_\mathrm{10}$ is apparent across the entire stellar-mass range, while the underestimation of SFR$_\mathrm{100}$ becomes progressively more severe toward the high-mass end.}
    \label{fig:sfr_dust}
\end{figure}

\subsubsection{Impact of intrinsic UV slope errors (SFH model) on colour excess, $A_V$, and SFR}

Because reconstructed SFHs rarely reproduce the true stellar mass-to-light ratio ($\Upsilon$) precisely, Eqs.~\ref{eq:mass_offset_1} to~\ref{eq:Efit} tend to  not be satisfied overall. In such cases, the colour excess can be accurately recovered only if the shape of the fitted intrinsic stellar continuum closely matches the true shape. Otherwise, the fitting behaves as when metallicity is treated as a free parameter with the normalised true SFH: if the fitted intrinsic SED becomes redder (bluer), the fitted attenuation curve correspondingly becomes shallower (steeper) to reproduce the colour of the input attenuated SED. This behaviour is supported by Fig.~\ref{fig:beta_and_color} (left panel), which displays the posterior distribution of the differences between the fitted and true colour excess ($\Delta (A_{1500}-A_{5500}$)) and between the fitted and true UV slope of the intrinsic stellar continuum ($\Delta \beta_\mathrm{int}$) for a randomly selected galaxy. Here, $\beta_\mathrm{int}$ is measured from the intrinsic stellar continuum across $1268$ to $2700\,\text{\AA}$ \citep[e.g.][]{Calzetti1994}. Our findings reveal that the difference in colour excess is consistently and negatively correlated with $\beta_\mathrm{int}$, irrespective of the SFH model used.

For most galaxies in our sample, the fitted $\beta_\mathrm{int}$ is also overestimated across all three SFH models. The constant SFH model exhibits the largest deviation ($\Delta \beta_{\rm{int}}\approx+0.31$), followed by the double power-law ($\Delta \beta_{\rm{int}} \approx+0.08$) and then the flexible model ($\Delta \beta_{\rm{int}} \approx+0.04$). This trend is directly connected to the distribution of the fitted colour excess. As illustrated in the middle panel of Fig.~\ref{fig:beta_and_color}, the constant SFH model substantially underestimates the colour excess ($\approx 0.32$ dex) because of the overestimation of $\beta_\mathrm{int}$. The normalised true SFH model yields the smallest offset in colour excess (0.04 dex), followed by the flexible (0.12 dex) and double power-law (0.15 dex) models. Because the colour excess is determined solely by $A_V$ under both the SMC and Calzetti laws, the fitted $A_V$ tends to be smallest when the constant SFH model is applied. This explains the fitted $A_V$ distribution in the right panel of Fig.~\ref{fig:beta_and_color}, where the constant SFH model produces the lowest $A_V$ across all attenuation laws, including Salim.

A shallower attenuation curve, such as the Calzetti law, requires a larger colour excess to reproduce the same level of reddening between two wavelengths (e.g. $1500\,\text{\AA}$ and $5500\,\text{\AA}$), particularly when fitting the full shape of the observed SED. This is analogous to fitting a curve through multiple points: a shallower attenuation curve constrained by fixed endpoints (i.e. the same colour excess) will inevitably fail to reproduce the intermediate fluxes accurately. Consequently, the fitting procedure favours a larger colour excess to better match the overall spectral shape, which in turn yields a steeper UV slope owing to the increased SFR. Figure~\ref{fig:sfr_dust} shows the differences between the true and fitted star formation rates as a function of stellar mass for different SFH models. As a direct consequence of the behaviour described above, the median $\Delta \mathrm{SFR}_{100}$ increases systematically from SMC to the Salim to the Calzetti attenuation curve, with values of 0.02, 0.09, and 0.19, respectively, while the overall mass-dependent trends remain  qualitatively similar. Because shallower attenuation curves (e.g. Calzetti) require large values of $A_V$ to reproduce a given colour excess, the inferred stellar mass increases progressively from SMC to Salim to Calzetti (0.20, 0.24, and 0.27, respectively; see Fig.~\ref{fig:att_stellar}).

In summary, because reconstructed SFHs are rarely sufficient to recover the true intrinsic continuum, errors in the fitted UV slope translate directly into dust biases: redder inferred slopes force shallower attenuation curves and yield underestimated colour excesses and $A_V$, especially for the constant SFH model. These systematics are further amplified, with shallower curves (e.g. Calzetti) requiring larger colour excesses and higher $A_V$, thereby boosting the recovered SFR and stellar mass.

\subsection{Impact of nebular emission} 

Because high-redshift galaxies are actively star-forming, nebular emission from ionised gas, such as H$\beta$ 4861, [O III] 4959/5007, H$\alpha$ 6563, and [N II] 6548/6584, can contribute substantially to broad-band fluxes in the UV-optical range \citep[e.g.][]{Anders2003,Zackrisson2008,Wilkins2013}. The intrinsic Ly$\alpha$ emission at $1216\,\text{\AA}$ is also very strong, but it is heavily attenuated by the neutral IGM at $z=6$ in \sphinx\ \citep{Garel2021}. The presence of strong emission lines reddens the UV-optical colours, causing stellar masses to be overestimated when nebular emission is neglected \citep{Schaerer2009, Salmon2015, Yuan2019, Miranda2025}.

\begin{figure}
\includegraphics[width=0.47\textwidth]{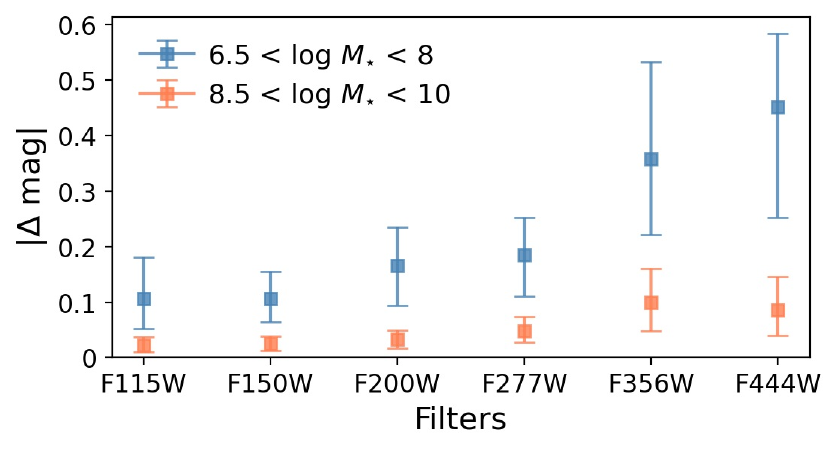}
\includegraphics[width=0.47\textwidth]{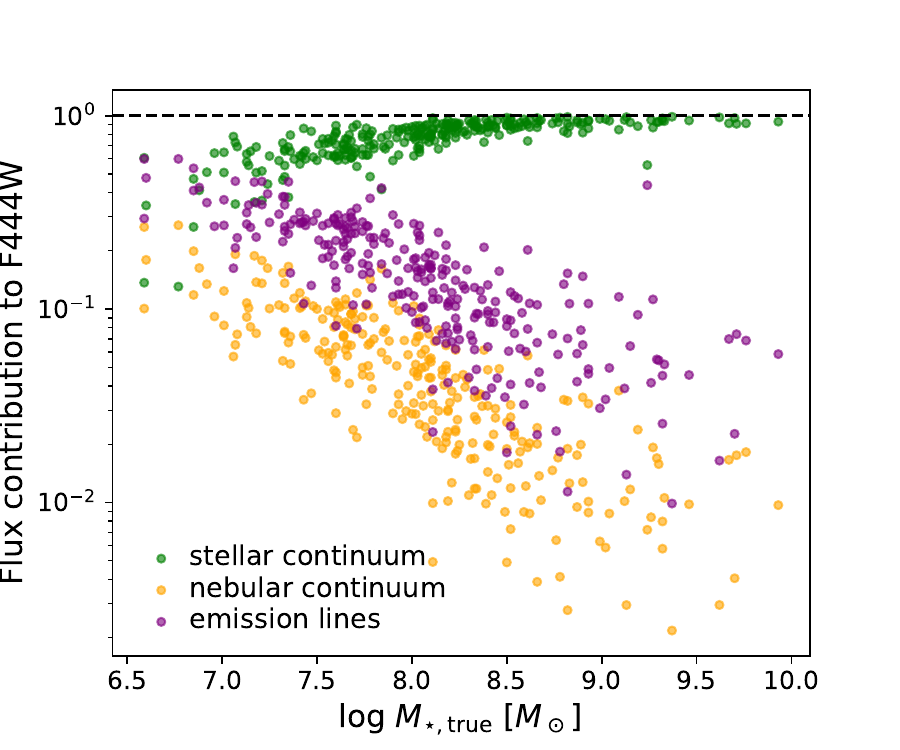}
    \caption{ Change in filter magnitudes when nebular emission is added to the stellar continuum for low- and high-mass galaxies, shown in the upper panel. Each point marks the median, with error bars indicating the 16th--84th percentile range. Both galaxy populations appear redder, mainly due to strong optical emission lines in the rest frame optical range, with the effect being more pronounced in low-mass systems. The lower panel presents the flux contributions of the stellar continuum, nebular continuum, and nebular emission lines to the F444W magnitude for each galaxy. In galaxies with $\log M_\mathrm{\star}/M_\odot\lesssim 7.3$, the emission-line contribution is comparable to that of the stellar continuum.}
    \label{fig:em_contribution}
\end{figure}

\begin{figure*}
    \centering
\includegraphics[width=\textwidth]{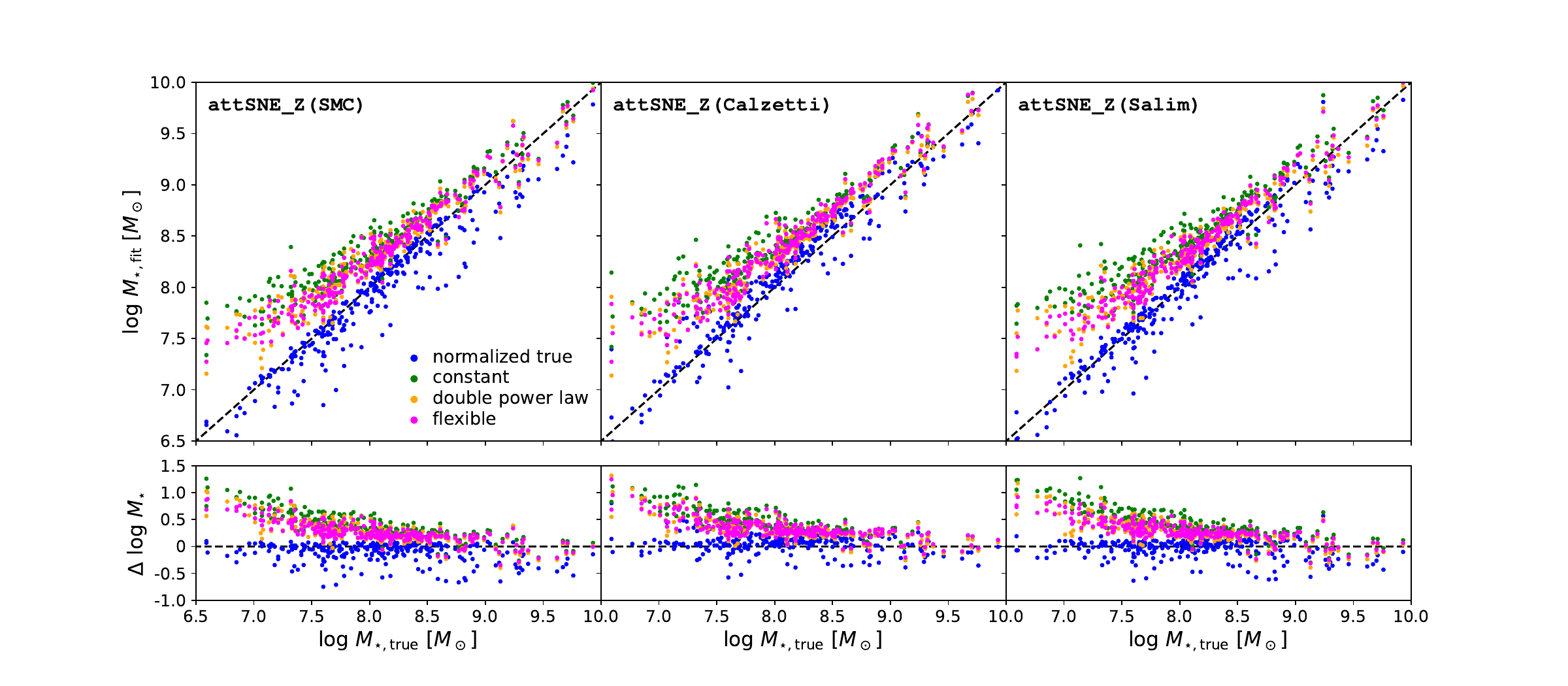}
    \caption{Same as Fig.~\ref{fig:att_stellar}, but with nebular emission included (i.e. the \texttt{attSNE\_Z} model). Across all dust attenuation laws and SFH models, the overestimation of $M_\star$ exceeds that in the \texttt{attS\_Z} case, especially in low-mass galaxies, where it reaches  $\sim$1.5 dex. Consistent with the \texttt{attS\_Z} results, the bias is smallest for the SMC law, largest for the Calzetti law, and intermediate for the Salim law.}
    \label{fig:full_metal_free}
\end{figure*}

For galaxies in our sample, including nebular emission indeed makes the input SED appear redder. To illustrate its impact on galaxy colour, we compared the filter magnitudes with and without nebular emission for massive ($8.5<\log M_\star/\msun<10$) and less massive ($6.5<\log M_\star/\msun<8$) sub-samples. As depicted in the upper panel of Fig.~\ref{fig:em_contribution}, the inclusion of nebular emission results in substantial flux increases in the F356W and F444W filters. This enhancement is especially pronounced in low-mass galaxies, leading to a higher inferred $M_\star$, as discussed further below. 

The lower panel of Fig.~\ref{fig:em_contribution} displays the relative contributions of the stellar continuum, nebular continuum, and emission lines to the F444W filter. In most galaxies, the stellar continuum dominates the total flux, whereas emission lines contribute comparably in low-mass, actively star-forming systems. Within the F444W bandpass, the dominant emission lines are H$\alpha$ $\lambda$6563 and [NII] $\lambda$6583, whereas in the F356W bandpass, H$\beta$ 4861, [O III] 4959/5007 dominate. The nebular continuum is generally approximately three times fainter than the line emission and therefore remains a sub-dominant component of the input SED.

To evaluate the impact of nebular emission on the inferred galaxy properties, we included the attenuated nebular component (both line and continuum) together with the attenuated stellar continuum in the input spectrum. The combined spectrum is then fitted using \bagpipes, with the nebular component enabled (\texttt{attSNE\_Z}). Figure~\ref{fig:full_metal_free} illustrates that when nebular emission is included under the normalised true SFH, stellar masses can still be recovered with a reasonable level of accuracy. The median differences in $\log M_\star$ are $-0.05$, $0.06$, and $0.01$ for the SMC, Calzetti, and Salim dust models, respectively; these values are smaller than the corresponding offsets without nebular emission ($0.04$, $0.19$, and $0.12$; see Fig.~\ref{fig:att_stellar}). In contrast, estimates deviate more strongly when using the other three SFH models, with offsets of 0.36, 0.26, and 0.26 dex for the constant, double power-law, and flexible models, respectively. This discrepancy is most evident in low-mass galaxies, where strong optical lines redden the SED and bias the posterior towards older stellar populations, producing mass overestimations of up to 1.31 dex. Conversely, in massive galaxies, the effect is weak, as the contribution of nebular emission to the total flux is negligible (see the lower panel of Fig.~\ref{fig:em_contribution}). Galaxies with $M_{\star,\mathrm{true}}>10^{8.5}\, M_\odot$ yield fitted masses consistent with those from \texttt{attS\_Z}. Although the nebular flux contribution is generally small, excluding it may not be justified, as it can still bias $M_\star$ estimates.

\begin{figure}
\includegraphics[width=0.47\textwidth]{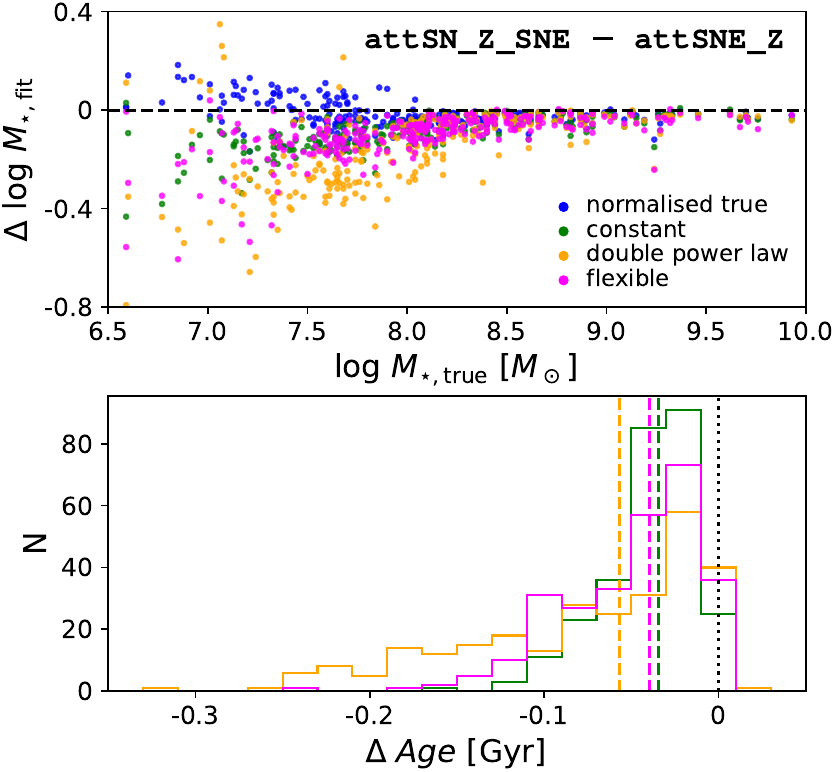}
    \caption{Impact of emission lines in the input spectrum on inferred galaxy properties. The upper panel presents the difference in the fitted $M_\star$ between \texttt{attSN\_Z\_SNE} (which excludes emission lines from the input spectra, but includes them in the fitting) and \texttt{attSNE\_Z}. Emission lines have the strongest effect in low-mass galaxies. The reduction in fitted $M_\star$ arises because \texttt{attSN\_Z\_SNE} infers younger stellar populations, producing a bluer continuum to match the observed colour in the absence of emission lines. The lower panel illustrates this change in stellar populations by illustrating the distribution of differences in mass-weighted age between \texttt{attSN\_Z\_SNE} and \texttt{attSNE\_Z} with the median marked by a dashed line.}
    \label{fig:full_Z_noEL}
\end{figure}

\begin{figure}
\includegraphics[width=0.47\textwidth]{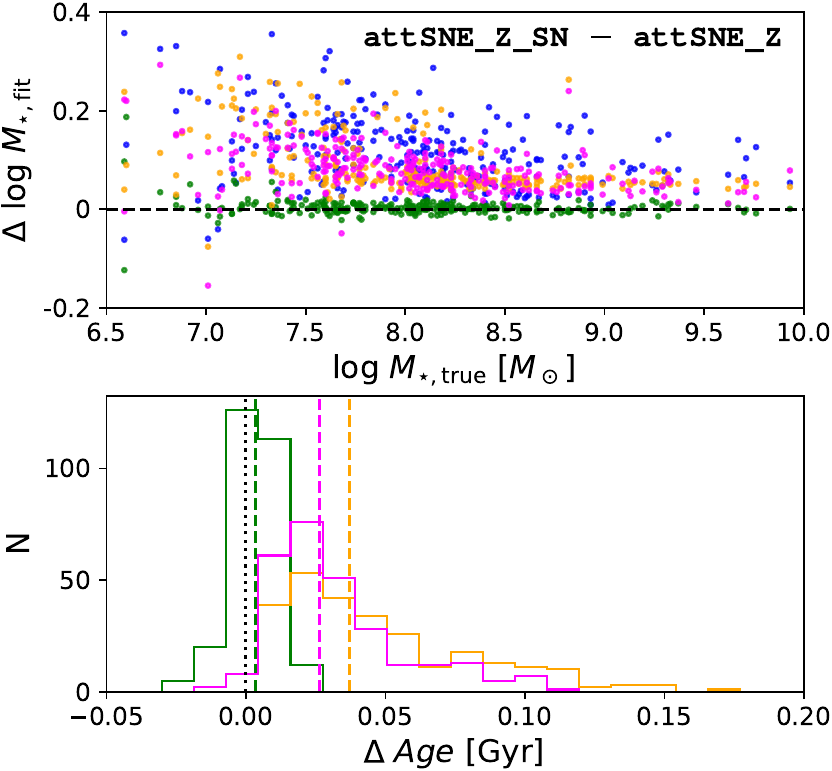}
    \caption{Same as Fig.~\ref{fig:full_Z_noEL} but for the \texttt{attSNE\_Z\_SN} model, which includes emission lines in the input spectra but neglects them in the fitting process. Here, the increase in the fitted $M_\star$ mainly reflects changes in the inferred stellar populations, except for the normalised true SFH model, which increases $A_\mathrm{V}$ to redden the model spectrum instead. 
    }
    \label{fig:full_Z_ignoEL}
\end{figure}

Because emission lines increase the total flux and redden the SED, the stellar population is inferred to be older and/or more dust-rich, which, in turn, raises the estimated stellar mass. To evaluate this effect, we repeated the fitting with input spectra that exclude emission lines (\texttt{attSN\_Z\_SNE}) and compared the results with the fiducial case (\texttt{attSNE\_Z}), thereby isolating the influence of intrinsic emission on mass inference. The upper panel of Fig.~\ref{fig:full_Z_noEL} indicates that the recovered mass remains consistent under the normalised true SFH model is used. In contrast, the other three models yield systematically lower masses, as the omission of emission lines produces a bluer SED that favours younger populations, as illustrated in the lower panel of Fig.~\ref{fig:full_Z_noEL}. This shift in stellar population leads to a pronounced reduction in the inferred stellar mass, particularly for low-mass galaxies. 

Conversely, when emission is included in the input SED, but not modelled during the fitting (\texttt{attSNE\_Z\_SN}), as in the study of \citet{Schaerer2009}, the excess optical flux is attributed to older stellar populations under the double power-law and flexible SFH models (Fig.~\ref{fig:full_Z_ignoEL}). The normalised true SFH model also returns higher $M_\star$ values, but in this case, the increase is driven by a larger fitted $A_\mathrm{V}$, as the stellar population is fixed. By comparison, the constant SFH model exhibits negligible variation in any fitted parameter, as it primarily reproduces the broadband colours through the stellar continuum alone, which limits the influence of emission lines on the inferred age or mass. Although including emission lines lowers the $\chi^2$ values, this improvement only accounts for only a small fraction of the posterior distributions.

In summary, strong emission lines substantially enhance the rest-frame optical flux, leading to an overestimation of the fitted $M_\star$ in the \texttt{attSNE\_Z} model compared to that in \texttt{attS\_Z}. This effect is most pronounced in low-mass galaxies, with biases reaching up to 0.8 dex. Neglecting emission-line modelling in SED fitting further exacerbates the bias, introducing an additional overestimation of up to 0.4 dex because emission-line optical flux is misattributed to the stellar continuum of older populations. By contrast, the constant SFH model yields few or no stars younger than 10 Myr irrespective of emission, indicating that it is the least reliable of the tested models for recovering recent SFRs. 

\begin{figure}
\includegraphics[width=0.47\textwidth]{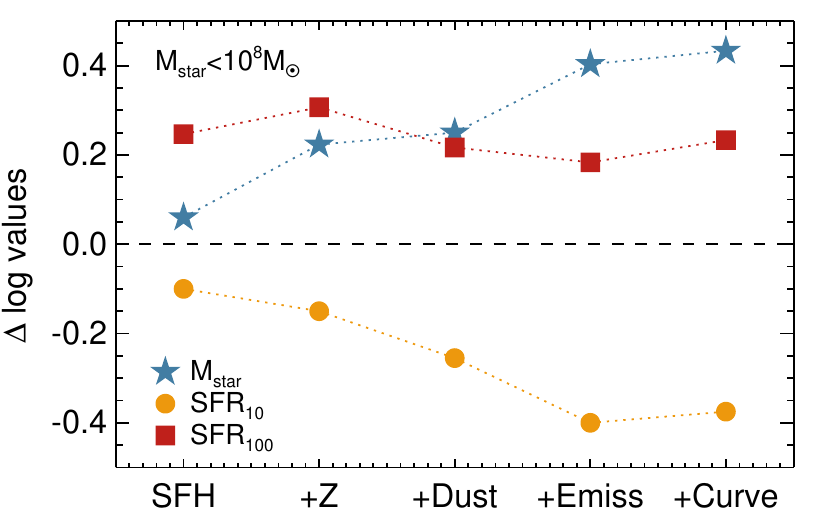}
\includegraphics[width=0.47\textwidth]{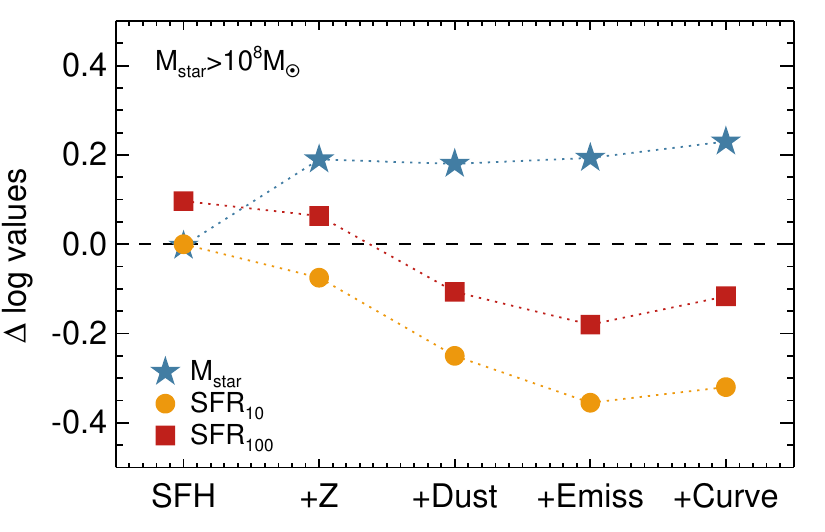}
    \caption{Summary of how different modelling assumptions influence the inferred physical quantities.
Top and bottom panels show the logarithmic offsets for low-mass ($M_\star < 10^8\,\msun$) and high-mass ($M_\star > 10^8\,\msun$) galaxies, respectively. Each point represents the mean offsets in the inferred physical properties obtained with the \texttt{intS\_Z$_\mathrm{true}$}, \texttt{intS\_Z}, and \texttt{attS\_Z} configurations; the two right-most points use the same \texttt{attSNE\_Z} configuration. The left-most point isolates the effects of the assumed SFH, while the points to the right (marked by plus signs on the x-axis) illustrate the cumulative impact of successively adding each modelling ingredient. 
    }
    \label{fig:summary_assumptions}
\end{figure}

\subsection{Summary of the impact of modelling assumptions}

Figure~\ref{fig:summary_assumptions} summarises the impact of various modelling assumptions adopted in the SED fitting. We first quantified the uncertainty associated with the choice of SFH by taking the mean logarithmic offsets in the derived physical quantities obtained using the \texttt{intS\_Z$_\mathrm{true}$} configuration. We then illustrate how the inferred values deviate from the true ones as additional modelling ingredients are introduced sequentially, as indicated along the x-axis. We emphasise that this figure is intended solely to provide a qualitative indication of the assumptions that influence the recovered galaxy properties most strongly. A detailed quantitative breakdown is presented in Appendix~\ref{sec:appendix_D}.

For less massive galaxies  ($M_\star<10^8\,\msun$, top panel), we find that the stellar mass estimates are most strongly affected by the age--metallicity degeneracy ($\sim$0.2 dex) and by emission line modelling ($\sim$0.2 dex). The short-term SFR is likewise most sensitive to the inclusion of emission lines, while dust modelling contributes comparably to systematic underestimation. In contrast, the long-term SFR is most strongly affected by the assumed SFHs. 

For more massive galaxies  ($M_\star>10^8\,\msun$, top panel), the stellar mass estimates remain primarily driven by the age--metallicity degeneracy, whereas the impact of the other ingredients is negligible. For both the short-term and long-term SFR estimates, dust modelling emerges as the dominant source of uncertainty, followed by emission line modelling.  

\section{Discussion}
\label{sec:discussion}

In the previous section, we detail how SFH, metallicity, dust, and emission influence SED fitting. Building on this, we go on to evaluate how choices in the observational setup and statistical treatment affect the accuracy and robustness of SED fitting, with particular attention to the role of adding a medium-band filter in recovering galaxy properties. We further consider how uncertainties in the inferred properties can propagate into the derived quantities, such as the SMF and the star formation main sequence.

\subsection{ Improving fitting accuracy } 

\subsubsection{Impact of adding a medium band on fitting results} 
\label{Sec:discussion_medium_band}

Figure~\ref{fig:full_metal_free} illustrates that stellar mass is considerably overestimated when strong emission lines are present in the input photometry (\texttt{attSNE\_Z}) compared to the case without emission (\texttt{attS\_Z}, Fig.~\ref{fig:att_stellar}). As noted earlier, this bias arises because the optical flux in the stellar continuum is boosted by demanding a larger contribution from older stellar populations. The inclusion of a medium-band filter without strong emission lines can therefore potentially mitigate this bias in the SED fitting \citep[e.g.][]{Roberts2021}. 

\begin{figure}
\includegraphics[width=0.47\textwidth]{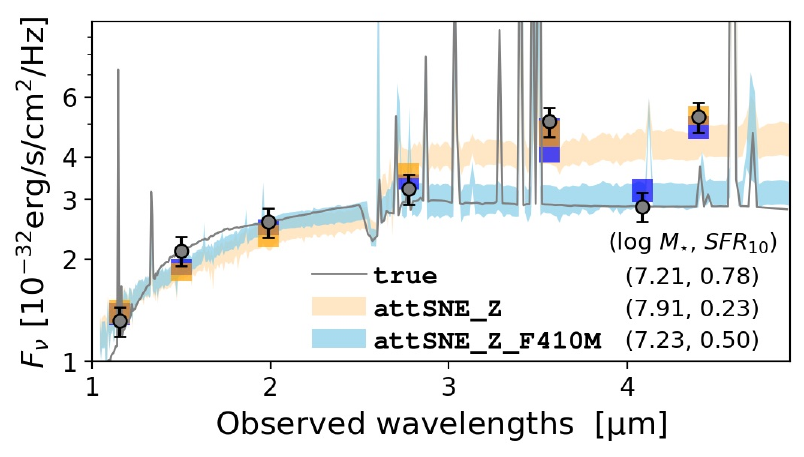}
\includegraphics[width=0.47\textwidth]{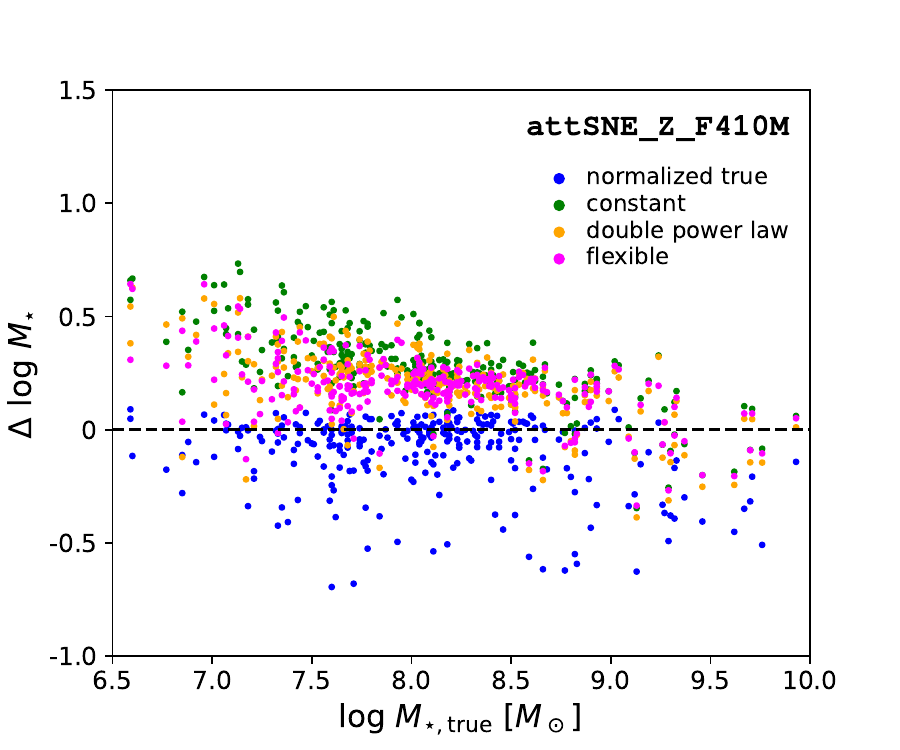}
    \caption{Upper panel:  16th--84th percentile distribution of the fitted SEDs is presented for an example galaxy using the \texttt{attSNE\_Z} (orange) and \texttt{attSNE\_Z\_F410M} (sky blue) models with the double power-law SFH. Comparing each shaded region with the true SED (gray) indicates that including F410M photometry provides tighter constraints on the SED fitting and consequently on $M_\star$ and $\mathrm{SFR}_\mathrm{10}$. Lower panel:  Difference between the true and fitted $M_\star$ is shown when the SMC law is applied with \texttt{attSNE\_Z\_F410M}. The inclusion of F410M photometry substantially reduces the overestimation of $M_\star$, particularly in low-mass galaxies.}
    \label{fig:f410}
\end{figure}

To test this, we recompute stellar masses after adding the F410M band, which is largely unaffected by strong emission lines at $z=6$, to the input photometry (\texttt{attSNE\_Z\_F410M}). The upper panel of Fig.~\ref{fig:f410} displays an example of the posterior distribution for a low-mass galaxy with $M_\star=10^{7.2}\,\msun$. We find that the addition of the extra band reduces the stellar mass offset from 0.7 dex in \texttt{attSNE\_Z} to nearly zero. Furthermore, the true SFR$_\mathrm{10}$ (0.78\,\msunyr) is more accurately recovered (0.23$\rightarrow$0.5\,\msunyr), as the contribution from older stellar populations is diminished.

A statistical comparison of the stellar mass offset is presented in the lower panel of Fig.~\ref{fig:f410}. The maximum offset in low-mass galaxies under \texttt{attSNE\_Z} is $1.26$ dex (Fig.~\ref{fig:full_metal_free}), but it decreases to $0.57$ dex when the F410M band is included. On average, $\DeltaMfit$ in low-mass galaxies with $M_{\star,\mathrm{true}} \lesssim 10^8\,\msun$ decreases from $0.53$ to $0.39$, $0.40$ to $0.27$, and $0.39$ to $0.24$ dex for the constant, double power-law, and flexible SFH models, respectively, relative to \texttt{attSNE\_Z}. By contrast, galaxies more massive than $\sim 10^{8}\,\msun$ are only marginally influenced by the addition of the medium band, again because the contribution from emission lines is negligible. Nevertheless, the medium band can still be useful when bright galaxies display strong emission lines.

This exercise highlights the importance of emission-line-free photometric bands in the optical range, in line with the findings of \citet{Roberts2021}. In practice, however, such medium bands may not be accessible for galaxies at higher redshift. For instance, the F410M band is likely contaminated by the [OII] $\lambda$3727, [OIII] $\lambda$5007 and Balmer lines in galaxies at $z\gtrsim7$. Moreover, galaxies at higher redshift are typically more actively star-forming and therefore more strongly influenced by emission lines. To evaluate whether these effects alter our conclusions, we repeated  the same analysis for a sample of 66 galaxies at $z=9$. Although we do not show them in this work, we did find that the trends observed at $z=6$ persist at $z=9$, indicating that our conclusions remain robust across these epochs. At $z=9$, however, the F410M filter is no longer free of strong emission lines, and thus less useful than at $z=6$. Consequently, the stellar mass of low-mass galaxies is overestimated by $\approx 0.1$ dex more than that at $z=6$. We also observe that the degree of mass overestimation from SED fitting decreases with increasing true stellar mass, confirming that the overall conclusions remain valid even at higher redshift ($z\sim9$).

\subsubsection{ Choice of summary statistics }

\begin{figure*}
    \centering
\includegraphics[width=\textwidth]{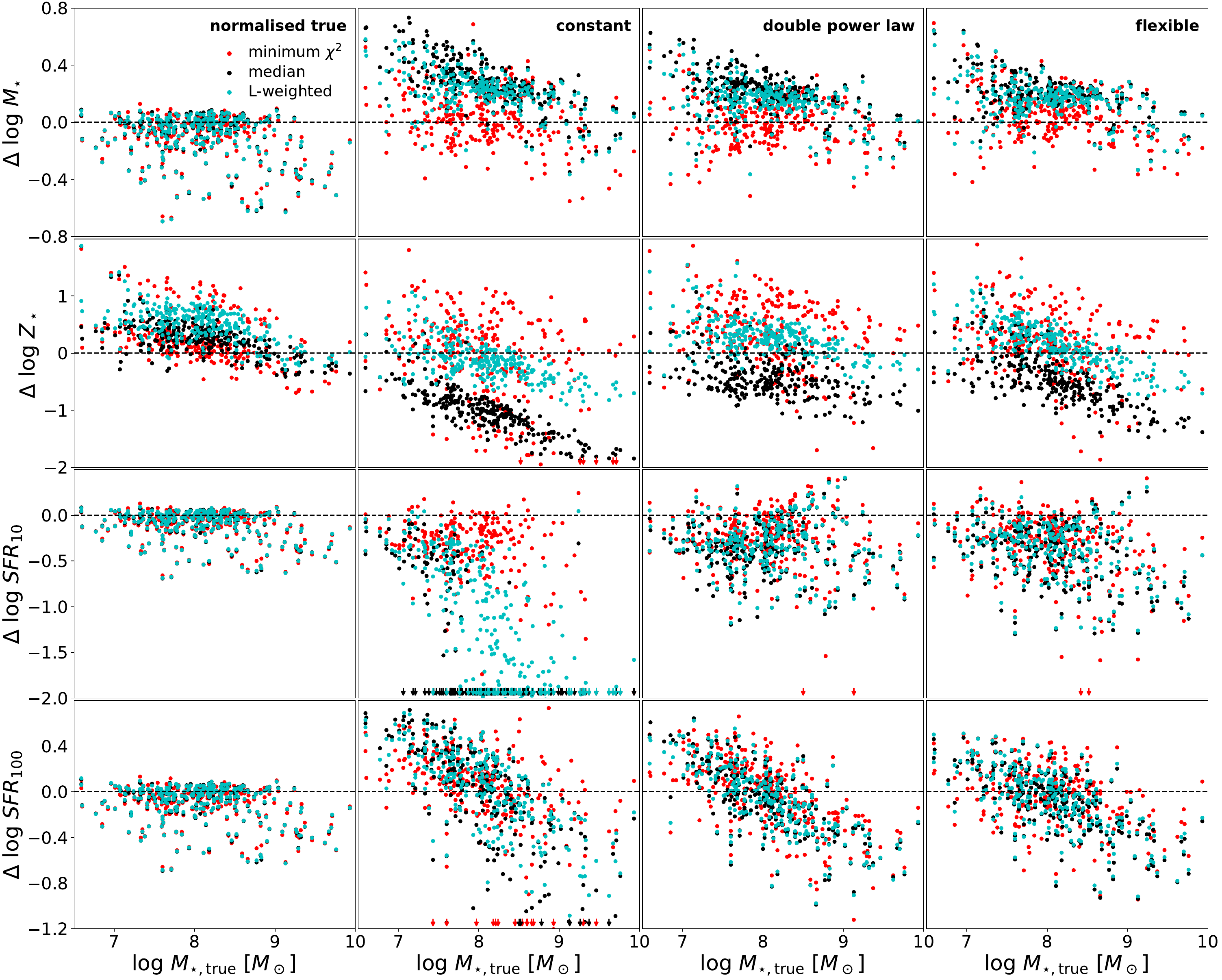}
    \caption{Impact of summary statistics on the recovered physical quantities in the \texttt{attSNE\_Z\_F410M} model. Each row displays the difference between the true and fitted values of $M_\star$, $Z_\star$, $\mathrm{SFR}_\mathrm{10}$, and $\mathrm{SFR}_\mathrm{100}$, while each column corresponds to a different SFH model. The fitted values are derived from the posterior distribution using one of the three statistics: minimum $\chi^2$ (red), median posterior (black), or likelihood-weighted mean (cyan). The minimum $\chi^2$ posterior yields the most accurate recovery of $M_\star$, whereas the likelihood-weighted mean performs best for $Z_\star$. The recovered SFRs are generally insensitive to the choice of the summary statistic, except in the constant SFH model.}
    \label{fig:median_best}
\end{figure*}

\begin{table*}
    \renewcommand{\arraystretch}{1.5}
    \centering
    \caption{Offsets in the inferred physical properties for different summary statistics. }
    \resizebox{\textwidth}{!}{    
    \begin{tabular}{clcccccc}
    \hline \hline
        Variable & SFH model & \multicolumn{2}{c}{Median} & \multicolumn{2}{c}{Minimum-$\chi^2$} & \multicolumn{2}{c}{L-weighted} \\
         &  & low($\le10^8M_\odot$) & high($>10^8M_\odot$) & low & high & low & high \\
    \hline
        $\Delta \log M_\mathrm{\star}$ & Normalised true & $-0.04^{+0.07}_{-0.15}$ & $-0.04^{+0.07}_{-0.25}$ & $-0.06^{+0.08}_{-0.15}$ & $-0.05^{+0.08}_{-0.24}$ & $\bf{-0.04^{+0.07}_{-0.14}}$ & $\bf{-0.04^{+0.07}_{-0.25}}$\\
        & Constant & $0.37^{+0.15}_{-0.11}$ & $0.23^{+0.07}_{-0.12}$ & $\bf{0.05^{+0.18}_{-0.17}}$ & $\bf{0.03^{+0.20}_{-0.16}}$ & $0.27^{+0.14}_{-0.13}$ & $0.20^{+0.06}_{-0.13}$\\
        & Double power & $0.27^{+0.13}_{-0.11}$ & $0.17^{+0.08}_{-0.12}$ & $\bf{-0.01^{+0.17}_{-0.16}}$ & $\bf{0.02^{+0.14}_{-0.15}}$ & $0.18^{+0.15}_{-0.14}$ & $0.15^{+0.07}_{-0.12}$\\
        & Flexible & $0.21^{+0.16}_{-0.09}$ & $0.19^{+0.05}_{-0.12}$ & $\bf{0.08^{+0.18}_{-0.15}}$ & $\bf{0.05^{+0.12}_{-0.15}}$ & $0.17^{+0.16}_{-0.11}$ & $0.17^{+0.05}_{-0.12}$\\
        \hline
        $\Delta \log \mathrm{SFR}_\mathrm{10}$ & Normalised true & $-0.05^{+0.06}_{-0.15}$ & $-0.05^{+0.07}_{-0.25}$ & $-0.06^{+0.07}_{-0.15}$ & $-0.06^{+0.08}_{-0.24}$ & $\bf{-0.04^{+0.07}_{-0.14}}$ & $\bf{-0.04^{+0.07}_{-0.25}}$\\
        & Double power & $-0.37^{+0.18}_{-0.21}$ & $-0.29^{+0.30}_{-0.33}$ & $\bf{-0.23^{+0.18}_{-0.18}}$ & $\bf{-0.21^{+0.23}_{-0.36}}$ & $-0.32^{+0.20}_{-0.21}$ & $-0.28^{+0.29}_{-0.31}$\\
        & Flexible & $-0.28^{+0.16}_{-0.26}$ & $-0.41^{+0.27}_{-0.40}$ & $\bf{-0.22^{+0.23}_{-0.26}}$ & $\bf{-0.30^{+0.20}_{-0.43}}$ & $-0.24^{+0.16}_{-0.27}$ & $-0.38^{+0.26}_{-0.37}$\\
        \hline
        $\Delta \log \mathrm{SFR}_\mathrm{100}$ & Normalised true & $\bf{-0.04^{+0.07}_{-0.15}}$ & $-0.05^{+0.08}_{-0.25}$ & $-0.07^{+0.09}_{-0.14}$ & $-0.06^{+0.08}_{-0.24}$ & $-0.04^{+0.07}_{-0.16}$ & $\bf{-0.04^{+0.07}_{-0.26}}$\\
        & Constant & $0.30^{+0.23}_{-0.26}$ & $-0.14^{+0.29}_{-0.46}$ & $\bf{0.16^{+0.21}_{-0.18}}$ & $\bf{-0.03^{+0.22}_{-0.38}}$ & $0.28^{+0.19}_{-0.24}$ & $-0.10^{+0.29}_{-0.38}$\\
        & Double power & $\bf{0.12^{+0.23}_{-0.23}}$ & $-0.21^{+0.22}_{-0.24}$ & $0.17^{+0.22}_{-0.18}$ & $\bf{-0.15^{+0.29}_{-0.27}}$ & $0.14^{+0.21}_{-0.22}$ & $-0.19^{+0.22}_{-0.25}$\\
        & Flexible & $0.08^{+0.19}_{-0.19}$ & $-0.16^{+0.22}_{-0.28}$ & $\bf{0.03^{+0.20}_{-0.23}}$ & $\bf{-0.02^{+0.17}_{-0.32}}$ & $0.10^{+0.17}_{-0.20}$ & $-0.14^{+0.22}_{-0.27}$\\
        \hline
    \end{tabular}
    }
    \tablefoot{The table shows the 16th, 50th and 84th percentiles for the fitting model of \texttt{attSNE\_Z\_F410M}.  $\Delta \mathrm{SFR}_{10}$ for the constant SFH model is not shown, as $\mathrm{SFR}_\mathrm{10,fit} =0$ in many galaxies. For each SFH model, we highlight the value that has the minimum offset.}
    \label{tab:summary_table1}
\end{table*}

In Bayesian SED fitting, consensus on the appropriate statistical criterion for defining the `best-fit' value is still lacking. The minimum $\chi^2$ solution is widely used, yet it does not necessarily represent the most probable outcome, as posterior distributions are often non-Gaussian and asymmetric. The posterior median is another common choice, but, as illustrated in the case of stellar metallicity in Fig.~\ref{fig:intr_stellar}, it can be biased by distribution shape and may not reliably trace the true value. Therefore, assessing which of the three widely used statistics (minimum $\chi^2$, posterior median, or likelihood ($\mathcal{L}$)-weighted mean) best recovers the underlying physical parameters is particularly instructive.

Figure~\ref{fig:median_best} presents the differences between the true and fitted values of $M_\star$, $Z_\star$, $\mathrm{SFR}_\mathrm{10}$, and $\mathrm{SFR}_\mathrm{100}$ in the \texttt{attSNE\_Z\_F410M} model across the three statistics. For the normalised true SFH case, $M_\star$ and the SFRs display no notable dependence on the choice of statistic. By contrast, $Z_\star$ is better reproduced by the posterior median than by $\mathcal{L}$-weighted mean. The minimum $\chi^2$ method often performs well, but also produces the largest offsets. 

In the constant SFH model, $M_\star$ is more biased when the posterior median or the $\mathcal{L}$-weighted mean is used, whereas the minimum $\chi^2$ estimate is less biased but exhibits greater scatter. A similar pattern emerges for $Z_\star$, except that the $\mathcal{L}$-weighted mean appears least biased. This apparent accuracy largely stems from the restricted prior range, $Z_\star/Z_\odot=[0.001, 1]$. As noted in Sect.~\ref{sec:3.1}, the fitted $Z_\star$ values cluster near the lower bound of this range, generating systematic bias when the posterior median is adopted. The effect is strongest for the constant SFH model and also influences the $\mathcal{L}$-weighted mean. Extending the prior range to $[10^{-5},\, 1]\,Z_\odot$ causes both the posterior median and the $\mathcal{L}$-weighted mean of $Z_\star$ to decline sharply (by nearly a factor of $50$). For SFR$_\mathrm{10}$, the minimum $\chi^2$ method yields closer agreement with the true value than the other two, although SFR$_\mathrm{10}$ remains generally underestimated by a (relatively low) factor. This underestimation indicates that most posteriors lack recent star formation, even when solutions with non-zero SFR in the past 10 Myr provide lower $\chi^2$ values. In contrast, SFR$_\mathrm{100}$ is less sensitive to the chosen statistic, although the minimum $\chi^2$ method still produces the largest scatter.

The double power law and flexible SFH models display similar behaviour. The minimum $\chi^2$ statistic more effectively recovers $M_\star$ than the other two measures, although the scatter is comparable to that of the constant SFH model. For $Z_\star$, the $\mathcal{L}$-weighted mean provides the most reasonable estimates, even when the prior range is extended. While the posterior median of $Z_\star$ is strongly influenced by the adopted prior, both the minimum $\chi^2$ and $\mathcal{L}$-weighted mean are less sensitive to this effect. For SFR$_\mathrm{10}$, the minimum $\chi^2$ achieves slightly better agreement with the true value, whereas SFR$_\mathrm{100}$ remains largely unaffected by the choice of summary statistics.

We attribute these performance differences to two main factors. Firstly, \bagpipes\ is a Bayesian inference code that explores the full prior space. Therefore, parameters that are weakly constrained (either because the model SED is relatively insensitive to them or because they suffer from strong degeneracies) can yield posterior distributions that are heavily shaped by the assumed priors. In such cases, summary statistics such as the median or $\mathcal{L}$-weighted mean become less reliable, whereas the minimum $\chi^2$ estimate often tracks the true value more closely. Secondly, for strongly constrained parameters, where small changes lead to large variations in the model SED, the minimum $\chi^2$ estimate is more susceptible to outliers. In these cases, the $\mathcal{L}$-weighted mean, which considers the entire posterior distribution, often provides a more reliable estimate.

In summary, the minimum $\chi^2$ statistic more accurately recovers $M_\star$ than the other two, albeit with substantial scatter. In contrast, the $\mathcal{L}$-weighted mean yields a more accurate estimate of $Z_\star$, although it remains sensitive to the prior range, particularly under the constant SFH model. SFR$_\mathrm{10}$ is more reliably recovered with the minimum $\chi^2$, especially when the constant SFH model is applied, whereas SFR$_\mathrm{100}$ appears largely insensitive to the choice of the summary statistic. Taken together, these results support the use of flexible or double power-law SFH models in combination with the minimum $\chi^2$ statistic for estimating stellar mass, while the $\mathcal{L}$-weighted mean might be preferable for recovering other parameters (see Table~\ref{tab:summary_table1} for detailed values).

\subsection{Impact on derived galaxy population statistics} 

Stellar mass and SFR are widely used to characterise galaxy populations across cosmic time and an accurate recovery of these properties from SED fitting is essential for constraining galaxy formation models. In this sub-section, we examine how the inferred stellar mass and SFR influence the SMF and star formation main sequence (SFMS). Here, $M_\mathrm{\star,curr}$ refers to the total stellar mass of a galaxy at $z=6$; namely, the mass that remains in stars after stellar mass loss has been accounted for, rather than the total mass ever formed in stars.

\subsubsection{Stellar mass functions}

\begin{figure}
\includegraphics[width=\linewidth]{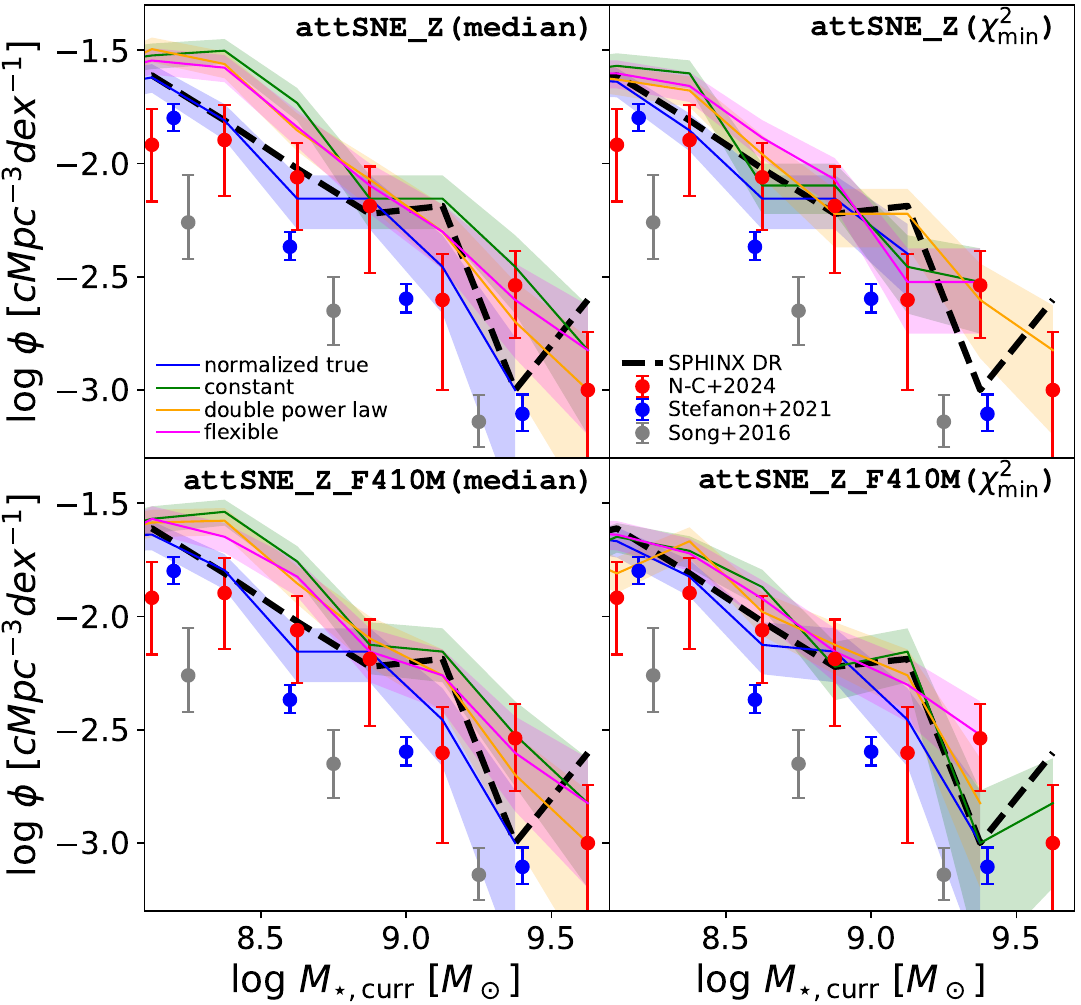}
    \caption{SMFs of star-forming galaxies at $z=6$. The dashed gray line denotes the true SMF from the \sphinx\ simulation, while the solid coloured lines show the SMFs derived from the \texttt{attSNE\_Z} (upper panel) and \texttt{attSNE\_Z\_F410M} (lower panel) models assuming the SMC dust law. Shaded regions mark Poisson uncertainties. The left panels use the posterior median, and the right panels use the minimum $\chi^2$ posterior. Adding the F410M photometry yields a modest improvement in agreement with the true SMF when the median posterior is adopted, whereas the minimum $\chi^2$ posterior produces a more accurate SMF, even without the extra band.}
    \label{fig:stellar_mass_function}
\end{figure}

Figure~\ref{fig:stellar_mass_function} displays the SMFs of star-forming galaxies at $z=6$, derived from the \sphinx\ data \citep[dashed black lines;][]{Katz2023}. The upper panels illustrate the SMFs obtained with the \texttt{attSNE\_Z} model and an SMC-type dust law, while the lower panels show results from \texttt{attSNE\_Z\_F410M}. The left and right panels present results based on the median posterior and the minimum $\chi^2$ estimate, respectively. For \texttt{attSNE\_Z} with the median posterior, the SMF shows an offset of $\sim0.3$ dex at $M_\mathrm{\star,curr} \sim 10^{8.5}M_\odot$ relative to the true SMF, arising from stellar mass overestimation in low-mass galaxies that shifts the function rightward. The SMFs from the \texttt{attSNE\_Z\_F410M} model are nearly identical, indicating that inclusion of the medium band has little influence on the overall shape. By contrast, the offset is notably reduced with the minimum $\chi^2$ estimate, implying that the SMF can be more reliably recovered using NGDEEP-like broad-band photometry. 

Having established a baseline for uncertainties in stellar mass estimates, we can then compare the SMFs from \sphinx\ with empirical determinations. Figure~\ref{fig:stellar_mass_function} displays SMFs from the studies of \citet{Song2016}, \citet{Stefanon2021}, and \citet{Navarro2024}, each derived via SED fitting of galaxy samples at $z\sim6$. The first two are based on HST data, whereas the latter relies on JWST observations. As reported by \citet{Katz2023}, \sphinx\ galaxies generally overproduce stars compared to \citet{Song2016} and \citet{Stefanon2021}. Because uncertainties in SED fitting typically bias stellar masses upward, this offset implies that the true discrepancy could be even larger. By contrast, the SMF obtained by \citet{Navarro2024} agrees more closely with \sphinx, suggesting possible consistency between the simulation and JWST-based measurements. Nevertheless, given the limited survey area of that study, wider-field observations will be crucial for establishing whether the apparent agreement is genuine or merely coincidental.

\begin{figure}
\includegraphics[width=\linewidth]{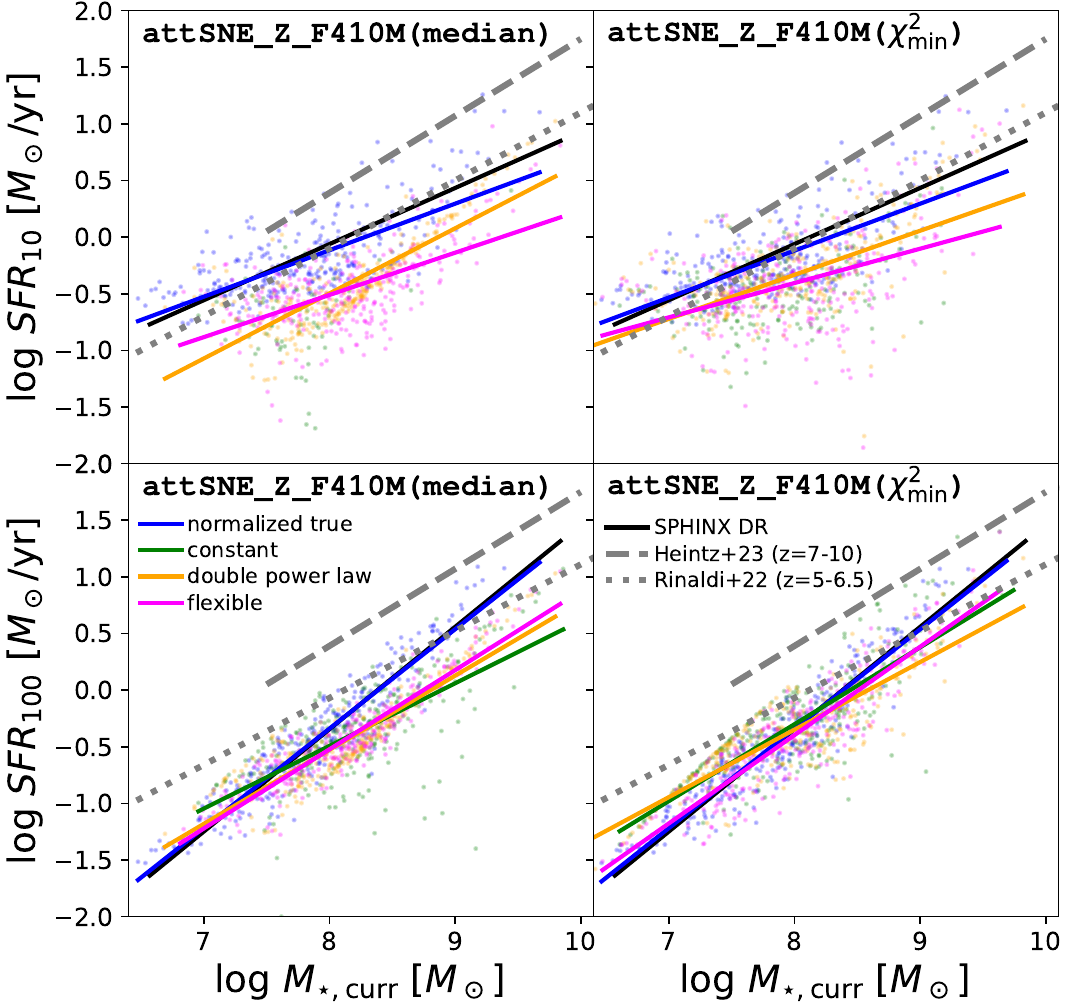}
    \caption{SFMSs of the sample galaxies at $z=6$. The upper and lower panels present the SFMSs constructed with SFR$_\mathrm{10}$ and SFR$_\mathrm{100}$, respectively, while the left and right panels show results obtained from the median posterior and the posterior with the minimum $\chi^2$, respectively. The solid black line marks the best fit to the true \sphinx\ data, and the solid coloured lines indicate best-fit relations derived with different SFH models in the \texttt{attSNE\_Z\_F410M} model. The best fit for the constant SFH model is excluded from the upper panels, as many galaxies return SFR$_\mathrm{10}$ values of zero. The minimum $\chi^2$ posterior more accurately reproduces the true SFMS from the simulation more closely, although it lies slightly below current high-$z$ observations \citep{Heintz2023,Rinaldi2023}.}
    \label{fig:main_sequence}
\end{figure}

\subsubsection{Star formation main sequence}

We also examined the SFMS to gain insight into ongoing star formation across the galaxy population. Figure~\ref{fig:main_sequence} illustrates the SFMSs derived from the \texttt{attSNE\_Z\_F410M} model and emphasises how parameter biases affect the inferred relation. The upper and lower panels correspond to SFR$_\mathrm{10}$ and SFR$_\mathrm{100}$, respectively, while the left and right panels contrast the median posterior with the minimum $\chi^2$. The SFMS from the \texttt{attSNE\_Z} model is not included, as its behaviour largely mirrors that of \texttt{attSNE\_Z\_F410M}, apart from the constant SFH case. This model consistently yields the least accurate SFR, irrespective of the addition of F410M photometry. 

The normalised true SFH model closely matches the true SFMS relation (solid black line), which represents the best-fit trend of the \sphinx\ simulation at $z=6$. This results from the fact that, under a true SFH, variations in stellar mass formed scale directly with the SFR, limiting deviations from the intrinsic relation. In contrast, the constant SFH model markedly underestimates SFR$_\mathrm{10}$, making it the least reliable model for recovering the SFMS, regardless of the posterior statistic applied. Accordingly, the best-fit line for the constant SFH model is not displayed in the upper panel of Fig.~\ref{fig:main_sequence}. The limitation is alleviated in the $M_\mathrm{\star,curr}$--SFR$_\mathrm{100}$ plane and is further suppressed when the minimum $\chi^2$ posterior is employed. The double power-law and flexible SFH models produce broadly consistent results. Using the median posterior, galaxies are inferred to form stars less actively on 10 Myr timescales than they actually do in \sphinx. This bias stems from the joint effect of $M_\star$ overestimation and SFR$_\mathrm{10}$ underestimation. Adopting the minimum $\chi^2$ posterior mitigates $M_\star$ overestimation and thereby reduces the deviation from the true SFMS, although an offset remains for galaxies with $M_\mathrm{\star,curr}\lesssim10^{8.5}M_\odot$.

Although the $M_\mathrm{\star,curr}$--$\mathrm{SFR}_{10}$ relation is sensitive to both the SFH model and the choice of summary statistic, the SFMS constructed from SFR$_\mathrm{100}$ reproduces the true relation more reliably. In the lower panel of Fig.~\ref{fig:main_sequence}, low-mass galaxies generally lie above the SFMS of the \sphinx\ data release (solid black line), whereas intermediate- and high-mass galaxies lie below it. This behaviour is a result of the systematic overestimation of SFR$_\mathrm{100}$ in low-mass galaxies and its underestimation in high-mass galaxies, as illustrated in the bottom row of Fig.~\ref{fig:median_best}. A comparable trend is evident in Figs.~\ref{fig:sfh_median} and ~\ref{fig:sfr_intr}: as $M_\star$ is overestimated owing to the enhanced contribution of older stellar populations, SFR$_\mathrm{100}$ is correspondingly underestimated. This effect is absent in low-mass galaxies, as most of their stars have formed within the past 100 Myr. Consequently, the reconstructed SFMS slopes are systematically shallower than the true relation, even when the minimum $\chi^2$ is adopted. Notably, recent observations by \citet{Heintz2023} and \citet{Rinaldi2023} report a slope consistent with that of our reconstructed SFMSs, although the \sphinx\ galaxies generally {are inferred to} form stars at a lower rate at fixed stellar mass. This offset again likely reflects the excessive star formation in the simulation, leading to a horizontal shift in the relation. More importantly, the shallower slope inferred from SED fitting underscores the need for accurate and flexible SFH models to reliably recover the SFMS at high redshift.

\begin{figure}
\includegraphics[width=\linewidth]{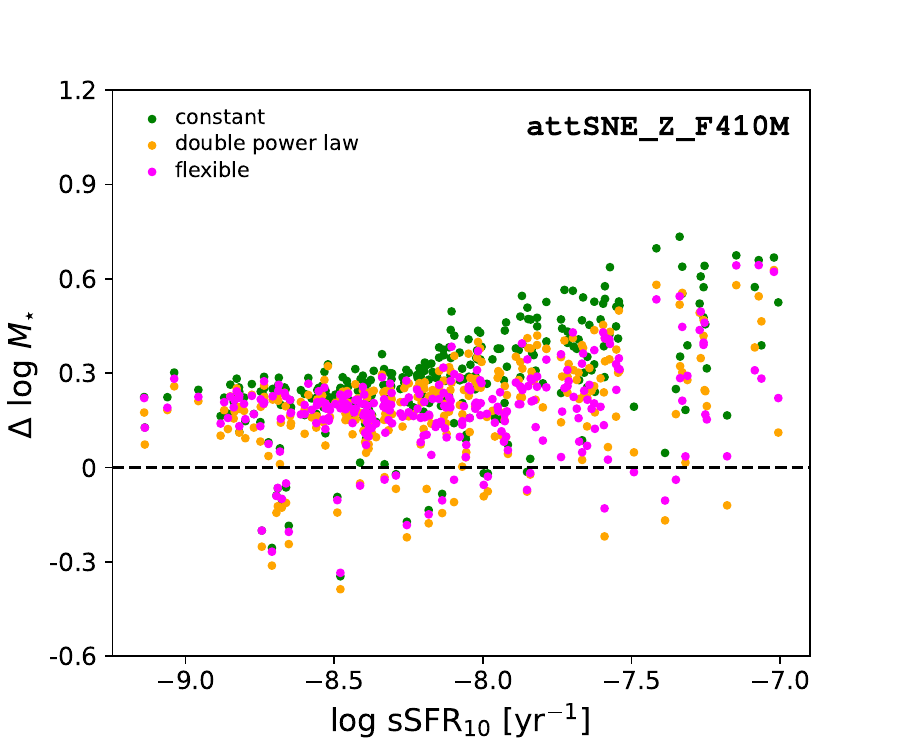}
    \caption{ Uncertainties in the inferred stellar mass as a function of specific SFR for the \texttt{attSNE\_Z\_F410M} model (median posterior). Different colours indicate results obtained with different SFH models, as shown in the legend. Stellar masses are increasingly overestimated at higher specific SFRs. }
    \label{fig:sSFR}
\end{figure}

\subsubsection{Overabundance of bright galaxies}

Finally, we discuss the implications of our results for the overabundance of bright galaxies observed at high redshift \citep{Lovell2022, Finkelstein2023, Harikane2024}, a phenomenon often regarded as a challenge to the $\Lambda$CDM framework. Figure~\ref{fig:sSFR} demonstrates that stellar masses inferred for actively star-forming galaxies can easily deviate by as much as a factor of two when the median posterior is adopted. This offset increases with specific SFR (sSFR), reaching $\sim$0.6 dex (approximately a factor of four) at $\mathrm{sSFR} \sim 10^{-7}\,\mathrm{yr^{-1}}$. If star formation in high-$z$ galaxies is more bursty than in \sphinx\ \citep[e.g.][]{Mauerhofer2025}, the SED fitting might considerably overestimate stellar masses because of short-lived starbursts and strong emission lines. However, this does not mean that the observed tension can be explained solely by SED-fitting uncertainties, as the comparison is performed directly in the UV rather than in stellar mass \citep[e.g.][]{Yung2024, Mauerhofer2025}. Instead, our findings reinforce previous arguments that the prevalence of UV-bright galaxies does not necessarily correspond to an overabundance of massive galaxies \citep{Katz2023, Yung2024}. By contrast, \citet{Narayanan2024} contended that massive systems ($\mstar \sim 10^9\,\msun$) are largely dominated in mass by old stars, while outshining from young stellar populations leads to a systematic underestimation of $M_\star$. Because the test simulations, such as SIMBA \citep{Dave2019} and SMUGGLE \citep{Marinacci2019}, and the SED-fitting methods \citep[\prospector,][]{Leja2017} differ, the origin of this discrepancy remains unresolved. Nevertheless, both studies emphasise the need for caution when estimating stellar masses and their uncertainties, before appealing to more exotic explanations for the excess of bright galaxies, such as non-standard cosmology.

\section{Summary and conclusions}
\label{sec:summary}

In this study, we investigated how assumptions made in the process of performaing an SED fitting, such as the SFH, dust, and nebular emission, affect the recovery of key galaxy physical parameters of galaxies when using JWST/NIRCam photometry. Using a suite of synthetic SEDs of star-forming galaxies at $z=6$ taken from the \sphinx\ simulation \citep{Katz2023} and applying a Bayesian fitting with \bagpipes\ \citep{Carnall2018}, we systematically analysed the impact of each component through a series of model variations. Our primary results are summarised below.

\begin{itemize}
\item In the absence of dust attenuation and nebular emission (i.e. the intrinsic stellar continuum), SED fitting with JWST/NIRCam photometry (covering rest-frame UV to optical bands) systematically overestimates stellar masses (Fig.~\ref{fig:intr_stellar}). Although the mean deviation is modest (0.20 dex), the stellar masses of low-mass galaxies ($M_\star\lesssim10^{7.5}M_\odot$) end up overestimated by up to 0.73 dex. This bias arises primarily from metallicity and age effects, as the fits tend to favour older stellar populations. Consequently, SFR$_\mathrm{10}$ is underestimated by an average of 0.11 dex for the double power-law and flexible SFH models, whereas the constant SFH model yields SFR$_\mathrm{10}=0$ in most cases (Fig.~\ref{fig:sfr_intr}). This trend worsens when the dust attenuation and nebular emission are included, highlighting how these components exacerbate degeneracies among stellar age, metallicity, and mass-to-light ratio.

\item By comparing the slope of the assumed model attenuation curve with that of the true curve from the simulated galaxy, we find that the stellar mass is overestimated when the model curve is shallower and underestimated when it is steeper (Fig.~\ref{fig:S-Av}). This bias persists even when the normalised true SFH is used because mismatches in the attenuation slope force the model to adjust $A_V$ to reproduce the observed colours, thereby yielding an incorrect total attenuation.

\item The choice of the SFH model influences the distribution of the intrinsic UV slope. The constant SFH model leads to the largest overestimation, followed by the double power-law and flexible SFH models. This, in turn, alters the distribution of the inferred colour excess and $A_V$ by causing both to be underestimated, which introduces systematic biases in stellar mass and SFR estimates (Fig.~\ref{fig:beta_and_color}). 

\item Strong emission lines at optical wavelengths redden the observed colours, particularly in low-mass galaxies ($\lesssim10^8 M_\odot$) where nebular contributions are most pronounced (Fig.~\ref{fig:em_contribution}). To reproduce these redder colours, the fitting procedure tends to favour older stellar populations, resulting in stellar masses being overestimated by more than $1$ dex. In contrast, for galaxies with stellar masses above $10^{8.5}\,M_\odot$, where the emission lines contribute only marginally to the total flux, this effect remains minimal.

\item By fitting  SEDs without emission lines in either the model or the input spectrum, we were able to isolate their effect. Stellar mass estimates can vary by up to 0.8 dex solely because of emission lines in the input spectrum (Fig.~\ref{fig:full_Z_noEL}), whereas neglecting them in SED fitting introduces biases of up to 0.3 dex (Fig.~\ref{fig:full_Z_ignoEL}), driven mainly by changes in the inferred stellar population.

\item Including photometric data largely free from strong emission-line contamination, such as the F410M filter, improves the recovery of young stellar populations by constraining the contribution of emission lines relative to the stellar continuum. The overestimation of stellar mass, which reached 1.26 dex for galaxies with $M_\star\lesssim10^8 M_\odot$, was reduced to within 0.7 dex, while the effect was less pronounced in more massive systems (Fig.~\ref{fig:f410}).

\item Furthermore, adopting either minimum $\chi^2$ values or likelihood-weighted means, rather than posterior medians, yields closer agreement with the true bulk galaxy properties. These approaches more reliably recovered key observables, including the SMF and the star-forming main sequence, even under complex modelling assumptions (Figs.~\ref{fig:stellar_mass_function} and~\ref{fig:main_sequence}).
\end{itemize}

Taken together, our analysis underscores the cumulative influence of model assumptions in SED fittings, particularly in regimes where bursty star formation, nebular emission, and dust attenuation are non-negligible. Future studies should account for these factors when interpreting photometric SEDs to better constrain galaxy formation at high redshift, especially for early galaxies, which are increasingly thought to have bursty SFHs. A promising avenue for improvement is machine-learning-based SED fitting trained on successful cosmological simulations, which could yield more accurate stellar mass estimates \citep[e.g.][]{Gilda2021}.

\begin{acknowledgements}
We thank Sukyoung Yi and Hyunmi Song for detailed comments on the manuscript. J.C. and T.K. are supported by the National Research Foundation of Korea (RS-2022-NR070872 and RS-2025-00516961) and also by the Yonsei Fellowship, funded by Lee Youn Jae. T.K. acted as the corresponding author. The large data transfer was supported by KREONET, which is managed and operated by KISTI. 
\end{acknowledgements}

\bibliographystyle{aa}
\bibliography{refs}

@article{Leung2025,
	adsurl = {https://ui.adsabs.harvard.edu/abs/2025ApJ...992...26L},
	archiveprefix = {arXiv},
	author = {{Leung}, Gene C.~K. and {Finkelstein}, Steven L. and {P{\'e}rez-Gonz{\'a}lez}, Pablo G. and {Morales}, Alexa M. and {Taylor}, Anthony J. and {Barro}, Guillermo and {Kocevski}, Dale D. and {Akins}, Hollis B. and {Carnall}, Adam C. and {Ch{\'a}vez Ortiz}, {\'O}scar A. and {Cleri}, Nikko J. and {Cullen}, Fergus and {Donnan}, Callum T. and {Dunlop}, James S. and {Ellis}, Richard S. and {Grogin}, Norman A. and {Hirschmann}, Michaela and {Koekemoer}, Anton M. and {Kokorev}, Vasily and {Lucas}, Ray A. and {McLeod}, Derek J. and {Papovich}, Casey and {Yung}, L.~Y. Aaron},
	doi = {10.3847/1538-4357/adfcce},
	eid = {26},
	eprint = {2411.12005},
	journal = {\apj},
	month = oct,
	number = {1},
	pages = {26},
	primaryclass = {astro-ph.GA},
	title = {{Exploring the Nature of Little Red Dots: Constraints on Active Galactic Nucleus and Stellar Contributions from PRIMER MIRI Imaging}},
	volume = {992},
	year = 2025}

@article{Shen2023,
	adsurl = {https://ui.adsabs.harvard.edu/abs/2023MNRAS.525.3254S},
	archiveprefix = {arXiv},
	author = {{Shen}, Xuejian and {Vogelsberger}, Mark and {Boylan-Kolchin}, Michael and {Tacchella}, Sandro and {Kannan}, Rahul},
	doi = {10.1093/mnras/stad2508},
	eprint = {2305.05679},
	journal = {\mnras},
	month = nov,
	number = {3},
	pages = {3254-3261},
	primaryclass = {astro-ph.GA},
	title = {{The impact of UV variability on the abundance of bright galaxies at z {\ensuremath{\geq}} 9}},
	volume = {525},
	year = 2023}

@article{Dave2019,
	adsurl = {https://ui.adsabs.harvard.edu/abs/2019MNRAS.486.2827D},
	archiveprefix = {arXiv},
	author = {{Dav{\'e}}, Romeel and {Angl{\'e}s-Alc{\'a}zar}, Daniel and {Narayanan}, Desika and {Li}, Qi and {Rafieferantsoa}, Mika H. and {Appleby}, Sarah},
	doi = {10.1093/mnras/stz937},
	eprint = {1901.10203},
	journal = {\mnras},
	month = jun,
	number = {2},
	pages = {2827-2849},
	primaryclass = {astro-ph.GA},
	title = {{SIMBA: Cosmological simulations with black hole growth and feedback}},
	volume = {486},
	year = 2019}

@article{Marinacci2019,
	adsurl = {https://ui.adsabs.harvard.edu/abs/2019MNRAS.489.4233M},
	archiveprefix = {arXiv},
	author = {{Marinacci}, Federico and {Sales}, Laura V. and {Vogelsberger}, Mark and {Torrey}, Paul and {Springel}, Volker},
	doi = {10.1093/mnras/stz2391},
	eprint = {1905.08806},
	journal = {\mnras},
	month = nov,
	number = {3},
	pages = {4233-4260},
	primaryclass = {astro-ph.GA},
	title = {{Simulating the interstellar medium and stellar feedback on a moving mesh: implementation and isolated galaxies}},
	volume = {489},
	year = 2019}

@inproceedings{Kroupa2026,
	adsurl = {https://ui.adsabs.harvard.edu/abs/2026enap....2..173K},
	archiveprefix = {arXiv},
	author = {{Kroupa}, Pavel and {Gjergo}, Eda and {Jerabkova}, Tereza and {Yan}, Zhiqiang},
	booktitle = {Encyclopedia of Astrophysics, Volume 2},
	doi = {10.1016/B978-0-443-21439-4.00035-3},
	eprint = {2410.07311},
	month = jan,
	pages = {173-210},
	primaryclass = {astro-ph.GA},
	title = {{The initial mass function of stars}},
	volume = {2},
	year = 2026}

@article{Cameron2024,
	adsurl = {https://ui.adsabs.harvard.edu/abs/2024MNRAS.534..523C},
	archiveprefix = {arXiv},
	author = {{Cameron}, Alex J. and {Katz}, Harley and {Witten}, Callum and {Saxena}, Aayush and {Laporte}, Nicolas and {Bunker}, Andrew J.},
	doi = {10.1093/mnras/stae1547},
	eprint = {2311.02051},
	journal = {\mnras},
	month = oct,
	number = {1},
	pages = {523-543},
	primaryclass = {astro-ph.GA},
	title = {{Nebular dominated galaxies: insights into the stellar initial mass function at high redshift}},
	volume = {534},
	year = 2024}

@article{Yung2024,
	adsurl = {https://ui.adsabs.harvard.edu/abs/2024MNRAS.527.5929Y},
	archiveprefix = {arXiv},
	author = {{Yung}, L.~Y. Aaron and {Somerville}, Rachel S. and {Finkelstein}, Steven L. and {Wilkins}, Stephen M. and {Gardner}, Jonathan P.},
	doi = {10.1093/mnras/stad3484},
	eprint = {2304.04348},
	journal = {\mnras},
	month = jan,
	number = {3},
	pages = {5929-5948},
	primaryclass = {astro-ph.GA},
	title = {{Are the ultra-high-redshift galaxies at z > 10 surprising in the context of standard galaxy formation models?}},
	volume = {527},
	year = 2024}

@article{Anders2003,
	adsurl = {https://ui.adsabs.harvard.edu/abs/2003A&A...401.1063A},
	archiveprefix = {arXiv},
	author = {{Anders}, P. and {Fritze-v. Alvensleben}, U.},
	doi = {10.1051/0004-6361:20030151},
	eprint = {astro-ph/0302146},
	journal = {\aap},
	month = apr,
	pages = {1063-1070},
	primaryclass = {astro-ph},
	title = {{Spectral and photometric evolution of young stellar populations: The impact of gaseous emission at various metallicities}},
	volume = {401},
	year = 2003}

@article{Naab2017,
	adsurl = {https://ui.adsabs.harvard.edu/abs/2017ARA&A..55...59N},
	archiveprefix = {arXiv},
	author = {{Naab}, Thorsten and {Ostriker}, Jeremiah P.},
	doi = {10.1146/annurev-astro-081913-040019},
	eprint = {1612.06891},
	journal = {\araa},
	month = aug,
	number = {1},
	pages = {59-109},
	primaryclass = {astro-ph.GA},
	title = {{Theoretical Challenges in Galaxy Formation}},
	volume = {55},
	year = 2017}

@article{Marshall2022,
	adsurl = {https://ui.adsabs.harvard.edu/abs/2022MNRAS.516.1047M},
	archiveprefix = {arXiv},
	author = {{Marshall}, Madeline A. and {Watts}, Katelyn and {Wilkins}, Stephen and {Di Matteo}, Tiziana and {Kuusisto}, Jussi K. and {Roper}, William J. and {Vijayan}, Aswin P. and {Ni}, Yueying and {Feng}, Yu and {Croft}, Rupert A.~C.},
	doi = {10.1093/mnras/stac2111},
	eprint = {2206.08941},
	journal = {\mnras},
	month = oct,
	number = {1},
	pages = {1047-1061},
	primaryclass = {astro-ph.GA},
	title = {{The BLUETIDES mock image catalogue: simulated observations of high-redshift galaxies and predictions for JWST imaging surveys}},
	volume = {516},
	year = 2022}

@article{Narayanan2024,
	adsurl = {https://ui.adsabs.harvard.edu/abs/2024ApJ...961...73N},
	archiveprefix = {arXiv},
	author = {{Narayanan}, Desika and {Lower}, Sidney and {Torrey}, Paul and {Brammer}, Gabriel and {Cui}, Weiguang and {Dav{\'e}}, Romeel and {Iyer}, Kartheik G. and {Li}, Qi and {Lovell}, Christopher C. and {Sales}, Laura V. and {Stark}, Daniel P. and {Marinacci}, Federico and {Vogelsberger}, Mark},
	doi = {10.3847/1538-4357/ad0966},
	eid = {73},
	eprint = {2306.10118},
	journal = {\apj},
	month = jan,
	number = {1},
	pages = {73},
	primaryclass = {astro-ph.GA},
	title = {{Outshining by Recent Star Formation Prevents the Accurate Measurement of High-z Galaxy Stellar Masses}},
	volume = {961},
	year = 2024}

@article{Sun2023,
	adsurl = {https://ui.adsabs.harvard.edu/abs/2023ApJ...955L..35S},
	archiveprefix = {arXiv},
	author = {{Sun}, Guochao and {Faucher-Gigu{\`e}re}, Claude-Andr{\'e} and {Hayward}, Christopher C. and {Shen}, Xuejian and {Wetzel}, Andrew and {Cochrane}, Rachel K.},
	doi = {10.3847/2041-8213/acf85a},
	eid = {L35},
	eprint = {2307.15305},
	journal = {\apjl},
	month = oct,
	number = {2},
	pages = {L35},
	primaryclass = {astro-ph.GA},
	title = {{Bursty Star Formation Naturally Explains the Abundance of Bright Galaxies at Cosmic Dawn}},
	volume = {955},
	year = 2023}

@article{Finkelstein2023,
	adsurl = {https://ui.adsabs.harvard.edu/abs/2023ApJ...946L..13F},
	archiveprefix = {arXiv},
	author = {{Finkelstein}, Steven L. and {Bagley}, Micaela B. and {Ferguson}, Henry C. and {Wilkins}, Stephen M. and {Kartaltepe}, Jeyhan S. and {Papovich}, Casey and {Yung}, L.~Y. Aaron and {Haro}, Pablo Arrabal and {Behroozi}, Peter and {Dickinson}, Mark and {Kocevski}, Dale D. and {Koekemoer}, Anton M. and {Larson}, Rebecca L. and {Le Bail}, Aur{\'e}lien and {Morales}, Alexa M. and {P{\'e}rez-Gonz{\'a}lez}, Pablo G. and {Burgarella}, Denis and {Dav{\'e}}, Romeel and {Hirschmann}, Michaela and {Somerville}, Rachel S. and {Wuyts}, Stijn and {Bromm}, Volker and {Casey}, Caitlin M. and {Fontana}, Adriano and {Fujimoto}, Seiji and {Gardner}, Jonathan P. and {Giavalisco}, Mauro and {Grazian}, Andrea and {Grogin}, Norman A. and {Hathi}, Nimish P. and {Hutchison}, Taylor A. and {Jha}, Saurabh W. and {Jogee}, Shardha and {Kewley}, Lisa J. and {Kirkpatrick}, Allison and {Long}, Arianna S. and {Lotz}, Jennifer M. and {Pentericci}, Laura and {Pierel}, Justin D.~R. and {Pirzkal}, Nor and {Ravindranath}, Swara and {Ryan}, Russell E. and {Trump}, Jonathan R. and {Yang}, Guang and {Bhatawdekar}, Rachana and {Bisigello}, Laura and {Buat}, V{\'e}ronique and {Calabr{\`o}}, Antonello and {Castellano}, Marco and {Cleri}, Nikko J. and {Cooper}, M.~C. and {Croton}, Darren and {Daddi}, Emanuele and {Dekel}, Avishai and {Elbaz}, David and {Franco}, Maximilien and {Gawiser}, Eric and {Holwerda}, Benne W. and {Huertas-Company}, Marc and {Jaskot}, Anne E. and {Leung}, Gene C.~K. and {Lucas}, Ray A. and {Mobasher}, Bahram and {Pandya}, Viraj and {Tacchella}, Sandro and {Weiner}, Benjamin J. and {Zavala}, Jorge A.},
	doi = {10.3847/2041-8213/acade4},
	eid = {L13},
	eprint = {2211.05792},
	journal = {\apjl},
	month = mar,
	number = {1},
	pages = {L13},
	primaryclass = {astro-ph.GA},
	title = {{CEERS Key Paper. I. An Early Look into the First 500 Myr of Galaxy Formation with JWST}},
	volume = {946},
	year = 2023}

@article{Calzetti2000,
	adsurl = {https://ui.adsabs.harvard.edu/abs/2000ApJ...533..682C},
	archiveprefix = {arXiv},
	author = {{Calzetti}, Daniela and {Armus}, Lee and {Bohlin}, Ralph C. and {Kinney}, Anne L. and {Koornneef}, Jan and {Storchi-Bergmann}, Thaisa},
	doi = {10.1086/308692},
	eprint = {astro-ph/9911459},
	journal = {\apj},
	month = apr,
	number = {2},
	pages = {682-695},
	primaryclass = {astro-ph},
	title = {{The Dust Content and Opacity of Actively Star-forming Galaxies}},
	volume = {533},
	year = 2000}

@article{delooze2014,
	adsurl = {https://ui.adsabs.harvard.edu/abs/2014A&A...568A..62D},
	archiveprefix = {arXiv},
	author = {{De Looze}, Ilse and {Cormier}, Diane and {Lebouteiller}, Vianney and {Madden}, Suzanne and {Baes}, Maarten and {Bendo}, George J. and {Boquien}, M{\'e}d{\'e}ric and {Boselli}, Alessandro and {Clements}, David L. and {Cortese}, Luca and {Cooray}, Asantha and {Galametz}, Maud and {Galliano}, Fr{\'e}d{\'e}ric and {Graci{\'a}-Carpio}, Javier and {Isaak}, Kate and {Karczewski}, Oskar {\L}. and {Parkin}, Tara J. and {Pellegrini}, Eric W. and {R{\'e}my-Ruyer}, Aur{\'e}lie and {Spinoglio}, Luigi and {Smith}, Matthew W.~L. and {Sturm}, Eckhard},
	doi = {10.1051/0004-6361/201322489},
	eid = {A62},
	eprint = {1402.4075},
	journal = {\aap},
	month = aug,
	pages = {A62},
	primaryclass = {astro-ph.GA},
	title = {{The applicability of far-infrared fine-structure lines as star formation rate tracers over wide ranges of metallicities and galaxy types}},
	volume = {568},
	year = 2014}

@article{Byler2017,
	adsurl = {https://ui.adsabs.harvard.edu/abs/2017ApJ...840...44B},
	archiveprefix = {arXiv},
	author = {{Byler}, Nell and {Dalcanton}, Julianne J. and {Conroy}, Charlie and {Johnson}, Benjamin D.},
	doi = {10.3847/1538-4357/aa6c66},
	eid = {44},
	eprint = {1611.08305},
	journal = {\apj},
	month = may,
	number = {1},
	pages = {44},
	primaryclass = {astro-ph.GA},
	title = {{Nebular Continuum and Line Emission in Stellar Population Synthesis Models}},
	volume = {840},
	year = 2017}

@article{Pacifici2023,
	adsurl = {https://ui.adsabs.harvard.edu/abs/2023ApJ...944..141P},
	archiveprefix = {arXiv},
	author = {{Pacifici}, Camilla and {Iyer}, Kartheik G. and {Mobasher}, Bahram and {da Cunha}, Elisabete and {Acquaviva}, Viviana and {Burgarella}, Denis and {Calistro Rivera}, Gabriela and {Carnall}, Adam C. and {Chang}, Yu-Yen and {Chartab}, Nima and {Cooke}, Kevin C. and {Fairhurst}, Ciaran and {Kartaltepe}, Jeyhan and {Leja}, Joel and {Ma{\l}ek}, Katarzyna and {Salmon}, Brett and {Torelli}, Marianna and {Vidal-Garc{\'\i}a}, Alba and {Boquien}, M{\'e}d{\'e}ric and {Brammer}, Gabriel G. and {Brown}, Michael J.~I. and {Capak}, Peter L. and {Chevallard}, Jacopo and {Circosta}, Chiara and {Croton}, Darren and {Davidzon}, Iary and {Dickinson}, Mark and {Duncan}, Kenneth J. and {Faber}, Sandra M. and {Ferguson}, Harry C. and {Fontana}, Adriano and {Guo}, Yicheng and {Haeussler}, Boris and {Hemmati}, Shoubaneh and {Jafariyazani}, Marziye and {Kassin}, Susan A. and {Larson}, Rebecca L. and {Lee}, Bomee and {Mantha}, Kameswara Bharadwaj and {Marchi}, Francesca and {Nayyeri}, Hooshang and {Newman}, Jeffrey A. and {Pandya}, Viraj and {Pforr}, Janine and {Reddy}, Naveen and {Sanders}, Ryan and {Shah}, Ekta and {Shahidi}, Abtin and {Stevans}, Matthew L. and {Triani}, Dian Puspita and {Tyler}, Krystal D. and {Vanderhoof}, Brittany N. and {de la Vega}, Alexander and {Wang}, Weichen and {Weston}, Madalyn E.},
	doi = {10.3847/1538-4357/acacff},
	eid = {141},
	eprint = {2212.01915},
	journal = {\apj},
	month = feb,
	number = {2},
	pages = {141},
	primaryclass = {astro-ph.GA},
	title = {{The Art of Measuring Physical Parameters in Galaxies: A Critical Assessment of Spectral Energy Distribution Fitting Techniques}},
	volume = {944},
	year = 2023}

@article{Chevallard2016,
	adsurl = {https://ui.adsabs.harvard.edu/abs/2016MNRAS.462.1415C},
	archiveprefix = {arXiv},
	author = {{Chevallard}, Jacopo and {Charlot}, St{\'e}phane},
	doi = {10.1093/mnras/stw1756},
	eprint = {1603.03037},
	journal = {\mnras},
	month = oct,
	number = {2},
	pages = {1415-1443},
	primaryclass = {astro-ph.GA},
	title = {{Modelling and interpreting spectral energy distributions of galaxies with BEAGLE}},
	volume = {462},
	year = 2016}

@article{Carnall2018,
	adsurl = {https://ui.adsabs.harvard.edu/abs/2018MNRAS.480.4379C},
	archiveprefix = {arXiv},
	author = {{Carnall}, A.~C. and {McLure}, R.~J. and {Dunlop}, J.~S. and {Dav{\'e}}, R.},
	doi = {10.1093/mnras/sty2169},
	eprint = {1712.04452},
	journal = {\mnras},
	month = nov,
	number = {4},
	pages = {4379-4401},
	primaryclass = {astro-ph.GA},
	title = {{Inferring the star formation histories of massive quiescent galaxies with BAGPIPES: evidence for multiple quenching mechanisms}},
	volume = {480},
	year = 2018}

@article{Rieke2023,
	adsurl = {https://ui.adsabs.harvard.edu/abs/2023ApJS..269...16R},
	archiveprefix = {arXiv},
	author = {{Rieke}, Marcia J. and {Robertson}, Brant and {Tacchella}, Sandro and {Hainline}, Kevin and {Johnson}, Benjamin D. and {Hausen}, Ryan and {Ji}, Zhiyuan and {Willmer}, Christopher N.~A. and {Eisenstein}, Daniel J. and {Pusk{\'a}s}, D{\'a}vid and {Alberts}, Stacey and {Arribas}, Santiago and {Baker}, William M. and {Baum}, Stefi and {Bhatawdekar}, Rachana and {Bonaventura}, Nina and {Boyett}, Kristan and {Bunker}, Andrew J. and {Cameron}, Alex J. and {Carniani}, Stefano and {Charlot}, Stephane and {Chevallard}, Jacopo and {Chen}, Zuyi and {Curti}, Mirko and {Curtis-Lake}, Emma and {Danhaive}, A. Lola and {DeCoursey}, Christa and {Dressler}, Alan and {Egami}, Eiichi and {Endsley}, Ryan and {Helton}, Jakob M. and {Hviding}, Raphael E. and {Kumari}, Nimisha and {Looser}, Tobias J. and {Lyu}, Jianwei and {Maiolino}, Roberto and {Maseda}, Michael V. and {Nelson}, Erica J. and {Rieke}, George and {Rix}, Hans-Walter and {Sandles}, Lester and {Saxena}, Aayush and {Sharpe}, Katherine and {Shivaei}, Irene and {Skarbinski}, Maya and {Smit}, Renske and {Stark}, Daniel P. and {Stone}, Meredith and {Suess}, Katherine A. and {Sun}, Fengwu and {Topping}, Michael and {{\"U}bler}, Hannah and {Villanueva}, Natalia C. and {Wallace}, Imaan E.~B. and {Williams}, Christina C. and {Willott}, Chris and {Whitler}, Lily and {Witstok}, Joris and {Woodrum}, Charity},
	doi = {10.3847/1538-4365/acf44d},
	eid = {16},
	eprint = {2306.02466},
	journal = {\apjs},
	month = nov,
	number = {1},
	pages = {16},
	primaryclass = {astro-ph.GA},
	title = {{JADES Initial Data Release for the Hubble Ultra Deep Field: Revealing the Faint Infrared Sky with Deep JWST NIRCam Imaging}},
	volume = {269},
	year = 2023}

@article{Kimm2014,
	adsurl = {https://ui.adsabs.harvard.edu/abs/2014ApJ...788..121K},
	archiveprefix = {arXiv},
	author = {{Kimm}, Taysun and {Cen}, Renyue},
	doi = {10.1088/0004-637X/788/2/121},
	eid = {121},
	eprint = {1405.0552},
	journal = {\apj},
	month = jun,
	number = {2},
	pages = {121},
	primaryclass = {astro-ph.GA},
	title = {{Escape Fraction of Ionizing Photons during Reionization: Effects due to Supernova Feedback and Runaway OB Stars}},
	volume = {788},
	year = 2014}

@article{Rosen1995,
	adsurl = {https://ui.adsabs.harvard.edu/abs/1995ApJ...440..634R},
	author = {{Rosen}, Alexander and {Bregman}, Joel N.},
	doi = {10.1086/175303},
	journal = {\apj},
	month = feb,
	pages = {634},
	title = {{Global Models of the Interstellar Medium in Disk Galaxies}},
	volume = {440},
	year = 1995}

@article{Tweed2009,
	adsurl = {https://ui.adsabs.harvard.edu/abs/2009A&A...506..647T},
	archiveprefix = {arXiv},
	author = {{Tweed}, D. and {Devriendt}, J. and {Blaizot}, J. and {Colombi}, S. and {Slyz}, A.},
	doi = {10.1051/0004-6361/200911787},
	eprint = {0902.0679},
	journal = {\aap},
	month = nov,
	number = {2},
	pages = {647-660},
	primaryclass = {astro-ph.CO},
	title = {{Building merger trees from cosmological N-body simulations. Towards improving galaxy formation models using subhaloes}},
	volume = {506},
	year = 2009}

@article{Garel2021,
	adsurl = {https://ui.adsabs.harvard.edu/abs/2021MNRAS.504.1902G},
	archiveprefix = {arXiv},
	author = {{Garel}, Thibault and {Blaizot}, J{\'e}r{\'e}my and {Rosdahl}, Joakim and {Michel-Dansac}, L{\'e}o and {Haehnelt}, Martin G. and {Katz}, Harley and {Kimm}, Taysun and {Verhamme}, Anne},
	doi = {10.1093/mnras/stab990},
	eprint = {2104.03339},
	journal = {\mnras},
	month = jun,
	number = {2},
	pages = {1902-1926},
	primaryclass = {astro-ph.GA},
	title = {{Ly {\ensuremath{\alpha}} as a tracer of cosmic reionization in the SPHINX radiation-hydrodynamics cosmological simulation}},
	volume = {504},
	year = 2021}

@article{Rosdahl2018,
	adsurl = {https://ui.adsabs.harvard.edu/abs/2018MNRAS.479..994R},
	archiveprefix = {arXiv},
	author = {{Rosdahl}, Joakim and {Katz}, Harley and {Blaizot}, J{\'e}r{\'e}my and {Kimm}, Taysun and {Michel-Dansac}, L{\'e}o and {Garel}, Thibault and {Haehnelt}, Martin and {Ocvirk}, Pierre and {Teyssier}, Romain},
	doi = {10.1093/mnras/sty1655},
	eprint = {1801.07259},
	journal = {\mnras},
	month = sep,
	number = {1},
	pages = {994-1016},
	primaryclass = {astro-ph.GA},
	title = {{The SPHINX cosmological simulations of the first billion years: the impact of binary stars on reionization}},
	volume = {479},
	year = 2018}

@article{Yung2019,
	adsurl = {https://ui.adsabs.harvard.edu/abs/2019MNRAS.483.2983Y},
	archiveprefix = {arXiv},
	author = {{Yung}, L.~Y. Aaron and {Somerville}, Rachel S. and {Finkelstein}, Steven L. and {Popping}, Gerg{\"o} and {Dav{\'e}}, Romeel},
	doi = {10.1093/mnras/sty3241},
	eprint = {1803.09761},
	journal = {\mnras},
	month = mar,
	number = {3},
	pages = {2983-3006},
	primaryclass = {astro-ph.GA},
	title = {{Semi-analytic forecasts for JWST - I. UV luminosity functions at z = 4-10}},
	volume = {483},
	year = 2019}

@article{Curti2023,
	adsurl = {https://ui.adsabs.harvard.edu/abs/2024A&A...684A..75C},
	archiveprefix = {arXiv},
	author = {{Curti}, Mirko and {Maiolino}, Roberto and {Curtis-Lake}, Emma and {Chevallard}, Jacopo and {Carniani}, Stefano and {D'Eugenio}, Francesco and {Looser}, Tobias J. and {Scholtz}, Jan and {Charlot}, Stephane and {Cameron}, Alex and {{\"U}bler}, Hannah and {Witstok}, Joris and {Boyett}, Kristian and {Laseter}, Isaac and {Sandles}, Lester and {Arribas}, Santiago and {Bunker}, Andrew and {Giardino}, Giovanna and {Maseda}, Michael V. and {Rawle}, Tim and {Rodr{\'\i}guez Del Pino}, Bruno and {Smit}, Renske and {Willott}, Chris J. and {Eisenstein}, Daniel J. and {Hausen}, Ryan and {Johnson}, Benjamin and {Rieke}, Marcia and {Robertson}, Brant and {Tacchella}, Sandro and {Williams}, Christina C. and {Willmer}, Christopher and {Baker}, William M. and {Bhatawdekar}, Rachana and {Egami}, Eiichi and {Helton}, Jakob M. and {Ji}, Zhiyuan and {Kumari}, Nimisha and {Perna}, Michele and {Shivaei}, Irene and {Sun}, Fengwu},
	doi = {10.1051/0004-6361/202346698},
	eid = {A75},
	eprint = {2304.08516},
	journal = {\aap},
	month = apr,
	pages = {A75},
	primaryclass = {astro-ph.GA},
	title = {{JADES: Insights into the low-mass end of the mass-metallicity-SFR relation at 3 < z < 10 from deep JWST/NIRSpec spectroscopy}},
	volume = {684},
	year = 2024}

@article{Ferland2017,
	adsurl = {https://ui.adsabs.harvard.edu/abs/2017RMxAA..53..385F},
	archiveprefix = {arXiv},
	author = {{Ferland}, G.~J. and {Chatzikos}, M. and {Guzm{\'a}n}, F. and {Lykins}, M.~L. and {van Hoof}, P.~A.~M. and {Williams}, R.~J.~R. and {Abel}, N.~P. and {Badnell}, N.~R. and {Keenan}, F.~P. and {Porter}, R.~L. and {Stancil}, P.~C.},
	doi = {10.48550/arXiv.1705.10877},
	eprint = {1705.10877},
	journal = {\rmxaa},
	month = oct,
	pages = {385-438},
	primaryclass = {astro-ph.GA},
	title = {{The 2017 Release Cloudy}},
	volume = {53},
	year = 2017}

@article{Gordon2003,
	adsurl = {https://ui.adsabs.harvard.edu/abs/2003ApJ...594..279G},
	archiveprefix = {arXiv},
	author = {{Gordon}, Karl D. and {Clayton}, Geoffrey C. and {Misselt}, K.~A. and {Landolt}, Arlo U. and {Wolff}, Michael J.},
	doi = {10.1086/376774},
	eprint = {astro-ph/0305257},
	journal = {\apj},
	month = sep,
	number = {1},
	pages = {279-293},
	primaryclass = {astro-ph},
	title = {{A Quantitative Comparison of the Small Magellanic Cloud, Large Magellanic Cloud, and Milky Way Ultraviolet to Near-Infrared Extinction Curves}},
	volume = {594},
	year = 2003}

@article{Michel-Dansac2020,
	adsurl = {https://ui.adsabs.harvard.edu/abs/2020A&A...635A.154M},
	archiveprefix = {arXiv},
	author = {{Michel-Dansac}, L. and {Blaizot}, J. and {Garel}, T. and {Verhamme}, A. and {Kimm}, T. and {Trebitsch}, M.},
	doi = {10.1051/0004-6361/201834961},
	eid = {A154},
	eprint = {2001.11252},
	journal = {\aap},
	month = mar,
	pages = {A154},
	primaryclass = {astro-ph.GA},
	title = {{RASCAS: RAdiation SCattering in Astrophysical Simulations}},
	volume = {635},
	year = 2020}

@article{Rosdahl2022,
	adsurl = {https://ui.adsabs.harvard.edu/abs/2022MNRAS.515.2386R},
	archiveprefix = {arXiv},
	author = {{Rosdahl}, Joakim and {Blaizot}, J{\'e}r{\'e}my and {Katz}, Harley and {Kimm}, Taysun and {Garel}, Thibault and {Haehnelt}, Martin and {Keating}, Laura C. and {Martin-Alvarez}, Sergio and {Michel-Dansac}, L{\'e}o and {Ocvirk}, Pierre},
	doi = {10.1093/mnras/stac1942},
	eprint = {2207.03232},
	journal = {\mnras},
	month = sep,
	number = {2},
	pages = {2386-2414},
	primaryclass = {astro-ph.GA},
	title = {{LyC escape from SPHINX galaxies in the Epoch of Reionization}},
	volume = {515},
	year = 2022}

@article{Song2016,
	author = {Song, Mimi and Finkelstein, Steven L. and Ashby, Matthew L. N. and Grazian, A. and Lu, Yu and Papovich, Casey and Salmon, Brett and Somerville, Rachel S. and Dickinson, Mark and Duncan, K. and Faber, Sandy M. and Fazio, Giovanni G. and Ferguson, Henry C. and Fontana, Adriano and Guo, Yicheng and Hathi, Nimish and Lee, Seong-Kook and Merlin, Emiliano and Willner, S. P.},
	doi = {10.3847/0004-637X/825/1/5},
	issn = {0004-637X},
	journal = {Astrophys. J.},
	month = {jul},
	title = {{The Evolution of the Galaxy Stellar Mass Function at z= 4-8: A Steepening Low-mass-end Slope with Increasing Redshift}},
	url = {http://arxiv.org/abs/1507.05636 http://dx.doi.org/10.3847/0004-637X/825/1/5},
	volume = {825},
	year = {2016}}

@article{Ma2016,
	adsurl = {https://ui.adsabs.harvard.edu/abs/2016MNRAS.456.2140M},
	archiveprefix = {arXiv},
	author = {{Ma}, Xiangcheng and {Hopkins}, Philip F. and {Faucher-Gigu{\`e}re}, Claude-Andr{\'e} and {Zolman}, Nick and {Muratov}, Alexander L. and {Kere{\v{s}}}, Du{\v{s}}an and {Quataert}, Eliot},
	doi = {10.1093/mnras/stv2659},
	eprint = {1504.02097},
	journal = {\mnras},
	month = feb,
	number = {2},
	pages = {2140-2156},
	primaryclass = {astro-ph.GA},
	title = {{The origin and evolution of the galaxy mass-metallicity relation}},
	volume = {456},
	year = 2016}

@article{Rosdahl2013,
	adsurl = {https://ui.adsabs.harvard.edu/abs/2013MNRAS.436.2188R},
	archiveprefix = {arXiv},
	author = {{Rosdahl}, J. and {Blaizot}, J. and {Aubert}, D. and {Stranex}, T. and {Teyssier}, R.},
	doi = {10.1093/mnras/stt1722},
	eprint = {1304.7126},
	journal = {\mnras},
	month = dec,
	number = {3},
	pages = {2188-2231},
	primaryclass = {astro-ph.CO},
	title = {{RAMSES-RT: radiation hydrodynamics in the cosmological context}},
	volume = {436},
	year = 2013}

@article{Kimm2017,
	adsurl = {https://ui.adsabs.harvard.edu/abs/2017MNRAS.466.4826K},
	archiveprefix = {arXiv},
	author = {{Kimm}, Taysun and {Katz}, Harley and {Haehnelt}, Martin and {Rosdahl}, Joakim and {Devriendt}, Julien and {Slyz}, Adrianne},
	doi = {10.1093/mnras/stx052},
	eprint = {1608.04762},
	journal = {\mnras},
	month = apr,
	number = {4},
	pages = {4826-4846},
	primaryclass = {astro-ph.GA},
	title = {{Feedback-regulated star formation and escape of LyC photons from mini-haloes during reionization}},
	volume = {466},
	year = 2017}

@article{Stanway2018,
	adsurl = {https://ui.adsabs.harvard.edu/abs/2018MNRAS.479...75S},
	archiveprefix = {arXiv},
	author = {{Stanway}, E.~R. and {Eldridge}, J.~J.},
	doi = {10.1093/mnras/sty1353},
	eprint = {1805.08784},
	journal = {\mnras},
	month = sep,
	number = {1},
	pages = {75-93},
	primaryclass = {astro-ph.GA},
	title = {{Re-evaluating old stellar populations}},
	volume = {479},
	year = 2018}

@article{Kimm2015,
	adsurl = {https://ui.adsabs.harvard.edu/abs/2015MNRAS.451.2900K},
	archiveprefix = {arXiv},
	author = {{Kimm}, Taysun and {Cen}, Renyue and {Devriendt}, Julien and {Dubois}, Yohan and {Slyz}, Adrianne},
	doi = {10.1093/mnras/stv1211},
	eprint = {1501.05655},
	journal = {\mnras},
	month = aug,
	number = {3},
	pages = {2900-2921},
	primaryclass = {astro-ph.GA},
	title = {{Towards simulating star formation in turbulent high-z galaxies with mechanical supernova feedback}},
	volume = {451},
	year = 2015}

@article{Kannan2020,
	adsurl = {https://ui.adsabs.harvard.edu/abs/2020MNRAS.499.5732K},
	archiveprefix = {arXiv},
	author = {{Kannan}, Rahul and {Marinacci}, Federico and {Vogelsberger}, Mark and {Sales}, Laura V. and {Torrey}, Paul and {Springel}, Volker and {Hernquist}, Lars},
	doi = {10.1093/mnras/staa3249},
	eprint = {1910.14041},
	journal = {\mnras},
	month = dec,
	number = {4},
	pages = {5732-5748},
	primaryclass = {astro-ph.GA},
	title = {{Simulating the interstellar medium of galaxies with radiative transfer, non-equilibrium thermochemistry, and dust}},
	volume = {499},
	year = 2020}

@article{Laursen_2009,
	adsurl = {https://ui.adsabs.harvard.edu/abs/2009ApJ...704.1640L},
	archiveprefix = {arXiv},
	author = {{Laursen}, Peter and {Sommer-Larsen}, Jesper and {Andersen}, Anja C.},
	doi = {10.1088/0004-637X/704/2/1640},
	eprint = {0907.2698},
	journal = {\apj},
	month = oct,
	number = {2},
	pages = {1640-1656},
	primaryclass = {astro-ph.CO},
	title = {{Ly{\ensuremath{\alpha}} Radiative Transfer with Dust: Escape Fractions from Simulated High-Redshift Galaxies}},
	volume = {704},
	year = 2009}

@article{Katz2023,
	adsurl = {https://ui.adsabs.harvard.edu/abs/2023OJAp....6E..44K},
	archiveprefix = {arXiv},
	author = {{Katz}, Harley and {Rosdahl}, Joki and {Kimm}, Taysun and {Blaizot}, Jeremy and {Choustikov}, Nicholas and {Farcy}, Marion and {Garel}, Thibault and {Haehnelt}, Martin G. and {Michel-Dansac}, Leo and {Ocvirk}, Pierre},
	doi = {10.21105/astro.2309.03269},
	eid = {44},
	eprint = {2309.03269},
	journal = {The Open Journal of Astrophysics},
	month = dec,
	pages = {44},
	primaryclass = {astro-ph.GA},
	title = {{The SPHINX Public Data Release: Forward Modelling High-Redshift JWST Observations with Cosmological Radiation Hydrodynamics Simulations}},
	volume = {6},
	year = 2023}

@article{Bruzual2003,
	adsurl = {https://ui.adsabs.harvard.edu/abs/2003MNRAS.344.1000B},
	archiveprefix = {arXiv},
	author = {{Bruzual}, G. and {Charlot}, S.},
	doi = {10.1046/j.1365-8711.2003.06897.x},
	eprint = {astro-ph/0309134},
	journal = {\mnras},
	month = oct,
	number = {4},
	pages = {1000-1028},
	primaryclass = {astro-ph},
	title = {{Stellar population synthesis at the resolution of 2003}},
	volume = {344},
	year = 2003}

@article{Kroupa2002,
	adsurl = {https://ui.adsabs.harvard.edu/abs/2002MNRAS.336.1188K},
	archiveprefix = {arXiv},
	author = {{Kroupa}, P. and {Boily}, C.~M.},
	doi = {10.1046/j.1365-8711.2002.05848.x},
	eprint = {astro-ph/0207514},
	journal = {\mnras},
	month = nov,
	number = {4},
	pages = {1188-1194},
	primaryclass = {astro-ph},
	title = {{On the mass function of star clusters}},
	volume = {336},
	year = 2002}

@article{Feroz2008,
	author = {F. Feroz and M. P. Hobson},
	doi = {10.1111/j.1365-2966.2007.12353.x},
	journal = {Monthly Notices of the Royal Astronomical Society},
	month = {jan},
	number = {2},
	pages = {449--463},
	publisher = {Oxford University Press ({OUP})},
	title = {Multimodal nested sampling: an efficient and robust alternative to Markov Chain Monte Carlo methods for astronomical data analyses},
	url = {https://doi.org/10.1111%2Fj.1365-2966.2007.12353.x},
	volume = {384},
	year = 2008}

@article{Feroz2009,
	author = {F. Feroz and M. P. Hobson and M. Bridges},
	doi = {10.1111/j.1365-2966.2009.14548.x},
	journal = {Monthly Notices of the Royal Astronomical Society},
	month = {oct},
	number = {4},
	pages = {1601--1614},
	publisher = {Oxford University Press ({OUP})},
	title = {{MultiNest}: an efficient and robust Bayesian inference tool for cosmology and particle physics},
	url = {https://doi.org/10.1111%2Fj.1365-2966.2009.14548.x},
	volume = {398},
	year = 2009}

@article{Salim2018,
	adsurl = {https://ui.adsabs.harvard.edu/abs/2018ApJ...859...11S},
	archiveprefix = {arXiv},
	author = {{Salim}, Samir and {Boquien}, M{\'e}d{\'e}ric and {Lee}, Janice C.},
	doi = {10.3847/1538-4357/aabf3c},
	eid = {11},
	eprint = {1804.05850},
	journal = {\apj},
	month = may,
	number = {1},
	pages = {11},
	primaryclass = {astro-ph.GA},
	title = {{Dust Attenuation Curves in the Local Universe: Demographics and New Laws for Star-forming Galaxies and High-redshift Analogs}},
	volume = {859},
	year = 2018}

@article{Leja2017,
	adsurl = {https://ui.adsabs.harvard.edu/abs/2017ApJ...837..170L},
	archiveprefix = {arXiv},
	author = {{Leja}, Joel and {Johnson}, Benjamin D. and {Conroy}, Charlie and {van Dokkum}, Pieter G. and {Byler}, Nell},
	doi = {10.3847/1538-4357/aa5ffe},
	eid = {170},
	eprint = {1609.09073},
	journal = {\apj},
	month = mar,
	number = {2},
	pages = {170},
	primaryclass = {astro-ph.GA},
	title = {{Deriving Physical Properties from Broadband Photometry with Prospector: Description of the Model and a Demonstration of its Accuracy Using 129 Galaxies in the Local Universe}},
	volume = {837},
	year = 2017}

@article{Leja2019,
	adsurl = {https://ui.adsabs.harvard.edu/abs/2019ApJ...876....3L},
	archiveprefix = {arXiv},
	author = {{Leja}, Joel and {Carnall}, Adam C. and {Johnson}, Benjamin D. and {Conroy}, Charlie and {Speagle}, Joshua S.},
	doi = {10.3847/1538-4357/ab133c},
	eid = {3},
	eprint = {1811.03637},
	journal = {\apj},
	month = may,
	number = {1},
	pages = {3},
	primaryclass = {astro-ph.GA},
	title = {{How to Measure Galaxy Star Formation Histories. II. Nonparametric Models}},
	volume = {876},
	year = 2019}

@article{Caputi2017,
	adsurl = {https://ui.adsabs.harvard.edu/abs/2017ApJ...849...45C},
	archiveprefix = {arXiv},
	author = {{Caputi}, K.~I. and {Deshmukh}, S. and {Ashby}, M.~L.~N. and {Cowley}, W.~I. and {Bisigello}, L. and {Fazio}, G.~G. and {Fynbo}, J.~P.~U. and {Le F{\`e}vre}, O. and {Milvang-Jensen}, B. and {Ilbert}, O.},
	doi = {10.3847/1538-4357/aa901e},
	eid = {45},
	eprint = {1705.06179},
	journal = {\apj},
	month = nov,
	number = {1},
	pages = {45},
	primaryclass = {astro-ph.GA},
	title = {{Star Formation in Galaxies at z {\ensuremath{\sim}} 4-5 from the SMUVS Survey: A Clear Starburst/Main-sequence Bimodality for H{\ensuremath{\alpha}} Emitters on the SFR-M* Plane}},
	volume = {849},
	year = 2017}

@article{Rinaldi2023,
	adsurl = {https://ui.adsabs.harvard.edu/abs/2022ApJ...930..128R},
	archiveprefix = {arXiv},
	author = {{Rinaldi}, Pierluigi and {Caputi}, Karina I. and {van Mierlo}, Sophie E. and {Ashby}, Matthew L.~N. and {Caminha}, Gabriel B. and {Iani}, Edoardo},
	doi = {10.3847/1538-4357/ac5d39},
	eid = {128},
	eprint = {2112.03935},
	journal = {\apj},
	month = may,
	number = {2},
	pages = {128},
	primaryclass = {astro-ph.GA},
	title = {{The Galaxy Starburst/Main-sequence Bimodality over Five Decades in Stellar Mass at z {\ensuremath{\approx}} 3-6.5}},
	volume = {930},
	year = 2022}

@article{Dressler2024,
	adsurl = {https://ui.adsabs.harvard.edu/abs/2024ApJ...964..150D},
	archiveprefix = {arXiv},
	author = {{Dressler}, Alan and {Rieke}, Marcia and {Eisenstein}, Daniel and {Stark}, Daniel P. and {Burns}, Chris and {Bhatawdekar}, Rachana and {Bonaventura}, Nina and {Boyett}, Kristan and {Bunker}, Andrew J. and {Carniani}, Stefano and {Charlot}, Stephane and {Hausen}, Ryan and {Misselt}, Karl and {Tacchella}, Sandro and {Willmer}, Christopher},
	doi = {10.3847/1538-4357/ad1923},
	eid = {150},
	eprint = {2306.02469},
	journal = {\apj},
	month = apr,
	number = {2},
	pages = {150},
	primaryclass = {astro-ph.GA},
	title = {{Building the First Galaxies{\textemdash}Chapter 2. Starbursts Dominate the Star Formation Histories of 6 < z < 12 Galaxies}},
	volume = {964},
	year = 2024}

@article{Dressler2023,
	adsurl = {https://ui.adsabs.harvard.edu/abs/2023ApJ...947L..27D},
	archiveprefix = {arXiv},
	author = {{Dressler}, Alan and {Vulcani}, Benedetta and {Treu}, Tommaso and {Rieke}, Marcia and {Burns}, Chris and {Calabr{\`o}}, Antonello and {Bonchi}, Andrea and {Castellano}, Marco and {Fontana}, Adriano and {Leethochawalit}, Nicha and {Mason}, Charlotte and {Merlin}, Emiliano and {Morishita}, Takahiro and {Paris}, Diego and {Bradac}, Marusa and {Mercurio}, Amata and {Nanayakkara}, Themiya and {Poggianti}, Bianca M. and {Santini}, Paola and {Wang}, Xin and {Misselt}, Karl and {Stark}, Daniel P. and {Willmer}, Christopher},
	doi = {10.3847/2041-8213/ac9ebb},
	eid = {L27},
	eprint = {2208.04292},
	journal = {\apjl},
	month = apr,
	number = {2},
	pages = {L27},
	primaryclass = {astro-ph.GA},
	title = {{Early Results from GLASS-JWST. XVII. Building the First Galaxies-Chapter 1. Star Formation Histories for 5 < z < 7 Galaxies}},
	volume = {947},
	year = 2023}

@article{Hu2023,
	adsurl = {https://ui.adsabs.harvard.edu/abs/2023ApJ...950..132H},
	archiveprefix = {arXiv},
	author = {{Hu}, Chia-Yu and {Smith}, Matthew C. and {Teyssier}, Romain and {Bryan}, Greg L. and {Verbeke}, Robbert and {Emerick}, Andrew and {Somerville}, Rachel S. and {Burkhart}, Blakesley and {Li}, Yuan and {Forbes}, John C. and {Starkenburg}, Tjitske},
	doi = {10.3847/1538-4357/accf9e},
	eid = {132},
	eprint = {2208.10528},
	journal = {\apj},
	month = jun,
	number = {2},
	pages = {132},
	primaryclass = {astro-ph.GA},
	title = {{Code Comparison in Galaxy-scale Simulations with Resolved Supernova Feedback: Lagrangian versus Eulerian Methods}},
	volume = {950},
	year = 2023}

@article{Qin2022,
	author = {Qin, Jianbo and Zheng, Xian Zhong and Fang, Min and Pan, Zhizheng and Wuyts, Stijn and Shi, Yong and Peng, Yingjie and Gonzalez, Valentino and Bian, Fuyan and Huang, Jia-Sheng and Gu, Qiu-Sheng and Liu, Wenhao and Tan, Qinghua and Shi, Dong Dong and Ren, Jian and Zhang, Yuheng and Qiao, Man and Wen, Run and Liu, Shuang},
	doi = {10.1093/mnras/stac132},
	eprint = {https://academic.oup.com/mnras/article-pdf/511/1/765/42371071/stac132.pdf},
	issn = {0035-8711},
	journal = {Monthly Notices of the Royal Astronomical Society},
	month = {01},
	number = {1},
	pages = {765-783},
	title = {Systematic biases in determining dust attenuation curves through galaxy SED fitting},
	url = {https://doi.org/10.1093/mnras/stac132},
	volume = {511},
	year = {2022}}

@article{Narayanan2018,
	author = {Narayanan, Desika and Conroy, Charlie and Dav{\'e}, Romeel and Johnson, Benjamin D. and Popping, Gerg{\"o}},
	doi = {10.3847/1538-4357/aaed25},
	journal = {The Astrophysical Journal},
	month = {dec},
	number = {1},
	pages = {70},
	publisher = {The American Astronomical Society},
	title = {A Theory for the Variation of Dust Attenuation Laws in Galaxies},
	url = {https://dx.doi.org/10.3847/1538-4357/aaed25},
	volume = {869},
	year = {2018}}

@article{Trayford2019,
	author = {Trayford, James W and Lagos, Claudia del P and Robotham, Aaron S G and Obreschkow, Danail},
	doi = {10.1093/mnras/stz3234},
	eprint = {https://academic.oup.com/mnras/article-pdf/491/3/3937/33728467/stz3234.pdf},
	issn = {0035-8711},
	journal = {Monthly Notices of the Royal Astronomical Society},
	month = {11},
	number = {3},
	pages = {3937-3951},
	title = {Fade to grey: systematic variation of galaxy attenuation curves with galaxy properties in the eagle simulations},
	url = {https://doi.org/10.1093/mnras/stz3234},
	volume = {491},
	year = {2019}}

@article{Bagley2024,
	adsurl = {https://ui.adsabs.harvard.edu/abs/2024ApJ...965L...6B},
	archiveprefix = {arXiv},
	author = {{Bagley}, Micaela B. and {Pirzkal}, Nor and {Finkelstein}, Steven L. and {Papovich}, Casey and {Berg}, Danielle A. and {Lotz}, Jennifer M. and {Leung}, Gene C.~K. and {Ferguson}, Henry C. and {Koekemoer}, Anton M. and {Dickinson}, Mark and {Kartaltepe}, Jeyhan S. and {Kocevski}, Dale D. and {Somerville}, Rachel S. and {Yung}, L.~Y. Aaron and {Backhaus}, Bren E. and {Casey}, Caitlin M. and {Castellano}, Marco and {Ch{\'a}vez Ortiz}, {\'O}scar A. and {Chworowsky}, Katherine and {Cox}, Isabella G. and {Dav{\'e}}, Romeel and {Davis}, Kelcey and {Estrada-Carpenter}, Vicente and {Fontana}, Adriano and {Fujimoto}, Seiji and {Gardner}, Jonathan P. and {Giavalisco}, Mauro and {Grazian}, Andrea and {Grogin}, Norman A. and {Hathi}, Nimish P. and {Hutchison}, Taylor A. and {Jaskot}, Anne E. and {Jung}, Intae and {Kewley}, Lisa J. and {Kirkpatrick}, Allison and {Larson}, Rebecca L. and {Matharu}, Jasleen and {Natarajan}, Priyamvada and {Pentericci}, Laura and {P{\'e}rez-Gonz{\'a}lez}, Pablo G. and {Ravindranath}, Swara and {Rothberg}, Barry and {Ryan}, Russell and {Shen}, Lu and {Simons}, Raymond C. and {Snyder}, Gregory F. and {Trump}, Jonathan R. and {Wilkins}, Stephen M.},
	doi = {10.3847/2041-8213/ad2f31},
	eid = {L6},
	eprint = {2302.05466},
	journal = {\apjl},
	month = apr,
	number = {1},
	pages = {L6},
	primaryclass = {astro-ph.GA},
	title = {{The Next Generation Deep Extragalactic Exploratory Public (NGDEEP) Survey}},
	volume = {965},
	year = 2024}

@article{Csizi2024,
	adsurl = {https://ui.adsabs.harvard.edu/abs/2024A&A...689A..37C},
	archiveprefix = {arXiv},
	author = {{Csizi}, B. and {Tortorelli}, L. and {Siudek}, M. and {Gr{\"u}n}, D. and {Renard}, P. and {Tallada-Cresp{\'\i}}, P. and {S{\'a}nchez}, E. and {Miquel}, R. and {Padilla}, C. and {Garc{\'\i}a-Bellido}, J. and {Gazta{\~n}aga}, E. and {Casas}, R. and {Serrano}, S. and {De Vicente}, J. and {Fernandez}, E. and {Eriksen}, M. and {Manzoni}, G. and {Baugh}, C.~M. and {Carretero}, J. and {Castander}, F.~J.},
	doi = {10.1051/0004-6361/202449838},
	eid = {A37},
	eprint = {2405.20385},
	journal = {\aap},
	month = sep,
	pages = {A37},
	primaryclass = {astro-ph.GA},
	title = {{The PAU Survey: Galaxy stellar population properties estimates with narrowband data}},
	volume = {689},
	year = 2024}

@article{Papovich2001,
	author = {Papovich, Casey and Dickinson, Mark and Ferguson, Henry C.},
	doi = {10.1086/322412},
	journal = {The Astrophysical Journal},
	month = {oct},
	number = {2},
	pages = {620},
	title = {The Stellar Populations and Evolution of Lyman Break Galaxies*},
	url = {https://dx.doi.org/10.1086/322412},
	volume = {559},
	year = {2001}}

@article{Zackrisson2008,
	author = {Zackrisson, E. and Bergvall, N. and Leitet, E.},
	doi = {10.1086/587030},
	journal = {The Astrophysical Journal},
	month = {feb},
	number = {1},
	pages = {L9},
	title = {The Impact of Nebular Emission on the Broadband Fluxes of High-Redshift Galaxies},
	url = {https://dx.doi.org/10.1086/587030},
	volume = {676},
	year = {2008}}

@article{Wilkins2013,
	adsurl = {https://ui.adsabs.harvard.edu/abs/2013MNRAS.435.2885W},
	archiveprefix = {arXiv},
	author = {{Wilkins}, Stephen M. and {Coulton}, William and {Caruana}, Joseph and {Croft}, Rupert and {di Matteo}, Tiziana and {Khandai}, Nishikanta and {Feng}, Yu and {Bunker}, Andrew and {Elbert}, Holly},
	doi = {10.1093/mnras/stt1471},
	eprint = {1308.6146},
	journal = {\mnras},
	month = nov,
	number = {4},
	pages = {2885-2895},
	primaryclass = {astro-ph.CO},
	title = {{Theoretical predictions for the effect of nebular emission on the broad-band photometry of high-redshift galaxies}},
	volume = {435},
	year = 2013}

@article{Miranda2025,
	adsurl = {https://ui.adsabs.harvard.edu/abs/2025A&A...694A.102M},
	archiveprefix = {arXiv},
	author = {{Miranda}, Henrique and {Pappalardo}, Ciro and {Afonso}, Jos{\'e} and {Papaderos}, Polychronis and {Lobo}, Catarina and {Paulino-Afonso}, Ana and {Carvajal}, Rodrigo and {Matute}, Israel and {Lagos}, Patricio and {Barbosa}, Davi},
	doi = {10.1051/0004-6361/202451648},
	eid = {A102},
	eprint = {2412.12060},
	journal = {\aap},
	month = feb,
	pages = {A102},
	primaryclass = {astro-ph.GA},
	title = {{Importance of modelling the nebular continuum in galaxy spectra}},
	volume = {694},
	year = 2025}

@article{Salmon2015,
	adsurl = {https://ui.adsabs.harvard.edu/abs/2015ApJ...799..183S},
	archiveprefix = {arXiv},
	author = {{Salmon}, Brett and {Papovich}, Casey and {Finkelstein}, Steven L. and {Tilvi}, Vithal and {Finlator}, Kristian and {Behroozi}, Peter and {Dahlen}, Tomas and {Dav{\'e}}, Romeel and {Dekel}, Avishai and {Dickinson}, Mark and {Ferguson}, Henry C. and {Giavalisco}, Mauro and {Long}, James and {Lu}, Yu and {Mobasher}, Bahram and {Reddy}, Naveen and {Somerville}, Rachel S. and {Wechsler}, Risa H.},
	doi = {10.1088/0004-637X/799/2/183},
	eid = {183},
	eprint = {1407.6012},
	journal = {\apj},
	month = feb,
	number = {2},
	pages = {183},
	primaryclass = {astro-ph.GA},
	title = {{The Relation between Star Formation Rate and Stellar Mass for Galaxies at 3.5 <= z <= 6.5 in CANDELS}},
	volume = {799},
	year = 2015}

@article{Yuan2019,
	author = {Yuan, Fang-Ting and Burgarella, Denis and Corre, David and Buat, Veronique and Boquien, M{\'e}d{\'e}ric and Shen, Shiyin},
	doi = {10.1051/0004-6361/201935975},
	issn = {1432-0746},
	journal = {\aap},
	month = nov,
	pages = {A123},
	publisher = {EDP Sciences},
	title = {Properties of LBGs with [OIII] detection at z ∼ 3.5: The importance of including nebular emission data in SED fitting},
	url = {http://dx.doi.org/10.1051/0004-6361/201935975},
	volume = {631},
	year = {2019}}

@article{Schaerer2009,
	adsurl = {https://ui.adsabs.harvard.edu/abs/2009A&A...502..423S},
	archiveprefix = {arXiv},
	author = {{Schaerer}, D. and {de Barros}, S.},
	doi = {10.1051/0004-6361/200911781},
	eprint = {0905.0866},
	journal = {\aap},
	month = aug,
	number = {2},
	pages = {423-426},
	primaryclass = {astro-ph.CO},
	title = {{The impact of nebular emission on the ages of z{\ensuremath{\approx}} 6 galaxies}},
	volume = {502},
	year = 2009}

@article{Carniani2024,
	adsurl = {https://ui.adsabs.harvard.edu/abs/2024Natur.633..318C},
	archiveprefix = {arXiv},
	author = {{Carniani}, Stefano and {Hainline}, Kevin and {D'Eugenio}, Francesco and {Eisenstein}, Daniel J. and {Jakobsen}, Peter and {Witstok}, Joris and {Johnson}, Benjamin D. and {Chevallard}, Jacopo and {Maiolino}, Roberto and {Helton}, Jakob M. and {Willott}, Chris and {Robertson}, Brant and {Alberts}, Stacey and {Arribas}, Santiago and {Baker}, William M. and {Bhatawdekar}, Rachana and {Boyett}, Kristan and {Bunker}, Andrew J. and {Cameron}, Alex J. and {Cargile}, Phillip A. and {Charlot}, St{\'e}phane and {Curti}, Mirko and {Curtis-Lake}, Emma and {Egami}, Eiichi and {Giardino}, Giovanna and {Isaak}, Kate and {Ji}, Zhiyuan and {Jones}, Gareth C. and {Kumari}, Nimisha and {Maseda}, Michael V. and {Parlanti}, Eleonora and {P{\'e}rez-Gonz{\'a}lez}, Pablo G. and {Rawle}, Tim and {Rieke}, George and {Rieke}, Marcia and {Del Pino}, Bruno Rodr{\'\i}guez and {Saxena}, Aayush and {Scholtz}, Jan and {Smit}, Renske and {Sun}, Fengwu and {Tacchella}, Sandro and {{\"U}bler}, Hannah and {Venturi}, Giacomo and {Williams}, Christina C. and {Willmer}, Christopher N.~A.},
	doi = {10.1038/s41586-024-07860-9},
	eprint = {2405.18485},
	journal = {\nat},
	month = sep,
	number = {8029},
	pages = {318-322},
	primaryclass = {astro-ph.GA},
	title = {{Spectroscopic confirmation of two luminous galaxies at a redshift of 14}},
	volume = {633},
	year = 2024}

@article{Lovell2022,
	author = {Lovell, Christopher C and Harrison, Ian and Harikane, Yuichi and Tacchella, Sandro and Wilkins, Stephen M},
	doi = {10.1093/mnras/stac3224},
	issn = {1365-2966},
	journal = {Monthly Notices of the Royal Astronomical Society},
	month = nov,
	number = {2},
	pages = {2511--2520},
	publisher = {Oxford University Press (OUP)},
	title = {Extreme value statistics of the halo and stellar mass distributions at high redshift: are JWST results in tension with ΛCDM?},
	url = {http://dx.doi.org/10.1093/mnras/stac3224},
	volume = {518},
	year = {2022}}

@article{Haskell2024,
	adsurl = {https://ui.adsabs.harvard.edu/abs/2024MNRAS.530L...7H},
	archiveprefix = {arXiv},
	author = {{Haskell}, P. and {Das}, S. and {Smith}, D.~J.~B. and {Cochrane}, R.~K. and {Hayward}, C.~C. and {Angl{\'e}s-Alc{\'a}zar}, D.},
	doi = {10.1093/mnrasl/slae019},
	eprint = {2310.16097},
	journal = {\mnras},
	month = may,
	number = {1},
	pages = {L7-L12},
	primaryclass = {astro-ph.GA},
	title = {{Beware the recent past: a bias in spectral energy distribution modelling due to bursty star formation}},
	volume = {530},
	year = 2024}

@article{Meldorf2024,
	adsurl = {https://ui.adsabs.harvard.edu/abs/2024MNRAS.531.3242M},
	archiveprefix = {arXiv},
	author = {{Meldorf}, C. and {Palmese}, A. and {Salim}, S.},
	doi = {10.1093/mnras/stae1373},
	eprint = {2308.13974},
	journal = {\mnras},
	month = jul,
	number = {3},
	pages = {3242-3255},
	primaryclass = {astro-ph.GA},
	title = {{Measuring the dust attenuation law of galaxies using photometric data}},
	volume = {531},
	year = 2024}

@article{Lower2020,
	adsurl = {https://ui.adsabs.harvard.edu/abs/2020ApJ...904...33L},
	archiveprefix = {arXiv},
	author = {{Lower}, Sidney and {Narayanan}, Desika and {Leja}, Joel and {Johnson}, Benjamin D. and {Conroy}, Charlie and {Dav{\'e}}, Romeel},
	doi = {10.3847/1538-4357/abbfa7},
	eid = {33},
	eprint = {2006.03599},
	journal = {\apj},
	month = nov,
	number = {1},
	pages = {33},
	primaryclass = {astro-ph.GA},
	title = {{How Well Can We Measure the Stellar Mass of a Galaxy: The Impact of the Assumed Star Formation History Model in SED Fitting}},
	volume = {904},
	year = 2020}

@article{Jain2024,
	adsurl = {https://ui.adsabs.harvard.edu/abs/2024MNRAS.527.3291J},
	archiveprefix = {arXiv},
	author = {{Jain}, Shweta and {Tacchella}, Sandro and {Mosleh}, Moein},
	doi = {10.1093/mnras/stad3333},
	eprint = {2310.18462},
	journal = {\mnras},
	month = jan,
	number = {2},
	pages = {3291-3305},
	primaryclass = {astro-ph.GA},
	title = {{The motivation for flexible star-formation histories from spatially resolved scales within galaxies}},
	volume = {527},
	year = 2024}

@article{Harvey2025,
	adsurl = {https://ui.adsabs.harvard.edu/abs/2025ApJ...978...89H},
	archiveprefix = {arXiv},
	author = {{Harvey}, Thomas and {Conselice}, Christopher J. and {Adams}, Nathan J. and {Austin}, Duncan and {Juod{\v{z}}balis}, Ignas and {Trussler}, James and {Li}, Qiong and {Ormerod}, Katherine and {Ferreira}, Leonardo and {Lovell}, Christopher C. and {Duan}, Qiao and {Westcott}, Lewi and {Harris}, Honor and {Bhatawdekar}, Rachana and {Coe}, Dan and {Cohen}, Seth H. and {Caruana}, Joseph and {Cheng}, Cheng and {Driver}, Simon P. and {Frye}, Brenda and {Furtak}, Lukas J. and {Grogin}, Norman A. and {Hathi}, Nimish P. and {Holwerda}, Benne W. and {Jansen}, Rolf A. and {Koekemoer}, Anton M. and {Marshall}, Madeline A. and {Nonino}, Mario and {Vijayan}, Aswin P. and {Wilkins}, Stephen M. and {Windhorst}, Rogier and {Willmer}, Christopher N.~A. and {Yan}, Haojing and {Zitrin}, Adi},
	doi = {10.3847/1538-4357/ad8c29},
	eid = {89},
	eprint = {2403.03908},
	journal = {\apj},
	month = jan,
	number = {1},
	pages = {89},
	primaryclass = {astro-ph.GA},
	title = {{EPOCHS. IV. SED Modeling Assumptions and Their Impact on the Stellar Mass Function at 6.5 {\ensuremath{\leq}} z {\ensuremath{\leq}} 13.5 Using PEARLS and Public JWST Observations}},
	volume = {978},
	year = 2025}

@article{daCunha2015,
	adsurl = {https://ui.adsabs.harvard.edu/abs/2015ApJ...806..110D},
	archiveprefix = {arXiv},
	author = {{da Cunha}, E. and {Walter}, F. and {Smail}, I.~R. and {Swinbank}, A.~M. and {Simpson}, J.~M. and {Decarli}, R. and {Hodge}, J.~A. and {Weiss}, A. and {van der Werf}, P.~P. and {Bertoldi}, F. and {Chapman}, S.~C. and {Cox}, P. and {Danielson}, A.~L.~R. and {Dannerbauer}, H. and {Greve}, T.~R. and {Ivison}, R.~J. and {Karim}, A. and {Thomson}, A.},
	doi = {10.1088/0004-637X/806/1/110},
	eid = {110},
	eprint = {1504.04376},
	journal = {\apj},
	month = jun,
	number = {1},
	pages = {110},
	primaryclass = {astro-ph.GA},
	title = {{An ALMA Survey of Sub-millimeter Galaxies in the Extended Chandra Deep Field South: Physical Properties Derived from Ultraviolet-to-radio Modeling}},
	volume = {806},
	year = 2015}

@article{Hayes2011,
	adsurl = {https://ui.adsabs.harvard.edu/abs/2011ApJ...730....8H},
	archiveprefix = {arXiv},
	author = {{Hayes}, Matthew and {Schaerer}, Daniel and {{\"O}stlin}, G{\"o}ran and {Mas-Hesse}, J. Miguel and {Atek}, Hakim and {Kunth}, Daniel},
	doi = {10.1088/0004-637X/730/1/8},
	eid = {8},
	eprint = {1010.4796},
	journal = {\apj},
	month = mar,
	number = {1},
	pages = {8},
	primaryclass = {astro-ph.CO},
	title = {{On the Redshift Evolution of the Ly{\ensuremath{\alpha}} Escape Fraction and the Dust Content of Galaxies}},
	volume = {730},
	year = 2011}

@article{Wilkins2020,
	adsurl = {https://ui.adsabs.harvard.edu/abs/2020MNRAS.493.6079W},
	archiveprefix = {arXiv},
	author = {{Wilkins}, Stephen M. and {Lovell}, Christopher C. and {Fairhurst}, Ciaran and {Feng}, Yu and {Di Matteo}, Tiziana and {Croft}, Rupert and {Kuusisto}, Jussi and {Vijayan}, Aswin P. and {Thomas}, Peter},
	doi = {10.1093/mnras/staa649},
	eprint = {1904.07504},
	journal = {\mnras},
	month = apr,
	number = {4},
	pages = {6079-6094},
	primaryclass = {astro-ph.GA},
	title = {{Nebular-line emission during the Epoch of Reionization}},
	volume = {493},
	year = 2020}

@inproceedings{Esther2015,
	adsurl = {https://ui.adsabs.harvard.edu/abs/2015hsa8.conf..263M},
	author = {{M{\'a}rmol-Queralt{\'o}}, E. and {McLure}, R. and {Cullen}, F.},
	booktitle = {Highlights of Spanish Astrophysics VIII},
	editor = {{Cenarro}, A.~J. and {Figueras}, F. and {Hern{\'a}ndez-Monteagudo}, C. and {Trujillo Bueno}, J. and {Valdivielso}, L.},
	month = may,
	pages = {263-267},
	title = {{Nebular emission lines in high redshift galaxies}},
	year = 2015}

@article{Traina2024,
	adsurl = {https://ui.adsabs.harvard.edu/abs/2024A&A...690A..84T},
	archiveprefix = {arXiv},
	author = {{Traina}, A. and {Magnelli}, B. and {Gruppioni}, C. and {Delvecchio}, I. and {Parente}, M. and {Calura}, F. and {Bisigello}, L. and {Feltre}, A. and {Pozzi}, F. and {Vallini}, L.},
	doi = {10.1051/0004-6361/202451113},
	eid = {A84},
	eprint = {2407.09607},
	journal = {\aap},
	month = oct,
	pages = {A84},
	primaryclass = {astro-ph.GA},
	title = {{A$^{3}$COSMOS: Dust mass function and dust mass density at 0.5 < z < 6}},
	volume = {690},
	year = 2024}

@article{Heintz2023,
	adsurl = {https://ui.adsabs.harvard.edu/abs/2023A&A...679A..91H},
	archiveprefix = {arXiv},
	author = {{Heintz}, K.~E. and {De Cia}, A. and {Th{\"o}ne}, C.~C. and {Krogager}, J. -K. and {Yates}, R.~M. and {Vejlgaard}, S. and {Konstantopoulou}, C. and {Fynbo}, J.~P.~U. and {Watson}, D. and {Narayanan}, D. and {Wilson}, S.~N. and {Arabsalmani}, M. and {Campana}, S. and {D'Elia}, V. and {De Pasquale}, M. and {Hartmann}, D.~H. and {Izzo}, L. and {Jakobsson}, P. and {Kouveliotou}, C. and {Levan}, A. and {Li}, Q. and {Malesani}, D.~B. and {Melandri}, A. and {Milvang-Jensen}, B. and {M{\o}ller}, P. and {Palazzi}, E. and {Palmerio}, J. and {Petitjean}, P. and {Pugliese}, G. and {Rossi}, A. and {Saccardi}, A. and {Salvaterra}, R. and {Savaglio}, S. and {Schady}, P. and {Stratta}, G. and {Tanvir}, N.~R. and {de Ugarte Postigo}, A. and {Vergani}, S.~D. and {Wiersema}, K. and {Wijers}, R.~A.~M.~J. and {Zafar}, T.},
	doi = {10.1051/0004-6361/202347418},
	eid = {A91},
	eprint = {2308.14812},
	journal = {\aap},
	month = nov,
	pages = {A91},
	primaryclass = {astro-ph.GA},
	title = {{The cosmic buildup of dust and metals. Accurate abundances from GRB-selected star-forming galaxies at 1.7 < z < 6.3}},
	volume = {679},
	year = 2023}

@article{Somerville2015,
	adsurl = {https://ui.adsabs.harvard.edu/abs/2015ARA&A..53...51S},
	archiveprefix = {arXiv},
	author = {{Somerville}, Rachel S. and {Dav{\'e}}, Romeel},
	doi = {10.1146/annurev-astro-082812-140951},
	eprint = {1412.2712},
	journal = {\araa},
	month = aug,
	pages = {51-113},
	primaryclass = {astro-ph.GA},
	title = {{Physical Models of Galaxy Formation in a Cosmological Framework}},
	volume = {53},
	year = 2015}

@article{Robertson2022,
	adsurl = {https://ui.adsabs.harvard.edu/abs/2022ARA&A..60..121R},
	archiveprefix = {arXiv},
	author = {{Robertson}, Brant E.},
	doi = {10.1146/annurev-astro-120221-044656},
	eprint = {2110.13160},
	journal = {\araa},
	month = aug,
	pages = {121-158},
	primaryclass = {astro-ph.CO},
	title = {{Galaxy Formation and Reionization: Key Unknowns and Expected Breakthroughs by the James Webb Space Telescope}},
	volume = {60},
	year = 2022}

@article{Mauerhofer2025,
	adsurl = {https://ui.adsabs.harvard.edu/abs/2025A&A...696A.157M},
	archiveprefix = {arXiv},
	author = {{Mauerhofer}, Valentin and {Dayal}, Pratika and {Haehnelt}, Martin G. and {Kimm}, Taysun and {Rosdahl}, Joakim and {Teyssier}, Romain},
	doi = {10.1051/0004-6361/202554042},
	eid = {A157},
	eprint = {2502.02647},
	journal = {\aap},
	month = apr,
	pages = {A157},
	primaryclass = {astro-ph.GA},
	title = {{Synergising semi-analytical models and hydrodynamical simulations to interpret JWST data from the first billion years}},
	volume = {696},
	year = 2025}

@article{Chemerynska2024,
	adsurl = {https://ui.adsabs.harvard.edu/abs/2024ApJ...976L..15C},
	archiveprefix = {arXiv},
	author = {{Chemerynska}, Iryna and {Atek}, Hakim and {Dayal}, Pratika and {Furtak}, Lukas J. and {Feldmann}, Robert and {Greene}, Jenny E. and {Maseda}, Michael V. and {Nanayakkara}, Themiya and {Oesch}, Pascal A. and {Fujimoto}, Seiji and {Labb{\'e}}, Ivo and {Bezanson}, Rachel and {Brammer}, Gabriel and {Cutler}, Sam E. and {Leja}, Joel and {Pan}, Richard and {Price}, Sedona H. and {Wang}, Bingjie and {Weaver}, John R. and {Whitaker}, Katherine E.},
	doi = {10.3847/2041-8213/ad8dc9},
	eid = {L15},
	eprint = {2407.17110},
	journal = {\apjl},
	month = nov,
	number = {1},
	pages = {L15},
	primaryclass = {astro-ph.GA},
	title = {{The Extreme Low-mass End of the Mass{\textendash}Metallicity Relation at z {\ensuremath{\sim}} 7}},
	volume = {976},
	year = 2024}

@article{Nakajima2023,
	adsurl = {https://ui.adsabs.harvard.edu/abs/2023ApJS..269...33N},
	archiveprefix = {arXiv},
	author = {{Nakajima}, Kimihiko and {Ouchi}, Masami and {Isobe}, Yuki and {Harikane}, Yuichi and {Zhang}, Yechi and {Ono}, Yoshiaki and {Umeda}, Hiroya and {Oguri}, Masamune},
	doi = {10.3847/1538-4365/acd556},
	eid = {33},
	eprint = {2301.12825},
	journal = {\apjs},
	month = dec,
	number = {2},
	pages = {33},
	primaryclass = {astro-ph.GA},
	title = {{JWST Census for the Mass-Metallicity Star Formation Relations at z = 4-10 with Self-consistent Flux Calibration and Proper Metallicity Calibrators}},
	volume = {269},
	year = 2023}

@article{Bethermin2020,
	adsurl = {https://ui.adsabs.harvard.edu/abs/2020Msngr.180...31B},
	author = {{B{\'e}thermin}, M. and {Dessauges-Zavadsky}, M. and {Faisst}, A.~L. and {Ginolfi}, M. and {Gruppioni}, C. and {Jones}, G.~C. and {Khusanova}, Y. and {Lemaux}, B. and {Capak}, P.~L. and {Cassata}, P. and {Le F{\`e}vre}, O. and {Schaerer}, D. and {Silverman}, J.~D. and {Yan}, L. and {Alpine Collaboration}},
	doi = {10.18727/0722-6691/5198},
	journal = {The Messenger},
	month = jun,
	pages = {31-36},
	title = {{The ALPINE-ALMA [CII] Survey: Exploring the Dark Side of Normal Galaxies at the End of Reionisation}},
	volume = {180},
	year = 2020}

@article{Smit2018,
	adsurl = {https://ui.adsabs.harvard.edu/abs/2018Natur.553..178S},
	archiveprefix = {arXiv},
	author = {{Smit}, Renske and {Bouwens}, Rychard J. and {Carniani}, Stefano and {Oesch}, Pascal A. and {Labb{\'e}}, Ivo and {Illingworth}, Garth D. and {van der Werf}, Paul and {Bradley}, Larry D. and {Gonzalez}, Valentino and {Hodge}, Jacqueline A. and {Holwerda}, Benne W. and {Maiolino}, Roberto and {Zheng}, Wei},
	doi = {10.1038/nature24631},
	eprint = {1706.04614},
	journal = {\nat},
	month = jan,
	number = {7687},
	pages = {178-181},
	primaryclass = {astro-ph.GA},
	title = {{Rotation in [C II]-emitting gas in two galaxies at a redshift of 6.8}},
	volume = {553},
	year = 2018}

@article{Jones2022,
	adsurl = {https://ui.adsabs.harvard.edu/abs/2022MNRAS.514.5706J},
	archiveprefix = {arXiv},
	author = {{Jones}, Gareth T. and {Stanway}, Elizabeth R. and {Carnall}, Adam C.},
	doi = {10.1093/mnras/stac1667},
	eprint = {2206.06379},
	journal = {\mnras},
	month = aug,
	number = {4},
	pages = {5706-5724},
	primaryclass = {astro-ph.GA},
	title = {{On the simultaneous modelling of dust and stellar populations for interpretation of galaxy properties}},
	volume = {514},
	year = 2022}

@article{Scoville2007,
	adsurl = {https://ui.adsabs.harvard.edu/abs/2007ApJS..172....1S},
	archiveprefix = {arXiv},
	author = {{Scoville}, N. and {Aussel}, H. and {Brusa}, M. and {Capak}, P. and {Carollo}, C.~M. and {Elvis}, M. and {Giavalisco}, M. and {Guzzo}, L. and {Hasinger}, G. and {Impey}, C. and {Kneib}, J. -P. and {LeFevre}, O. and {Lilly}, S.~J. and {Mobasher}, B. and {Renzini}, A. and {Rich}, R.~M. and {Sanders}, D.~B. and {Schinnerer}, E. and {Schminovich}, D. and {Shopbell}, P. and {Taniguchi}, Y. and {Tyson}, N.~D.},
	doi = {10.1086/516585},
	eprint = {astro-ph/0612305},
	journal = {\apjs},
	month = sep,
	number = {1},
	pages = {1-8},
	primaryclass = {astro-ph},
	title = {{The Cosmic Evolution Survey (COSMOS): Overview}},
	volume = {172},
	year = 2007}

@article{Cochrane2025,
	adsurl = {https://ui.adsabs.harvard.edu/abs/2025ApJ...978L..42C},
	archiveprefix = {arXiv},
	author = {{Cochrane}, R.~K. and {Katz}, H. and {Begley}, R. and {Hayward}, C.~C. and {Best}, P.~N.},
	doi = {10.3847/2041-8213/ad9a4d},
	eid = {L42},
	eprint = {2412.02622},
	journal = {\apjl},
	month = jan,
	number = {2},
	pages = {L42},
	primaryclass = {astro-ph.GA},
	title = {{High-z Stellar Masses Can Be Recovered Robustly with JWST Photometry}},
	volume = {978},
	year = 2025}

@article{Laporte2017,
	adsurl = {https://ui.adsabs.harvard.edu/abs/2017ApJ...837L..21L},
	archiveprefix = {arXiv},
	author = {{Laporte}, N. and {Ellis}, R.~S. and {Boone}, F. and {Bauer}, F.~E. and {Qu{\'e}nard}, D. and {Roberts-Borsani}, G.~W. and {Pell{\'o}}, R. and {P{\'e}rez-Fournon}, I. and {Streblyanska}, A.},
	doi = {10.3847/2041-8213/aa62aa},
	eid = {L21},
	eprint = {1703.02039},
	journal = {\apjl},
	month = mar,
	number = {2},
	pages = {L21},
	primaryclass = {astro-ph.GA},
	title = {{Dust in the Reionization Era: ALMA Observations of a z = 8.38 Gravitationally Lensed Galaxy}},
	volume = {837},
	year = 2017}

@article{McLure2011,
	adsurl = {https://ui.adsabs.harvard.edu/abs/2011MNRAS.418.2074M},
	archiveprefix = {arXiv},
	author = {{McLure}, R.~J. and {Dunlop}, J.~S. and {de Ravel}, L. and {Cirasuolo}, M. and {Ellis}, R.~S. and {Schenker}, M. and {Robertson}, B.~E. and {Koekemoer}, A.~M. and {Stark}, D.~P. and {Bowler}, R.~A.~A.},
	doi = {10.1111/j.1365-2966.2011.19626.x},
	eprint = {1102.4881},
	journal = {\mnras},
	month = dec,
	number = {3},
	pages = {2074-2105},
	primaryclass = {astro-ph.CO},
	title = {{A robust sample of galaxies at redshifts 6.0<z<8.7: stellar populations, star formation rates and stellar masses}},
	volume = {418},
	year = 2011}

@article{Jolly2021,
	adsurl = {https://ui.adsabs.harvard.edu/abs/2021A&A...652A.128J},
	archiveprefix = {arXiv},
	author = {{Jolly}, Jean-Baptiste and {Knudsen}, Kirsten and {Laporte}, Nicolas and {Richard}, Johan and {Fujimoto}, Seiji and {Kohno}, Kotaro and {Ao}, Yiping and {Bauer}, Franz E. and {Egami}, Eiichi and {Espada}, Daniel and {Dessauges-Zavadsky}, Miroslava and {Magdis}, Georgios and {Schaerer}, Daniel and {Sun}, Fengwu and {Valentino}, Francesco and {Wang}, Wei-Hao and {Zitrin}, Adi},
	doi = {10.1051/0004-6361/202140878},
	eid = {A128},
	eprint = {2106.09085},
	journal = {\aap},
	month = aug,
	pages = {A128},
	primaryclass = {astro-ph.GA},
	title = {{ALMA Lensing Cluster Survey: A spectral stacking analysis of [C II] in lensed z {\ensuremath{\sim}} 6 galaxies}},
	volume = {652},
	year = 2021}

@article{Dekel2023,
	adsurl = {https://ui.adsabs.harvard.edu/abs/2023MNRAS.523.3201D},
	archiveprefix = {arXiv},
	author = {{Dekel}, Avishai and {Sarkar}, Kartick C. and {Birnboim}, Yuval and {Mandelker}, Nir and {Li}, Zhaozhou},
	doi = {10.1093/mnras/stad1557},
	eprint = {2303.04827},
	journal = {\mnras},
	month = aug,
	number = {3},
	pages = {3201-3218},
	primaryclass = {astro-ph.GA},
	title = {{Efficient formation of massive galaxies at cosmic dawn by feedback-free starbursts}},
	volume = {523},
	year = 2023}

@article{Roberts2021,
	adsurl = {https://ui.adsabs.harvard.edu/abs/2021ApJ...910...86R},
	archiveprefix = {arXiv},
	author = {{Roberts-Borsani}, Guido and {Treu}, Tommaso and {Mason}, Charlotte and {Schmidt}, Kasper B. and {Jones}, Tucker and {Fontana}, Adriano},
	doi = {10.3847/1538-4357/abe45b},
	eid = {86},
	eprint = {2102.04469},
	journal = {\apj},
	month = apr,
	number = {2},
	pages = {86},
	primaryclass = {astro-ph.GA},
	title = {{Improving z {\ensuremath{\sim}} 7-11 Galaxy Property Estimates with JWST/NIRCam Medium-band Photometry}},
	volume = {910},
	year = 2021}

@article{Ferrara2023,
	adsurl = {https://ui.adsabs.harvard.edu/abs/2023MNRAS.522.3986F},
	archiveprefix = {arXiv},
	author = {{Ferrara}, Andrea and {Pallottini}, Andrea and {Dayal}, Pratika},
	doi = {10.1093/mnras/stad1095},
	eprint = {2208.00720},
	journal = {\mnras},
	month = jul,
	number = {3},
	pages = {3986-3991},
	primaryclass = {astro-ph.GA},
	title = {{On the stunning abundance of super-early, luminous galaxies revealed by JWST}},
	volume = {522},
	year = 2023}

@article{Sabti2024,
	adsurl = {https://ui.adsabs.harvard.edu/abs/2024PhRvL.132f1002S},
	archiveprefix = {arXiv},
	author = {{Sabti}, Nashwan and {Mu{\~n}oz}, Julian B. and {Kamionkowski}, Marc},
	doi = {10.1103/PhysRevLett.132.061002},
	eid = {061002},
	eprint = {2305.07049},
	journal = {\prl},
	month = feb,
	number = {6},
	pages = {061002},
	primaryclass = {astro-ph.CO},
	title = {{Insights from HST into Ultramassive Galaxies and Early-Universe Cosmology}},
	volume = {132},
	year = 2024}

@article{Harikane2024,
	adsurl = {https://ui.adsabs.harvard.edu/abs/2024ApJ...960...56H},
	archiveprefix = {arXiv},
	author = {{Harikane}, Yuichi and {Nakajima}, Kimihiko and {Ouchi}, Masami and {Umeda}, Hiroya and {Isobe}, Yuki and {Ono}, Yoshiaki and {Xu}, Yi and {Zhang}, Yechi},
	doi = {10.3847/1538-4357/ad0b7e},
	eid = {56},
	eprint = {2304.06658},
	journal = {\apj},
	month = jan,
	number = {1},
	pages = {56},
	primaryclass = {astro-ph.GA},
	title = {{Pure Spectroscopic Constraints on UV Luminosity Functions and Cosmic Star Formation History from 25 Galaxies at z $_{spec}$ = 8.61-13.20 Confirmed with JWST/NIRSpec}},
	volume = {960},
	year = 2024}

@article{Navarro2024,
	adsurl = {https://ui.adsabs.harvard.edu/abs/2024ApJ...961..207N},
	archiveprefix = {arXiv},
	author = {{Navarro-Carrera}, Rafael and {Rinaldi}, Pierluigi and {Caputi}, Karina I. and {Iani}, Edoardo and {Kokorev}, Vasily and {van Mierlo}, Sophie E.},
	doi = {10.3847/1538-4357/ad0df6},
	eid = {207},
	eprint = {2305.16141},
	journal = {\apj},
	month = feb,
	number = {2},
	pages = {207},
	primaryclass = {astro-ph.GA},
	title = {{Constraints on the Faint End of the Galaxy Stellar Mass Function at z ≃ 4{\textendash}8 from Deep JWST Data}},
	volume = {961},
	year = 2024}

@article{Stefanon2021,
	adsurl = {https://ui.adsabs.harvard.edu/abs/2021ApJ...922...29S},
	archiveprefix = {arXiv},
	author = {{Stefanon}, Mauro and {Bouwens}, Rychard J. and {Labb{\'e}}, Ivo and {Illingworth}, Garth D. and {Gonzalez}, Valentino and {Oesch}, Pascal A.},
	doi = {10.3847/1538-4357/ac1bb6},
	eid = {29},
	eprint = {2103.16571},
	journal = {\apj},
	month = nov,
	number = {1},
	pages = {29},
	primaryclass = {astro-ph.GA},
	title = {{Galaxy Stellar Mass Functions from z 10 to z 6 using the Deepest Spitzer/Infrared Array Camera Data: No Significant Evolution in the Stellar-to-halo Mass Ratio of Galaxies in the First Gigayear of Cosmic Time}},
	volume = {922},
	year = 2021}

@article{Worthey1994,
	adsurl = {https://ui.adsabs.harvard.edu/abs/1994ApJS...95..107W},
	author = {{Worthey}, Guy},
	doi = {10.1086/192096},
	journal = {\apjs},
	month = nov,
	pages = {107},
	title = {{Comprehensive Stellar Population Models and the Disentanglement of Age and Metallicity Effects}},
	volume = {95},
	year = 1994}

@article{Gallazzi2005,
	adsurl = {https://ui.adsabs.harvard.edu/abs/2005MNRAS.362...41G},
	archiveprefix = {arXiv},
	author = {{Gallazzi}, Anna and {Charlot}, St{\'e}phane and {Brinchmann}, Jarle and {White}, Simon D.~M. and {Tremonti}, Christy A.},
	doi = {10.1111/j.1365-2966.2005.09321.x},
	eprint = {astro-ph/0506539},
	journal = {\mnras},
	month = sep,
	number = {1},
	pages = {41-58},
	primaryclass = {astro-ph},
	title = {{The ages and metallicities of galaxies in the local universe}},
	volume = {362},
	year = 2005}

@article{Conroy2013,
	adsurl = {https://ui.adsabs.harvard.edu/abs/2013ARA&A..51..393C},
	archiveprefix = {arXiv},
	author = {{Conroy}, Charlie},
	doi = {10.1146/annurev-astro-082812-141017},
	eprint = {1301.7095},
	journal = {\araa},
	month = aug,
	number = {1},
	pages = {393-455},
	primaryclass = {astro-ph.CO},
	title = {{Modeling the Panchromatic Spectral Energy Distributions of Galaxies}},
	volume = {51},
	year = 2013}

@article{Abel2002,
	adsurl = {https://ui.adsabs.harvard.edu/abs/2002Sci...295...93A},
	archiveprefix = {arXiv},
	author = {{Abel}, Tom and {Bryan}, Greg L. and {Norman}, Michael L.},
	doi = {10.1126/science.1063991},
	eprint = {astro-ph/0112088},
	journal = {Science},
	month = jan,
	number = {5552},
	pages = {93-98},
	primaryclass = {astro-ph},
	title = {{The Formation of the First Star in the Universe}},
	volume = {295},
	year = 2002}

@article{Bromm2002,
	adsurl = {https://ui.adsabs.harvard.edu/abs/2002ApJ...564...23B},
	archiveprefix = {arXiv},
	author = {{Bromm}, Volker and {Coppi}, Paolo S. and {Larson}, Richard B.},
	doi = {10.1086/323947},
	eprint = {astro-ph/0102503},
	journal = {\apj},
	month = jan,
	number = {1},
	pages = {23-51},
	primaryclass = {astro-ph},
	title = {{The Formation of the First Stars. I. The Primordial Star-forming Cloud}},
	volume = {564},
	year = 2002}

@article{Dayal2018,
	adsurl = {https://ui.adsabs.harvard.edu/abs/2018PhR...780....1D},
	archiveprefix = {arXiv},
	author = {{Dayal}, Pratika and {Ferrara}, Andrea},
	doi = {10.1016/j.physrep.2018.10.002},
	eprint = {1809.09136},
	journal = {\physrep},
	month = dec,
	pages = {1-64},
	primaryclass = {astro-ph.GA},
	title = {{Early galaxy formation and its large-scale effects}},
	volume = {780},
	year = 2018}

@article{Finlator2018,
	adsurl = {https://ui.adsabs.harvard.edu/abs/2018MNRAS.480.2628F},
	archiveprefix = {arXiv},
	author = {{Finlator}, Kristian and {Keating}, Laura and {Oppenheimer}, Benjamin D. and {Dav{\'e}}, Romeel and {Zackrisson}, Erik},
	doi = {10.1093/mnras/sty1949},
	eprint = {1805.00099},
	journal = {\mnras},
	month = oct,
	number = {2},
	pages = {2628-2649},
	primaryclass = {astro-ph.CO},
	title = {{Reionization in Technicolor}},
	volume = {480},
	year = 2018}

@article{Rinaldi2023Ha,
	adsurl = {https://ui.adsabs.harvard.edu/abs/2023ApJ...952..143R},
	archiveprefix = {arXiv},
	author = {{Rinaldi}, P. and {Caputi}, K.~I. and {Costantin}, L. and {Gillman}, S. and {Iani}, E. and {P{\'e}rez-Gonz{\'a}lez}, P.~G. and {{\"O}stlin}, G. and {Colina}, L. and {Greve}, T.~R. and {Noorgard-Nielsen}, H.~U. and {Wright}, G.~S. and {Alonso-Herrero}, A. and {{\'A}lvarez-M{\'a}rquez}, J. and {Eckart}, A. and {Garc{\'\i}a-Mar{\'\i}n}, M. and {Hjorth}, J. and {Ilbert}, O. and {Kendrew}, S. and {Labiano}, A. and {Le F{\`e}vre}, O. and {Pye}, J. and {Tikkanen}, T. and {Walter}, F. and {van der Werf}, P. and {Ward}, M. and {Annunziatella}, M. and {Azzollini}, R. and {Bik}, A. and {Boogaard}, L. and {Bosman}, S.~E.~I. and {Crespo G{\'o}mez}, A. and {Jermann}, I. and {Langeroodi}, D. and {Melinder}, J. and {Meyer}, R.~A. and {Moutard}, T. and {Peissker}, F. and {Topinka}, M. and {van Dishoeck}, E. and {G{\"u}del}, M. and {Henning}, Th. and {Lagage}, P. -O. and {Ray}, T. and {Vandenbussche}, B. and {Waelkens}, C. and {Navarro-Carrera}, R. and {Kokorev}, V.},
	doi = {10.3847/1538-4357/acdc27},
	eid = {143},
	eprint = {2301.10717},
	journal = {\apj},
	month = aug,
	number = {2},
	pages = {143},
	primaryclass = {astro-ph.GA},
	title = {{MIDIS: Strong (H{\ensuremath{\beta}}+[O III]) and H{\ensuremath{\alpha}} Emitters at Redshift z ≃ 7-8 Unveiled with JWST NIRCam and MIRI Imaging in the Hubble eXtreme Deep Field}},
	volume = {952},
	year = 2023}

@article{Narayanan2025,
	adsurl = {https://ui.adsabs.harvard.edu/abs/2025ApJ...982....7N},
	archiveprefix = {arXiv},
	author = {{Narayanan}, Desika and {Stark}, Daniel P. and {Finkelstein}, Steven L. and {Torrey}, Paul and {Li}, Qi and {Cullen}, Fergus and {Topping}, Micheal W. and {Marinacci}, Federico and {Sales}, Laura V. and {Shen}, Xuejian and {Vogelsberger}, Mark},
	doi = {10.3847/1538-4357/adb41c},
	eid = {7},
	eprint = {2408.13312},
	journal = {\apj},
	month = mar,
	number = {1},
	pages = {7},
	primaryclass = {astro-ph.GA},
	title = {{The Ultraviolet Slopes of Early Universe Galaxies: The Impact of Bursty Star Formation, Dust, and Nebular Continuum Emission}},
	volume = {982},
	year = 2025}

@article{Katz2024,
	adsurl = {https://ui.adsabs.harvard.edu/abs/2025OJAp....8E.104K},
	author = {{Katz}, Harley and {Cameron}, Alex J. and {Saxena}, Aayush and {Barrufet}, Laia and {Choustikov}, Nichloas and {Cleri}, Nikko J. and {de Graff}, Anna and {Ellis}, Richard S. and {Fosbury}, Robert A.~E. and {Heintz}, Kasper E. and {Maseda}, Michael and {Matthee}, Jorryt and {McConachie}, Ian and {Oesch}, Pascal A.},
	doi = {10.33232/001c.142570},
	eid = {104},
	journal = {The Open Journal of Astrophysics},
	month = jul,
	pages = {104},
	title = {{21 Balmer Jump Street: The Nebular Continuum at High Redshift and Implications for the Bright Galaxy Problem, UV Continuum Slopes, and Early Stellar Populations}},
	volume = {8},
	year = 2025}

@article{Calzetti1994,
	adsurl = {https://ui.adsabs.harvard.edu/abs/1994ApJ...429..582C},
	author = {{Calzetti}, Daniela and {Kinney}, Anne L. and {Storchi-Bergmann}, Thaisa},
	doi = {10.1086/174346},
	journal = {\apj},
	month = jul,
	pages = {582},
	title = {{Dust Extinction of the Stellar Continua in Starburst Galaxies: The Ultraviolet and Optical Extinction Law}},
	volume = {429},
	year = 1994}

@article{Gilda2021,
	adsurl = {https://ui.adsabs.harvard.edu/abs/2021ApJ...916...43G},
	archiveprefix = {arXiv},
	author = {{Gilda}, Sankalp and {Lower}, Sidney and {Narayanan}, Desika},
	doi = {10.3847/1538-4357/ac0058},
	eid = {43},
	eprint = {2101.04687},
	journal = {\apj},
	month = jul,
	number = {1},
	pages = {43},
	primaryclass = {astro-ph.GA},
	title = {{MIRKWOOD: Fast and Accurate SED Modeling Using Machine Learning}},
	volume = {916},
	year = 2021}

\appendix

\section{Impact of SED template and SFH model choice on fitting}
\label{sec:appendix_A}

As indicated in the main text, we adopt the BPASS SPS model in \bagpipes, as the input SEDs from \sphinx\ are likewise generated with this model. In contrast, \citet{Cochrane2025} employed the \citet[][BC03]{Bruzual2003} model to evaluate the accuracy of recovering stellar masses and SFRs from \sphinx\ SEDs. Therefore, examining how the choice of SED template influences the inferred galaxy properties is necessary.

Figures~\ref{fig:A1} and \ref{fig:A2} illustrate the impact of the SPS template on SED fitting results, highlighting the comparison between BPASS and BC03. We perform intrinsic stellar continuum fitting with several SFH models and compare the resulting $M_\star$ values with those obtained using BPASS (i.e. the \texttt{intS\_Z$_\mathrm{true}$} model in Table~\ref{tab:fitting_sets}). Figure~\ref{fig:A1} displays the difference between the true and fitted $M_\star$ values when each SPS template is applied with the true normalised SFH model, and the inset depicts an example of the intrinsic stellar continuum spanning rest-frame NUV to optical wavelengths. We find that the recovered $M_\star$ with BC03 is slightly larger (by $\sim 10\%$) than that with BPASS. This difference arises from the lower flux at rest-frame optical wavelengths in the stellar continuum when BC03 is applied, as can be seen in the inset of Fig.~\ref{fig:A1}. As shown by \citet{Stanway2018}, BPASS yields slightly more luminous and redder spectra at intermediate ages, driven by differences in the treatment of AGB stars, stellar atmosphere models, and binary evolution. Consequently, when $Z_\star$ is allowed to vary, the fitted $Z_\star$ with BC03 becomes larger to match the redder input colour of the \sphinx\ galaxy, which is based on the BPASS. 

A higher $Z_\star$ generally produces lower fluxes across all wavelengths, which can drive an increase in the fitted $M_\star$ to match the input flux. This preference for higher $Z_\star$ in BC03-based SED fitting persists across different SFH models and remains evident even with more complex approaches, such as \texttt{attSNE\_Z\_F410M}, although the trend weakens as model complexity increases.
This mitigation is illustrated in Fig.~\ref{fig:A2}, which presents the median posterior for \texttt{attSNE\_Z\_F410M} (cf. Fig.~\ref{fig:median_best}). 
Because BPASS generates more UV flux than BC03, the associated emission-line luminosity is brighter for the same SFH, which yields a redder colour (cf., Fig.~\ref{fig:em_contribution}). As the F410M photometry disentangles emission-line contributions from the optical continuum, a higher SFR$_\mathrm{10}$ is required to reproduce the emission-line luminosity when BC03 is employed. At the same time, because the stellar continuum is redder in BPASS (inset of Fig.~\ref{fig:A1}), a higher metallicity is still necessary. While the first effect tends to reduce $M_\star$, the second drives it upward. Moreover, an increase in SFR$_\mathrm{10}$ steepens the UV slope, necessitating a higher colour excess to reproduce the input UV colour. Because the SMC-type dust law raises the colour excess only through a higher $A_V$, the fitted $M_\star$ must also increase to compensate for the added dust extinction. Consequently, when BC03 is adopted, the increase in SFR$_\mathrm{10}$ partly counterbalances the higher $Z_\star$ and $A_V$, yielding a modest overestimation of the fitted $M_\star$. We find that the increases in these three parameters are broadly consistent across other complex SFH models (Fig.~\ref{fig:A2}). Nevertheless, using BC03 introduces only minor changes in the derived stellar mass, and we confirm that these differences are not insufficient to affect our main conclusions.

\begin{figure}
    \centering
\includegraphics[width=0.47\textwidth]{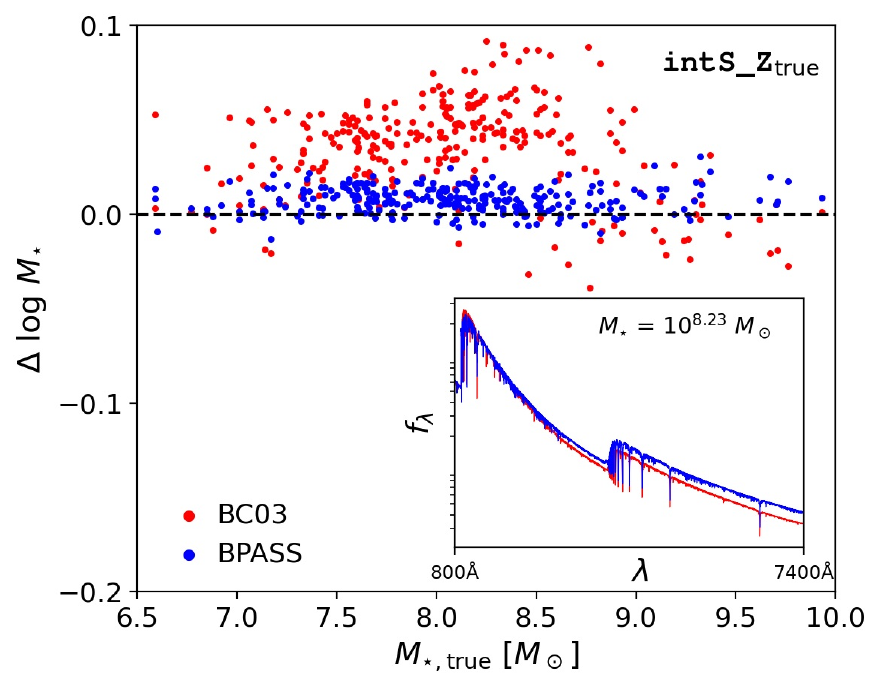}
    \caption{Difference between the true and fitted $M_\star$ for \sphinx\ galaxies at $z=6$ when the normalised true SFH model is fitted with \texttt{intS\_Z$_\mathrm{true}$}. Using BC03 template is used (red) yields systematically larger $M_\star$ values than BPASS (blue). This offset arises because BC03 predicts lower rest-frame optical fluxes than BPASS, even under the same SFH is used. The inset presents an example SED fit with both templates, with the x-axis showing to observed wavelengths from 0.9 to 5.0 microns.}
    \label{fig:A1}
\end{figure}

\begin{figure*}
    \centering
\includegraphics[width=\textwidth]{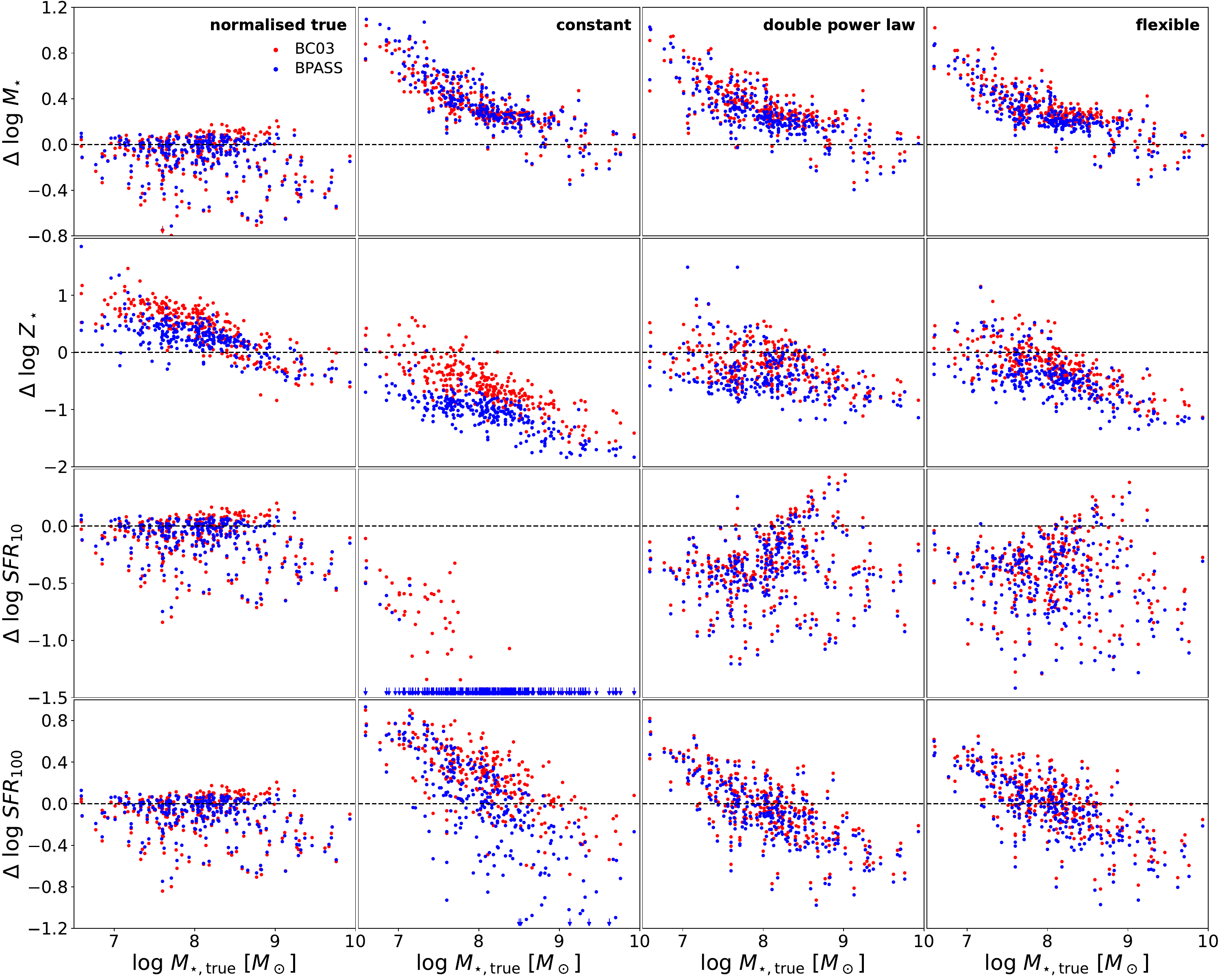}
    \caption{Same as Fig.~\ref{fig:median_best}, but with the BC03 template used for SED fitting. Only median posteriors are displayed. On average, the fitted $M_\star$ and SFRs differ by $\sim 0.04$ dex between templates, whereas the inferred metallicities differ by up to a factor of three. Notably, SFR$_\mathrm{10}$ remains severely underestimated when the constant SFH is fitted with BC03. }
    \label{fig:A2}
\end{figure*}

\section{Mass offset for the true normalise SFH model}
\label{sec:appendix_B}

In Sect.~\ref{sec:dust_impact}, we demonstrate a tight correlation between the UV-optical slope $S$ and $\DeltaMfit$ under the assumption of the true normalised SFH model. Here, we explain our derivation of Eq.~\ref{eq:mass_offset_1} in more detail.

We consider a galaxy for which the intrinsic and attenuated stellar continuum fluxes are denoted by $F(\lambda)$ and $f(\lambda)$, respectively. These two quantities are related by dust attenuation, $A(\lambda)$, as follows:
\begin{equation}
A(\lambda)=-2.5 \log \frac{f(\lambda)}{F(\lambda)}=A_\mathrm{V}\,s(\lambda),
        \label{eq:B1}
\end{equation}
where $s(\lambda)$ denotes the normalised dust attenuation curve (or slope), and $A_V=A(5500\,\text{\AA}$) represents the attenuation at $5500\,\text{\AA}$. As the shape of the intrinsic stellar continuum is determined by the stellar population, we can express $F(\lambda)$ as $F(\lambda)=M_{\star}\,\Upsilon(\lambda)$, where $\Upsilon(\lambda)$ denotes the stellar mass-to-light ratio. Equation~\ref{eq:B1} can then be re-expressed as
\begin{equation}
    \log M_{\star}+\log \Upsilon(\lambda)-\frac{A_\mathrm{V}\, s(\lambda)}{2.5}=\log f(\lambda).
        \label{eq:B2}
\end{equation}
Now consider a fitted model galaxy with the same stellar population but subject to SMC-type dust attenuation, as
\begin{equation}
    \log M_{\star, \mathrm{fit}}+\log \Upsilon(\lambda)-\frac{A_\mathrm{V,fit} \, s_\mathrm{SMC}(\lambda)}{2.5}=\log f_\mathrm{fit}(\lambda),
        \label{eq:B3}
\end{equation}
where $M_{\star,\mathrm{fit}}$, $A_{V, \mathrm{fit}}$ and $f_\mathrm{fit}(\lambda)$ denote the fitted stellar mass, attenuation, and attenuated flux, respectively; and $s_\mathrm{SMC}(\lambda)$ is the SMC attenuation curve. If the fitted model perfectly reproduces the attenuated SED of the galaxy, then $f(\lambda)=f_\mathrm{fit}(\lambda)$ is true for all wavelengths. Combining Eqs.~\ref{eq:B2} and~\ref{eq:B3}, we obtain
\begin{align}
    \log M_{\star,\mathrm{fit}}-\log M_{\star}&=\frac{A_{V, \mathrm{fit}} \, s_\mathrm{SMC}(\lambda)}{2.5}-\frac{A_V \, s(\lambda)}{2.5},\notag \\
    &=\frac{A_V}{2.5}\left[\frac{A_{V, \mathrm{fit}}}{A_V}\, s_\mathrm{SMC}(\lambda)-s(\lambda)\right].
        \label{eq:B4}
\end{align}
At $\lambda=5500\,\text{\AA}$, Eq.~\ref{eq:B4} becomes
\begin{equation}
    \Delta\log M_{\star,\mathrm{fit}}=\frac{A_V}{2.5}\left(\frac{A_{V,\mathrm{fit}}}{A_V}-1\right).
        \label{eq:mass_offset_2}
\end{equation}
Thus, when $A_{V,\mathrm{fit}}$ is larger (smaller) than the true $A_V$, $M_\star$ is overestimated (underestimated, respectively).

\section{Fitting results with redshift as a free parameter}
\label{sec:appendix_C}

As detailed in the main text, the redshift is fixed to its true value to focus on uncertainties in galactic properties, including SFHs. However, the redshift of observed galaxies is typically not known a priori. To assess the effect of redshift uncertainties, we repeat the analysis with the redshift treated as a free parameter, adopting a prior range of 0--10 and using the \texttt{attSNE\_Z\_F410M} prescriptions.

\begin{figure*}
    \centering
\includegraphics[width=0.48\textwidth]{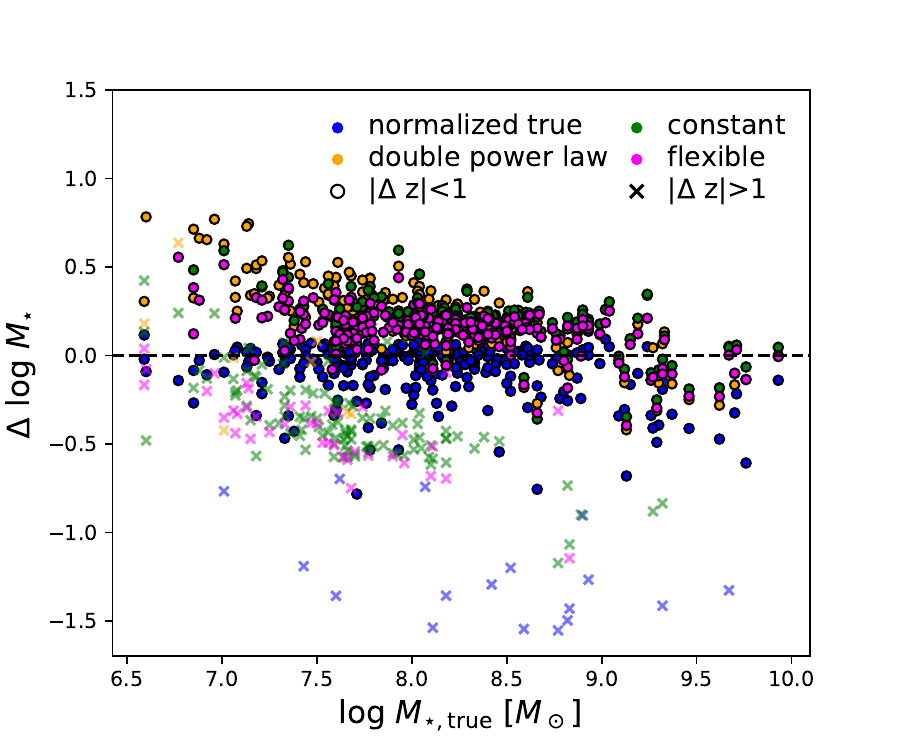}
\includegraphics[width=0.47\textwidth]{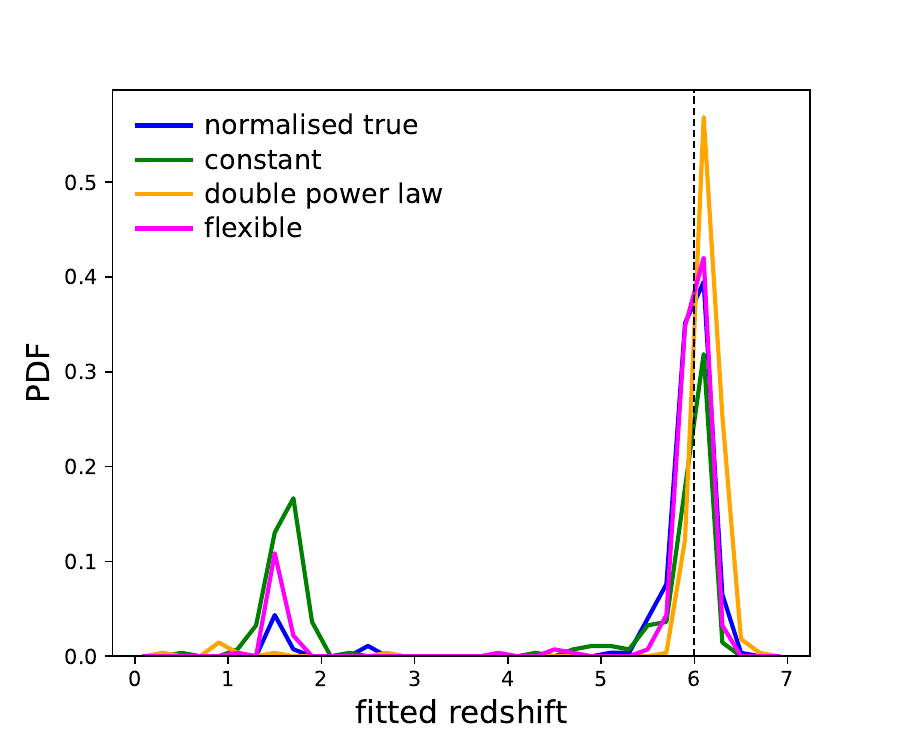}
\caption{
Impact of redshift determination on the inferred stellar mass offset (left). Different colours denote the SFH models listed in the legend. Stellar masses are recovered reasonably well when the estimated redshift lies within $\left|\Delta z\right| < 1$ (circles), whereas greater scatter arises when $\left|\Delta z\right| > 1$ (crosses). The corresponding redshift distributions are displayed in the right panel, with the true redshift ($z=6$) indicated by the dashed line.
}
\label{fig:zfree}
\end{figure*}

Figure~\ref{fig:zfree} presents the fitting results for stellar mass (left) and for redshift (right). The recovered stellar masses remain broadly consistent with those from the \texttt{attSNE\_Z\_F410M} model (lower panel of Fig.~\ref{fig:f410}), when the recovered redshift deviates by less than unity from the true value (i.e. $5\la z \la 7$). Redshift recovery is most reliable with the double power-law ($97\%$), followed by the normalised true ($94\%$), flexible ($85\%$) and constant SFH models ($59\%$). Galaxies wherein the redshift is not recovered typically yield  $z\sim1.5$, as the Balmer break in the fitted SED resembles the Gunn--Peterson trough in the true SED. Our simulated samples are prone to this redshift degeneracy because the filter does not cover the UV region of the SED where IGM attenuation is strong. This degeneracy can be mitigated by employing a filter that covers  shorter wavelengths, such as F070W. Alternatively, adopting the minimum-$\chi^2$ posterior can mitigate this issue, because posteriors with $z\sim1.5$ generally yield larger $\chi^2$ values than those with $z\sim6$.

\section{Uncertainties in derived physical properties from SED fitting}
\label{sec:appendix_D}

We summarise the uncertainties in various physical properties derived from SED fitting in Table~\ref{tab:summary_table1}--\ref{tab:summary_table3}.

\begin{table*}
\caption{Offsets in the inferred physical properties from different SED fitting models for low-mass ($<10^8\,\msun$) galaxies. }
    \renewcommand{\arraystretch}{1.5}
    \centering
    \resizebox{\textwidth}{!}{
    \begin{tabular}{llccccccc}
    \hline \hline
        Variable & SFH model & \texttt{intS\_Z$_\mathrm{true}$} & \texttt{intS\_Z} & \texttt{attS\_Z$_\mathrm{true}$} & \texttt{attS\_Z} & \texttt{attSNE\_Z$_\mathrm{true}$} & \texttt{attSNE\_Z} & \texttt{attSNE\_Z\_F410M}\\
        \hline  
        $\Delta \log M_\mathrm{\star}$ & Normalised true & $0.01\pm0.01$ & $0.01\pm0.01$ & $0.02\pm0.10$ & $0.05\pm0.11$ & $-0.07\pm0.15$ & $-0.05\pm0.16$ & $-0.04\pm0.15$\\
          & Constant & $0.07\pm0.21$ & $0.27\pm0.16$ & $0.12\pm0.17$ & $0.29\pm0.15$ & $0.34\pm0.23$ & $0.50\pm0.22$ & $0.37\pm0.13$\\
          & Double power law & $0.03\pm0.18$ & $0.20\pm0.15$ & $0.09\pm0.13$ & $0.24\pm0.14$ & $0.26\pm0.22$ & $0.36\pm0.19$ & $0.27\pm0.14$\\
          & Flexible & $0.09\pm0.17$ & $0.21\pm0.15$ & $0.12\pm0.13$ & $0.23\pm0.13$ & $0.25\pm0.21$ & $\bf{0.35\pm0.17}$ & $\bf{0.21\pm0.14}$\\
        \hline
        $\Delta \log \mathrm{SFR}_\mathrm{10}$ & Normalised true & $0.00\pm0.01$ & $0.01\pm0.01$ & $0.01\pm0.10$ & $0.04\pm0.11$ & $-0.07\pm0.15$ & $-0.05\pm0.16$ & $-0.05\pm0.15$\\
          & Double power law & $-0.14\pm0.12$ & $-0.20\pm0.11$ & $-0.27\pm0.27$ & $-0.31\pm0.27$ & $-0.41\pm0.26$ & $-0.42\pm0.25$ & $-0.37\pm0.24$\\
          & Flexible & $-0.07\pm0.16$ & $-0.10\pm0.12$ & $-0.15\pm0.28$ & $-0.20\pm0.30$ & $-0.35\pm0.27$ & $\bf{-0.38\pm0.26}$ & $\bf{-0.27\pm0.24}$\\
        \hline
        $\Delta \log \mathrm{SFR}_\mathrm{100}$ & Normalised true & $0.01\pm0.01$ & $0.01\pm0.01$ & $0.02\pm0.10$ & $0.05\pm0.11$ & $-0.07\pm0.15$ & $-0.05\pm0.16$ & $-0.04\pm0.15$\\
          & Constant & $0.36\pm0.20$ & $0.47\pm0.16$ & $0.37\pm0.21$ & $0.38\pm0.28$ & $0.38\pm0.26$ & $0.34\pm0.34$ & $0.31\pm0.31$\\
          & Double power law & $0.25\pm0.17$ & $0.28\pm0.17$ & $0.18\pm0.21$ & $0.15\pm0.24$ & $0.09\pm0.26$ & $\bf{0.08\pm0.25}$ & $0.12\pm0.24$\\
          & Flexible & $0.13\pm0.14$ & $0.17\pm0.14$ & $0.09\pm0.18$ & $0.13\pm0.19$ & $0.12\pm0.22$ & $0.13\pm0.23$ & $\bf{0.08\pm0.20}$\\
        \hline
    \end{tabular}
    }
    \tablefoot{The table shows the 16th, 50th, and 84th percentiles. $\Delta \mathrm{SFR}_\mathrm{10}$ for the constant SFH model is not shown, as $\mathrm{SFR}_\mathrm{10,fit} =0$ in many galaxies. Values with the minimum offset in \texttt{attSNE\_Z} and \texttt{attSNE\_Z\_F410M} are shown in bold.
    }
    \label{tab:summary_table2}
\end{table*}

\begin{table*}
\caption{ Same as Table~\ref{tab:summary_table2} but for high-mass ($>10^8\,\msun$) galaxies.}
    \renewcommand{\arraystretch}{1.5}
    \centering
    \resizebox{\textwidth}{!}{
    \begin{tabular}{llccccccc}
    \hline \hline
        Variable & SFH model & \texttt{intS\_Z$_\mathrm{true}$} & \texttt{intS\_Z} & \texttt{attS\_Z$_\mathrm{true}$} & \texttt{attS\_Z} & \texttt{attSNE\_Z$_\mathrm{true}$} & \texttt{attSNE\_Z} & \texttt{attSNE\_Z\_F410M}\\
        \hline  
        $\Delta \log M_\mathrm{\star}$ & Normalised true & $0.01\pm0.01$ & $0.02\pm0.01$ & $-0.01\pm0.14$ & $0.02\pm0.15$ & $-0.08\pm0.18$ & $-0.06\pm0.19$ & $-0.04\pm0.17$\\
          & Constant & $-0.07\pm0.08$ & $0.19\pm0.07$ & $-0.05\pm0.13$ & $0.17\pm0.12$ & $0.01\pm0.16$ & $0.23\pm0.15$ & $0.22\pm0.14$\\
          & Double power law & $-0.00\pm0.09$ & $0.20\pm0.06$ & $0.05\pm0.12$ & $0.19\pm0.12$ & $0.08\pm0.15$ & $0.17\pm0.15$ & $0.17\pm0.14$\\
          & Flexible & $0.06\pm0.06$ & $0.18\pm0.06$ & $0.05\pm0.11$ & $0.18\pm0.12$ & $0.05\pm0.15$ & $\bf{0.18\pm0.13}$ & $\bf{0.19\pm0.12}$\\
        \hline
        $\Delta \log \mathrm{SFR}_\mathrm{10}$ & Normalised true & $0.00\pm0.01$ & $0.01\pm0.01$ & $-0.01\pm0.14$ & $0.02\pm0.15$ & $-0.08\pm0.18$ & $-0.06\pm0.19$ & $-0.05\pm0.17$\\
          & Double power law & $-0.02\pm0.14$ & $-0.09\pm0.14$ & $-0.20\pm0.31$ & $-0.23\pm0.31$ & $-0.29\pm0.31$ & $\bf{-0.30\pm0.31}$ & $\bf{-0.29\pm0.31}$\\
          & Flexible & $0.02\pm0.12$ & $-0.06\pm0.11$ & $-0.14\pm0.33$ & $-0.27\pm0.35$ & $-0.28\pm0.32$ & $-0.41\pm0.34$ & $-0.41\pm0.34$\\
        \hline
        $\Delta \log \mathrm{SFR}_\mathrm{100}$ & Normalised true & $0.01\pm0.01$ & $0.01\pm0.01$ & $-0.01\pm0.14$ & $0.02\pm0.15$ & $-0.08\pm0.18$ & $-0.06\pm0.19$ & $-0.05\pm0.17$\\
          & Constant & $0.24\pm0.16$ & $0.20\pm0.19$ & $0.11\pm0.28$ & $-0.10\pm0.39$ & $0.07\pm0.27$ & $\bf{-0.15\pm0.41}$ & $\bf{-0.13\pm0.42}$\\
          & Double power law & $0.05\pm0.17$ & $0.00\pm0.18$ & $-0.11\pm0.25$ & $-0.12\pm0.24$ & $-0.20\pm0.24$ & $-0.21\pm0.25$ & $-0.21\pm0.25$\\
          & Flexible & $-0.00\pm0.14$ & $-0.01\pm0.15$ & $-0.06\pm0.24$ & $-0.10\pm0.26$ & $-0.13\pm0.24$ & $-0.17\pm0.26$ & $-0.16\pm0.26$\\
        \hline
    \end{tabular}
    }
    
    \label{tab:summary_table3}
\end{table*}

\label{LastPage}

\end{document}